\documentclass[fleqn,usenatbib]{mnras}

\usepackage[T1]{fontenc}

\DeclareRobustCommand{\VAN}[3]{#2}
\let\VANthebibliography\thebibliography
\def\thebibliography{\DeclareRobustCommand{\VAN}[3]{##3}\VANthebibliography}

\usepackage[utf8]{inputenc}
\usepackage{xcolor}
\usepackage{graphicx}
\usepackage{siunitx}
\usepackage[acronym,nomain]{glossaries}        
\usepackage{amsmath}
\usepackage{amssymb}
\usepackage{array}
\usepackage{lineno}
\usepackage{multirow}
\usepackage{enumitem}
\usepackage{xspace}  
\usepackage{aas_macros}  
\usepackage{newtxtext,newtxmath}

\newacronym{lmc}{LMC}{Large Magellanic Cloud}
\newacronym{snr}{SNR}{supernova remnant}
\newacronym{isnr}{iSNR}{interacting supernova remnant}
\newacronym{sn}{SN}{supernova}
\newacronym{sne}{SNe}{supernovae}
\newacronym{sb}{SB}{superbubble}
\newacronym[plural=PWNe,firstplural=pulsar wind nebulae (PWNe)]{pwn}{PWN}{pulsar wind nebula}
\newacronym{vhe}{VHE}{very-high-energy}
\newacronym{he}{HE}{high-energy}
\newacronym{cta}{CTA}{Cherenkov Telescope Array}
\newacronym{sfh}{SFH}{Star Formation History}
\newacronym{sfr}{SFR}{star-forming region}
\newacronym{ms}{MS}{Magellanic System}
\newacronym{smc}{SMC}{Small Magellanic Cloud}
\newacronym{mw}{MW}{Milky Way}
\newacronym{ism}{ISM}{interstellar medium}
\newacronym{ulirgs}{ULIRGs}{Ultraluminous Infrared Galaxies}
\newacronym{ic}{IC}{Inverse Compton}
\newacronym{cmb}{CMB}{Cosmic Microwave Background}
\newacronym{sm}{SM}{Standard Model}
\newacronym{dm}{DM}{Dark Matter}
\newacronym{wimp}{WIMPs}{Weakly Interacting Massive Particles}
\newacronym{gc}{GC}{Galactic Center}
\newacronym{iact}{IACT}{Imaging Atmospheric Cherenkov Telescope}
\newacronym{ksp}{KSP}{Key Science Project}
\newacronym{fov}{FoV}{Field of View}
\newacronym{roi}{ROI}{Region of Interest}
\newacronym{irf}{IRF}{Instrument Response Function}
\newacronym{cl}{CL}{Confidence Level}
\newacronym{mc}{MC}{Monte Carlo}
\newacronym{nfw}{NFW}{Navarro-Frenk-White}
\newacronym{cr}{CR}{cosmic-ray}
\newacronym{crs}{CRs}{cosmic rays}
\newacronym{dsphe}{DS}{dwarf spheroidals}
\newacronym{isrf}{ISRF}{interstellar radiation field}
\newacronym{ts}{TS}{test statistic}
\newacronym{sed}{SED}{spectral energy distribution}
\newacronym{dsa}{DSA}{diffusive shock acceleration}
\newacronym{hawc}{HAWC}{High-Altitude Water Cherenkov Observatory}
\newacronym{lhaaso}{LHAASO}{Large High-Altitude Air Shower Observatory}
\newacronym{askap}{ASKAP}{Australian Square Kilometre Array Pathfinder}

\newcommand{\sigv}{\langle \sigma v \rangle}
\newcommand{\mdm}{m_{\chi}}

\newcommand{\RomNumCap}[1]{\MakeUppercase{\romannumeral #1}}
\newcommand{\dop}[1]{\ensuremath{\operatorname{d}\!{#1}}}

\newcommand{\dunit}{\,cm$^{2}$\,s$^{-1}$\xspace}

\newcommand{\feunit}{\,erg\,cm$^{-2}$\,s$^{-1}$\xspace}

\newcommand{\kms}{\,km\,s$^{-1}$\xspace}
\newcommand{\punit}{\,erg\,s$^{-1}$\xspace}

\newcommand{\eunit}{\,erg\xspace}
\newcommand{\vunit}{\,cm$^{-3}$\xspace}

\newcommand{\nunit}{\,H\,cm$^{-3}$\xspace}
\newcommand{\cunit}{\,H\,cm$^{-2}$\xspace}
\newcommand{\bunit}{\,$\mu$G\xspace}
\newcommand{\msol}{\,M$_{\odot}$\xspace}

\newcommand{\snrate}{\,SN\,yr$^{-1}$\xspace}

\newcommand{\mev}{\,MeV\xspace}
\newcommand{\gev}{\,GeV\xspace}
\newcommand{\tev}{\,TeV\xspace}
\newcommand{\pev}{\,PeV\xspace}
\def\deg{\ensuremath{^\circ}}

\title[Sensitivity of the CTA to emission from the LMC]{\centering Sensitivity of the Cherenkov Telescope Array \\to TeV photon emission from the Large Magellanic Cloud}

\author[A.~Acharyya et al]{\parbox{\textwidth}{\large\centering%
  A.~Acharyya$^{\ref{AFFIL::UAlabamaTuscaloosa}}$
  R.~Adam$^{\ref{AFFIL::OCotedAzur},\ref{AFFIL::LLREcolePolytechnique}}$
  A.~Aguasca-Cabot$^{\ref{AFFIL::ICCUB}}$
  I.~Agudo$^{\ref{AFFIL::IAACSIC}}$
  A.~Aguirre-Santaella$^{\ref{AFFIL::IFTUAMCSIC}}$
  J.~Alfaro$^{\ref{AFFIL::UPontificiaCatolicadeChile}}$
  R.~Aloisio$^{\ref{AFFIL::GSSIandINFNAquila}}$
  R.~Alves~Batista$^{\ref{AFFIL::IFTUAMCSIC}}$
  E.~Amato$^{\ref{AFFIL::OArcetri}}$
  E.~O.~Ang\"uner$^{\ref{AFFIL::Tubitak}}$
  C.~Aramo$^{\ref{AFFIL::INFNNapoli}}$
  C.~Arcaro$^{\ref{AFFIL::UPadovaandINFN}}$
  K.~Asano$^{\ref{AFFIL::UTokyoICRR}}$
  J.~Aschersleben$^{\ref{AFFIL::UGroningen}}$
  H.~Ashkar$^{\ref{AFFIL::LLREcolePolytechnique}}$
  M.~Backes$^{\ref{AFFIL::UNamibia},\ref{AFFIL::NWU}}$
  A.~Baktash$^{\ref{AFFIL::UHamburg}}$
  C.~Balazs$^{\ref{AFFIL::UMonash}}$
  M.~Balbo$^{\ref{AFFIL::UGenevaISDC}}$
  J.~Ballet$^{\ref{AFFIL::CEAIRFUDAp}}$
  A.~Bamba$^{\ref{AFFIL::UTokyoGSS},\ref{AFFIL::UTokyoRCEUSS}}$
  A.~Baquero~Larriva$^{\ref{AFFIL::UCMAltasEnergias},\ref{AFFIL::UAzuay}}$
  V.~Barbosa~Martins$^{\ref{AFFIL::DESY}}$
  U.~Barres~de~Almeida$^{\ref{AFFIL::CBPF}}$
  J.~A.~Barrio$^{\ref{AFFIL::UCMAltasEnergias}}$
  D.~Bastieri$^{\ref{AFFIL::UPadovaandINFN}}$
  P.~Batista$^{\ref{AFFIL::DESY}}$
  I.~Batkovic$^{\ref{AFFIL::UPadovaandINFN}}$
  J.~R.~Baxter$^{\ref{AFFIL::UTokyoICRR}}$
  J.~Becerra~Gonz\'alez$^{\ref{AFFIL::IAC}}$
  J.~Becker~Tjus$^{\ref{AFFIL::UBochum}}$
  W.~Benbow$^{\ref{AFFIL::CfAHarvardSmithsonian}}$
  E.~Bernardini$^{\ref{AFFIL::UPadovaandINFN}}$
  M.~I.~Bernardos~Mart{\'\i}n$^{\ref{AFFIL::IAACSIC}}$
  J.~Bernete~Medrano$^{\ref{AFFIL::CIEMAT}}$
  A.~Berti$^{\ref{AFFIL::MPP}}$
  B.~Bertucci$^{\ref{AFFIL::UPerugiaandINFN}}$
  V.~Beshley$^{\ref{AFFIL::IAPMMLviv}}$
  P.~Bhattacharjee$^{\ref{AFFIL::LAPPUSavoieMontBlanc}}$
  S.~Bhattacharyya$^{\ref{AFFIL::UNovaGoricaCAC}}$
  C.~Bigongiari$^{\ref{AFFIL::ORoma}}$
  A.~Biland$^{\ref{AFFIL::ETHZurich}}$
  E.~Bissaldi$^{\ref{AFFIL::INFNBari},\ref{AFFIL::PolitecnicoBari}}$
  F.~Bocchino$^{\ref{AFFIL::OPalermo}}$
  P.~Bordas$^{\ref{AFFIL::ICCUB}}$
  J.~Borkowski$^{\ref{AFFIL::NicolausCopernicusAstronomicalCenter}}$
  E.~Bottacini$^{\ref{AFFIL::UPadovaandINFN}}$
  M.~B\"ottcher$^{\ref{AFFIL::NWU}}$
  F.~Bradascio$^{\ref{AFFIL::CEAIRFUDPhP}}$
  A.~M.~Brown$^{\ref{AFFIL::UDurham}}$
  A.~Bulgarelli$^{\ref{AFFIL::OASBologna}}$
  L.~Burmistrov$^{\ref{AFFIL::UGenevaDPNC}}$
  S.~Caroff$^{\ref{AFFIL::LAPPUSavoieMontBlanc}}$
  A.~Carosi$^{\ref{AFFIL::ORoma}}$
  E.~Carqu{\'\i}n$^{\ref{AFFIL::UTecnicaFedericoSantaMaria}}$
  S.~Casanova$^{\ref{AFFIL::IFJ}}$
  E.~Cascone$^{\ref{AFFIL::OCapodimonte}}$
  F.~Cassol$^{\ref{AFFIL::CPPMUAixMarseille}}$
  M.~Cerruti$^{\ref{AFFIL::APCUParisCite}}$
  P.~Chadwick$^{\ref{AFFIL::UDurham}}$
  S.~Chaty$^{\ref{AFFIL::APCUParisCite}}$
  A.~Chen$^{\ref{AFFIL::UWitwatersrand}}$
  A.~Chiavassa$^{\ref{AFFIL::INFNTorino},\ref{AFFIL::UTorino}}$
  L.~Chytka$^{\ref{AFFIL::UOlomouc}}$
  V.~Conforti$^{\ref{AFFIL::OASBologna}}$
  J.~Cortina$^{\ref{AFFIL::CIEMAT}}$
  A.~Costa$^{\ref{AFFIL::OCatania}}$
  H.~Costantini$^{\ref{AFFIL::CPPMUAixMarseille}}$
  G.~Cotter$^{\ref{AFFIL::UOxford}}$
  S.~Crestan$^{\ref{AFFIL::IASFMilano}}$
  P.~Cristofari$^{\ref{AFFIL::ObservatoiredeParis}}$
  F.~D'Ammando$^{\ref{AFFIL::RadioastronomiaINAF}}$
  M.~Dalchenko$^{\ref{AFFIL::UGenevaDPNC}}$
  F.~Dazzi$^{\ref{AFFIL::INAF}}$
  A.~De~Angelis$^{\ref{AFFIL::UPadovaandINFN}}$
  V.~De~Caprio$^{\ref{AFFIL::OCapodimonte}}$
  E.~M.~de~Gouveia~Dal~Pino$^{\ref{AFFIL::IAGUSaoPaulo}}$
  D.~De~Martino$^{\ref{AFFIL::OCapodimonte}}$
  M.~de~Naurois$^{\ref{AFFIL::LLREcolePolytechnique}}$
  V.~de~Souza$^{\ref{AFFIL::IFSCUSaoPaulo}}$
  M.~V.~del~Valle$^{\ref{AFFIL::IAGUSaoPaulo}}$
  A.~G.~Delgado~Giler$^{\ref{AFFIL::IFSCUSaoPaulo},\ref{AFFIL::UGroningen}}$
  C.~Delgado$^{\ref{AFFIL::CIEMAT}}$
  D.~della~Volpe$^{\ref{AFFIL::UGenevaDPNC}}$
  D.~Depaoli$^{\ref{AFFIL::MPIK}}$
  T.~Di~Girolamo$^{\ref{AFFIL::UNapoli},\ref{AFFIL::INFNNapoli}}$
  A.~Di~Piano$^{\ref{AFFIL::OASBologna}}$
  F.~Di~Pierro$^{\ref{AFFIL::INFNTorino}}$
  R.~Di~Tria$^{\ref{AFFIL::UandINFNBari}}$
  L.~Di~Venere$^{\ref{AFFIL::INFNBari}}$
  S.~Diebold$^{\ref{AFFIL::IAAT}}$
  M.~Doro$^{\ref{AFFIL::UPadovaandINFN}}$
  D.~Dumora$^{\ref{AFFIL::LP2I}}$
  V.~V.~Dwarkadas$^{\ref{AFFIL::UChicagoDAA}}$
  C.~Eckner$^{\ref{AFFIL::LAPPUSavoieMontBlanc},\ref{AFFIL::LAPTh}}$
  K.~Egberts$^{\ref{AFFIL::UPotsdam}}$
  G.~Emery$^{\ref{AFFIL::CPPMUAixMarseille}}$
  J.~Escudero$^{\ref{AFFIL::IAACSIC}}$
  D.~Falceta-Goncalves$^{\ref{AFFIL::EACHUSaoPaulo}}$
  E.~Fedorova$^{\ref{AFFIL::ORoma},\ref{AFFIL::AstObsofUKyiv}}$
  S.~Fegan$^{\ref{AFFIL::LLREcolePolytechnique}}$
  Q.~Feng$^{\ref{AFFIL::CfAHarvardSmithsonian}}$
  D.~Ferenc$^{\ref{AFFIL::UCaliforniaDavis}}$
  G.~Ferrand$^{\ref{AFFIL::RIKEN}}$
  E.~Fiandrini$^{\ref{AFFIL::UPerugiaandINFN}}$
  M.~Filipovic$^{\ref{AFFIL::UWesternSydney}}$
  V.~Fioretti$^{\ref{AFFIL::OASBologna}}$
  L.~Foffano$^{\ref{AFFIL::IAPS}}$
  G.~Fontaine$^{\ref{AFFIL::LLREcolePolytechnique}}$
  Y.~Fukui$^{\ref{AFFIL::UNagoya}}$
  D.~Gaggero$^{\ref{AFFIL::INFNPisa}}$
  G.~Galanti$^{\ref{AFFIL::IASFMilano}}$
  G.~Galaz$^{\ref{AFFIL::UPontificiaCatolicadeChile}}$
  S.~Gallozzi$^{\ref{AFFIL::ORoma}}$
  V.~Gammaldi$^{\ref{AFFIL::IFTUAMCSIC}}$
  M.~Garczarczyk$^{\ref{AFFIL::DESY}}$
  C.~Gasbarra$^{\ref{AFFIL::INFNRomaTorVergata}}$
  D.~Gasparrini$^{\ref{AFFIL::INFNRomaTorVergata}}$
  A.~Ghalumyan$^{\ref{AFFIL::NSLAlikhanyan}}$
  M.~Giarrusso$^{\ref{AFFIL::INFNCatania}}$
  G.~Giavitto$^{\ref{AFFIL::DESY}}$
  N.~Giglietto$^{\ref{AFFIL::INFNBari},\ref{AFFIL::PolitecnicoBari}}$
  F.~Giordano$^{\ref{AFFIL::UandINFNBari}}$
  A.~Giuliani$^{\ref{AFFIL::IASFMilano}}$
  J.-F.~Glicenstein$^{\ref{AFFIL::CEAIRFUDPhP}}$
  P.~Goldoni$^{\ref{AFFIL::APCUParisCite}}$
  J.~Goulart~Coelho$^{\ref{AFFIL::UFPR},\ref{AFFIL::UFES}}$
  J.~Granot$^{\ref{AFFIL::OpenUniversityofIsrael},\ref{AFFIL::GWUWashingtonDC}}$
  D.~Green$^{\ref{AFFIL::MPP}}$
  J.~G.~Green$^{\ref{AFFIL::MPP}}$
  M.-H.~Grondin$^{\ref{AFFIL::LP2I}}$
  O.~Gueta$^{\ref{AFFIL::DESY}}$
  D.~Hadasch$^{\ref{AFFIL::UTokyoICRR}}$
  P.~Hamal$^{\ref{AFFIL::FZU}}$
  T.~Hassan$^{\ref{AFFIL::CIEMAT}}$
  K.~Hayashi$^{\ref{AFFIL::NITIchinoseki},\ref{AFFIL::UTokyoICRR}}$
  M.~Heller$^{\ref{AFFIL::UGenevaDPNC}}$
  S.~Hern\'andez~Cadena$^{\ref{AFFIL::UNAMMexico}}$
  N.~Hiroshima$^{\ref{AFFIL::RIKEN}}$
  B.~Hnatyk$^{\ref{AFFIL::AstObsofUKyiv}}$
  R.~Hnatyk$^{\ref{AFFIL::AstObsofUKyiv}}$
  W.~Hofmann$^{\ref{AFFIL::MPIK}}$
  J.~Holder$^{\ref{AFFIL::UDelaware}}$
  M.~Holler$^{\ref{AFFIL::UInnsbruck}}$
  D.~Horan$^{\ref{AFFIL::LLREcolePolytechnique}}$
  P.~Horvath$^{\ref{AFFIL::UOlomouc}}$
  M.~Hrabovsky$^{\ref{AFFIL::UOlomouc}}$
  M.~H\"utten$^{\ref{AFFIL::UTokyoICRR}}$
  M.~Iarlori$^{\ref{AFFIL::UandINFNAquila}}$
  T.~Inada$^{\ref{AFFIL::UTokyoICRR}}$
  F.~Incardona$^{\ref{AFFIL::OCatania}}$
  S.~Inoue$^{\ref{AFFIL::RIKEN}}$
  F.~Iocco$^{\ref{AFFIL::UNapoli},\ref{AFFIL::INFNNapoli}}$
  M.~Jamrozy$^{\ref{AFFIL::UJagiellonian}}$
  W.~Jin$^{\ref{AFFIL::UAlabamaTuscaloosa}}$
  I.~Jung-Richardt$^{\ref{AFFIL::UErlangenECAP}}$
  J.~Jury\v{s}ek$^{\ref{AFFIL::FZU}}$
  D.~Kantzas$^{\ref{AFFIL::LAPPUSavoieMontBlanc}}$
  V.~Karas$^{\ref{AFFIL::ASU}}$
  H.~Katagiri$^{\ref{AFFIL::UIbaraki}}$
  D.~Kerszberg$^{\ref{AFFIL::IFAEBIST}}$
  J.~Kn\"odlseder$^{\ref{AFFIL::IRAPUToulouse}}$
  N.~Komin$^{\ref{AFFIL::UWitwatersrand}}$
  P.~Kornecki$^{\ref{AFFIL::ObservatoiredeParis}}$
  K.~Kosack$^{\ref{AFFIL::CEAIRFUDAp}}$
  G.~Kowal$^{\ref{AFFIL::EACHUSaoPaulo}}$
  H.~Kubo$^{\ref{AFFIL::UTokyoICRR}}$
  A.~Lamastra$^{\ref{AFFIL::ORoma}}$
  J.~Lapington$^{\ref{AFFIL::ULeicester}}$
  M.~Lemoine-Goumard$^{\ref{AFFIL::LP2I}}$
  J.-P.~Lenain$^{\ref{AFFIL::LPNHEUSorbonne}}$
  F.~Leone$^{\ref{AFFIL::UCatania}}$
  G.~Leto$^{\ref{AFFIL::OCatania}}$
  F.~Leuschner$^{\ref{AFFIL::IAAT}}$
  E.~Lindfors$^{\ref{AFFIL::UTurku}}$
  T.~Lohse$^{\ref{AFFIL::UBerlin}}$
  S.~Lombardi$^{\ref{AFFIL::ORoma}}$
  F.~Longo$^{\ref{AFFIL::UandINFNTrieste}}$
  R.~L\'opez-Coto$^{\ref{AFFIL::IAACSIC}}$
  A.~L\'opez-Oramas$^{\ref{AFFIL::IAC}}$
  S.~Loporchio$^{\ref{AFFIL::INFNBari}}$
  P.~L.~Luque-Escamilla$^{\ref{AFFIL::UJaen}}$
  O.~Macias$^{\ref{AFFIL::UAmsterdam}}$
  P.~Majumdar$^{\ref{AFFIL::SahaInstitute}}$
  D.~Mandat$^{\ref{AFFIL::FZU}}$
  S.~Mangano$^{\ref{AFFIL::CIEMAT}}$
  G.~Manic\`o$^{\ref{AFFIL::INFNCatania}}$
  M.~Mariotti$^{\ref{AFFIL::UPadovaandINFN}}$
  P.~Marquez$^{\ref{AFFIL::IFAEBIST}}$
  G.~Marsella$^{\ref{AFFIL::UPalermo},\ref{AFFIL::INFNCatania}}$
  J.~Mart{\'\i}$^{\ref{AFFIL::UJaen}}$
  P.~Martin$^{\ref{AFFIL::IRAPUToulouse}}$
  M.~Mart{\'\i}nez$^{\ref{AFFIL::IFAEBIST}}$
  D.~Mazin$^{\ref{AFFIL::UTokyoICRR},\ref{AFFIL::MPP}}$
  S.~Menchiari$^{\ref{AFFIL::USienaandINFN}}$
  D.~M.-A.~Meyer$^{\ref{AFFIL::UPotsdam}}$
  D.~Miceli$^{\ref{AFFIL::UPadovaandINFN}}$
  M.~Miceli$^{\ref{AFFIL::UPalermo},\ref{AFFIL::OPalermo}}$
  J.~Micha{\l}owski$^{\ref{AFFIL::IFJ}}$
  A.~Mitchell$^{\ref{AFFIL::UErlangenECAP}}$
  R.~Moderski$^{\ref{AFFIL::NicolausCopernicusAstronomicalCenter}}$
  L.~Mohrmann$^{\ref{AFFIL::MPIK}}$
  M.~Molero$^{\ref{AFFIL::IAC}}$
  E.~Molina$^{\ref{AFFIL::ICCUB}}$
  T.~Montaruli$^{\ref{AFFIL::UGenevaDPNC}}$
  A.~Moralejo$^{\ref{AFFIL::IFAEBIST}}$
  D.~Morcuende$^{\ref{AFFIL::UCMAltasEnergias}}$
  A.~Morselli$^{\ref{AFFIL::INFNRomaTorVergata}}$
  E.~Moulin$^{\ref{AFFIL::CEAIRFUDPhP}}$
  V.~Moya$^{\ref{AFFIL::UCMAltasEnergias}}$
  R.~Mukherjee$^{\ref{AFFIL::BarnardCollegeColumbiaUniversity}}$
  K.~Munari$^{\ref{AFFIL::OCatania}}$
  A.~Muraczewski$^{\ref{AFFIL::NicolausCopernicusAstronomicalCenter}}$
  S.~Nagataki$^{\ref{AFFIL::RIKEN}}$
  T.~Nakamori$^{\ref{AFFIL::UYamagata}}$
  A.~Nayak$^{\ref{AFFIL::UDurham}}$
  J.~Niemiec$^{\ref{AFFIL::IFJ}}$
  M.~Nievas$^{\ref{AFFIL::IAC}}$
  M.~Niko{\l}ajuk$^{\ref{AFFIL::UBiaystok}}$
  K.~Nishijima$^{\ref{AFFIL::UTokai}}$
  K.~Noda$^{\ref{AFFIL::UTokyoICRR}}$
  D.~Nosek$^{\ref{AFFIL::UPrague}}$
  B.~Novosyadlyj$^{\ref{AFFIL::AstObsofULviv}}$
  S.~Nozaki$^{\ref{AFFIL::MPP}}$
  M.~Ohishi$^{\ref{AFFIL::UTokyoICRR}}$
  S.~Ohm$^{\ref{AFFIL::DESY}}$
  A.~Okumura$^{\ref{AFFIL::UNagoyaISEE},\ref{AFFIL::UNagoyaKMI}}$
  B.~Olmi$^{\ref{AFFIL::OPalermo},\ref{AFFIL::OArcetri}}$
  R.~A.~Ong$^{\ref{AFFIL::UCLA}}$
  M.~Orienti$^{\ref{AFFIL::RadioastronomiaINAF}}$
  R.~Orito$^{\ref{AFFIL::UTokushima}}$
  M.~Orlandini$^{\ref{AFFIL::OASBologna}}$
  E.~Orlando$^{\ref{AFFIL::UandINFNTrieste}}$
  S.~Orlando$^{\ref{AFFIL::OPalermo}}$
  M.~Ostrowski$^{\ref{AFFIL::UJagiellonian}}$
  I.~Oya$^{\ref{AFFIL::CTAOHeidelberg}}$
  A.~Pagliaro$^{\ref{AFFIL::IASFPalermo}}$
  M.~Palatka$^{\ref{AFFIL::FZU}}$
  F.~R.~Pantaleo$^{\ref{AFFIL::INFNBari},\ref{AFFIL::PolitecnicoBari}}$
  R.~Paoletti$^{\ref{AFFIL::USienaandINFN}}$
  J.~M.~Paredes$^{\ref{AFFIL::ICCUB}}$
  N.~Parmiggiani$^{\ref{AFFIL::OASBologna}}$
  B.~Patricelli$^{\ref{AFFIL::ORoma},\ref{AFFIL::UPisa}}$
  M.~Pech$^{\ref{AFFIL::FZU}}$
  M.~Pecimotika$^{\ref{AFFIL::URijeka}}$
  M.~Persic$^{\ref{AFFIL::OPadova},\ref{AFFIL::UandINFNTrieste}}$
  O.~Petruk$^{\ref{AFFIL::IAPMMLviv}}$
  E.~Pierre$^{\ref{AFFIL::LPNHEUSorbonne}}$
  E.~Pietropaolo$^{\ref{AFFIL::UandINFNAquila}}$
  G.~Pirola$^{\ref{AFFIL::MPP}}$
  M.~Pohl$^{\ref{AFFIL::UPotsdam}}$
  E.~Prandini$^{\ref{AFFIL::UPadovaandINFN}}$
  C.~Priyadarshi$^{\ref{AFFIL::IFAEBIST}}$
  G.~P\"uhlhofer$^{\ref{AFFIL::IAAT}}$
  M.~L.~Pumo$^{\ref{AFFIL::UCatania},\ref{AFFIL::INFNCatania}}$
  M.~Punch$^{\ref{AFFIL::APCUParisCite}}$
  F.~S.~Queiroz$^{\ref{AFFIL::URioGrandedoNorteIIP},\ref{AFFIL::URioGrandedoNortePhys}}$
  A.~Quirrenbach$^{\ref{AFFIL::LSW}}$
  S.~Rain\`o$^{\ref{AFFIL::UandINFNBari}}$
  R.~Rando$^{\ref{AFFIL::UPadovaandINFN}}$
  S.~Razzaque$^{\ref{AFFIL::UJohannesburg}}$
  A.~Reimer$^{\ref{AFFIL::UInnsbruck}}$
  O.~Reimer$^{\ref{AFFIL::UInnsbruck}}$
  T.~Reposeur$^{\ref{AFFIL::LP2I}}$
  M.~Rib\'o$^{\ref{AFFIL::ICCUB}}$
  T.~Richtler$^{\ref{AFFIL::UdeConcepcion}}$
  J.~Rico$^{\ref{AFFIL::IFAEBIST}}$
  F.~Rieger$^{\ref{AFFIL::MPIK}}$
  M.~Rigoselli$^{\ref{AFFIL::IASFMilano}}$
  V.~Rizi$^{\ref{AFFIL::UandINFNAquila}}$
  E.~Roache$^{\ref{AFFIL::CfAHarvardSmithsonian}}$
  G.~Rodriguez~Fernandez$^{\ref{AFFIL::INFNRomaTorVergata}}$
  P.~Romano$^{\ref{AFFIL::OBrera}}$
  G.~Romeo$^{\ref{AFFIL::OCatania}}$
  J.~Rosado$^{\ref{AFFIL::UCMAltasEnergias}}$
  A.~Rosales~de~Leon$^{\ref{AFFIL::UDurham}}$
  B.~Rudak$^{\ref{AFFIL::NicolausCopernicusAstronomicalCenter}}$
  C.~Rulten$^{\ref{AFFIL::UDurham}}$
  I.~Sadeh$^{\ref{AFFIL::DESY}}$
  T.~Saito$^{\ref{AFFIL::UTokyoICRR}}$
  M.~S\'anchez-Conde$^{\ref{AFFIL::IFTUAMCSIC}}$
  H.~Sano$^{\ref{AFFIL::UTokyoICRR}}$
  A.~Santangelo$^{\ref{AFFIL::IAAT}}$
  R.~Santos-Lima$^{\ref{AFFIL::IAGUSaoPaulo}}$
  S.~Sarkar$^{\ref{AFFIL::UOxford}}$
  F.~G.~Saturni$^{\ref{AFFIL::ORoma}}$
  A.~Scherer$^{\ref{AFFIL::UPontificiaCatolicadeChile}}$
  P.~Schovanek$^{\ref{AFFIL::FZU}}$
  F.~Schussler$^{\ref{AFFIL::CEAIRFUDPhP}}$
  U.~Schwanke$^{\ref{AFFIL::UBerlin}}$
  O.~Sergijenko$^{\ref{AFFIL::AstObsofUKyiv},\ref{AFFIL::ObsNASUkraine},\ref{AFFIL::AGHCracowSTC}}$
  M.~Servillat$^{\ref{AFFIL::ObservatoiredeParis}}$
  H.~Siejkowski$^{\ref{AFFIL::CYFRONETAGH}}$
  C.~Siqueira$^{\ref{AFFIL::IFSCUSaoPaulo}}$
  S.~Spencer$^{\ref{AFFIL::UErlangenECAP},\ref{AFFIL::UOxford}}$
  A.~Stamerra$^{\ref{AFFIL::ORoma},\ref{AFFIL::CTAOBologna}}$
  S.~Stani\v{c}$^{\ref{AFFIL::UNovaGoricaCAC}}$
  C.~Steppa$^{\ref{AFFIL::UPotsdam}}$
  T.~Stolarczyk$^{\ref{AFFIL::CEAIRFUDAp}}$
  Y.~Suda$^{\ref{AFFIL::UHiroshima}}$
  T.~Tavernier$^{\ref{AFFIL::FZU}}$
  M.~Teshima$^{\ref{AFFIL::MPP}}$
  L.~Tibaldo$^{\ref{AFFIL::IRAPUToulouse}}$
  D.~F.~Torres$^{\ref{AFFIL::ICECSIC}}$
  N.~Tothill$^{\ref{AFFIL::UWesternSydney}}$
  M.~Vacula$^{\ref{AFFIL::UOlomouc}}$
  B.~Vallage$^{\ref{AFFIL::CEAIRFUDPhP}}$
  P.~Vallania$^{\ref{AFFIL::INFNTorino},\ref{AFFIL::OTorino}}$
  C.~van~Eldik$^{\ref{AFFIL::UErlangenECAP}}$
  M.~V\'azquez~Acosta$^{\ref{AFFIL::IAC}}$
  M.~Vecchi$^{\ref{AFFIL::UGroningen}}$
  S.~Ventura$^{\ref{AFFIL::USienaandINFN}}$
  S.~Vercellone$^{\ref{AFFIL::OBrera}}$
  A.~Viana$^{\ref{AFFIL::IFSCUSaoPaulo}}$
  C.~F.~Vigorito$^{\ref{AFFIL::INFNTorino},\ref{AFFIL::UTorino}}$
  J.~Vink$^{\ref{AFFIL::UAmsterdam}}$
  V.~Vitale$^{\ref{AFFIL::INFNRomaTorVergata}}$
  V.~Vodeb$^{\ref{AFFIL::UNovaGoricaCAC}}$
  S.~Vorobiov$^{\ref{AFFIL::UNovaGoricaCAC}}$
  T.~Vuillaume$^{\ref{AFFIL::LAPPUSavoieMontBlanc}}$
  S.~J.~Wagner$^{\ref{AFFIL::LSW}}$
  R.~Walter$^{\ref{AFFIL::UGenevaISDC}}$
  M.~White$^{\ref{AFFIL::UAdelaide}}$
  A.~Wierzcholska$^{\ref{AFFIL::IFJ}}$
  M.~Will$^{\ref{AFFIL::MPP}}$
  R.~Yamazaki$^{\ref{AFFIL::UAoyamaGakuin}}$
  L.~Yang$^{\ref{AFFIL::UJohannesburg},\ref{AFFIL::USunYatsen}}$
  T.~Yoshikoshi$^{\ref{AFFIL::UTokyoICRR}}$
  M.~Zacharias$^{\ref{AFFIL::LSW}}$
  G.~Zaharijas$^{\ref{AFFIL::UNovaGoricaCAC}}$
  D.~Zavrtanik$^{\ref{AFFIL::UNovaGoricaCAC}}$
  M.~Zavrtanik$^{\ref{AFFIL::UNovaGoricaCAC}}$
  A.~A.~Zdziarski$^{\ref{AFFIL::NicolausCopernicusAstronomicalCenter}}$
  V.~I.~Zhdanov$^{\ref{AFFIL::AstObsofUKyiv}}$
  K.~Zi\k{e}tara$^{\ref{AFFIL::UJagiellonian}}$
  M.~\v{Z}ivec$^{\ref{AFFIL::UNovaGoricaCAC}}$
\newline\newline
\emph{Affiliations can be found at the end of the article}}\\
\vspace{2.5cm}
}
\newpage

\date{Accepted 2023 May 22. Received 2023 May 09; in original form 2023 February 19}

\pubyear{2023}

\begin{document}
\label{firstpage}
\pagerange{\pageref{firstpage}--\pageref{lastpage}}
\maketitle

\begin{abstract}
A deep survey of the Large Magellanic Cloud at $\sim 0.1-100$\tev photon energies with the Cherenkov Telescope Array is planned. 
We assess the detection prospects based on a model for the emission of the galaxy, comprising the four known TeV emitters, mock populations of sources, and interstellar emission on galactic scales. We also assess the detectability of 30~Doradus and SN~1987A, and the constraints that can be derived on the nature of dark matter.
The survey will allow for fine spectral studies of N~157B, N~132D, LMC P3, and 30~Doradus C, and half a dozen other sources should be revealed, mainly pulsar-powered objects. 
The remnant from SN~1987A could be detected if it produces cosmic-ray nuclei with a flat power-law spectrum at high energies, or with a steeper index $2.3-2.4$ pending a flux increase by a factor $>3-4$ over $\sim2015-2035$. 
Large-scale interstellar emission remains mostly out of reach of the survey if its $>10$\gev spectrum has a soft photon index $\sim2.7$, but degree-scale $0.1-10$\tev pion-decay emission could be detected if the cosmic-ray spectrum hardens above $>$100\gev. 
The 30 Doradus star-forming region is detectable if acceleration efficiency is on the order of $1-10$\% of the mechanical luminosity and diffusion is suppressed by two orders of magnitude within $<100$\,pc. 
Finally, the survey could probe the canonical velocity-averaged cross section for self-annihilation of weakly interacting massive particles for cuspy Navarro-Frenk-White profiles.
\end{abstract}




\section{Introduction} \label{sec:intro}

It seems quite rare for a spiral galaxy like our \gls{mw} to be orbited by two star-forming satellites with the size and proximity of the Magellanic Clouds \citep{James:2011, Liu:2011}. It is even more valuable that one of the two is a disk that can be observed at high Galactic latitudes and under favorable inclination \citep{Subramanian:2010,Jacyszyn:2016}. The \gls{lmc} is an extraordinary opportunity for virtually all fields in astrophysics and constitutes a very convenient bridge between detailed studies of the \gls{mw} and surveys of far more distant galaxies.

In the field of high-energy astrophysics, the \gls{lmc} is one of the rare external star-forming galaxies on which spatially resolved studies can be carried out. At both GeV and TeV energies, with the performances of current instruments, only the Magellanic Clouds and Andromeda can be spatially resolved at a level allowing meaningful studies \citep{Abdo:2010d,Abdo:2010e,Ackermann:2016,Ackermann:2017,Acero:2009,Abdalla:2018d}. The \gls{lmc} is among the most interesting because of its proximity and large angular size, low inclination, and relatively high star formation activity. The \gls{lmc} is also home to unique and extraordinary objects -- the most luminous $\rm H\RomNumCap{2}$ region of the Local Group, 30 Doradus, the most powerful pulsar, PSR~J0537-6910, the remnant of the most nearby core-collapse supernova of modern times, SN~1987A -- all of which are either confirmed or expected particle accelerators and gamma-ray emitters. 

The current \gls{he} and \gls{vhe} picture of the \gls{lmc} was revealed by Fermi-LAT and H.E.S.S. observations and features five point sources, three of which are detected in both the GeV and TeV domains: the pulsar PSR~J0537-6910 and its nebula, the supernova remnant N~132D, and the gamma-ray binary LMC P3; the other two are the pulsar PSR~J0540-6919, whose pulsed magnetospheric emission is detected only in the GeV range, and the superbubble 30 Doradus C, detected only in the TeV range \citep{Ackermann:2015,Ackermann:2016,Abramowski:2015}. The \gls{lmc} also exhibits galaxy-scale diffuse emission that is most likely interstellar in origin and arises from the galactic population of \glspl{cr}, on top of which kpc-scale emission components of uncertain origin were observed from regions seemingly devoid of gas \citep{Ackermann:2016}. These extended signals, however, were only detected in the $\sim$100\mev-100\gev range and crucial higher-energy information is missing to build a complete and coherent picture of \glspl{cr} in the \gls{lmc}. Emission in the $\sim$100\gev-100\tev range probes more energetic \glspl{cr} and an earlier stage of their life cycle because the bulk of higher-energy \glspl{cr} can escape the system more easily through more efficient diffusion\footnote{The time scale to diffuse over 1\,kpc is on the order of 1\,Myr for 10\gev particles, and on the order of 100\,kyr for 10\tev particles, assuming a diffusion coefficient as introduced in Eq. \ref{eq:diffcoef}.}.

The future of \gls{vhe} gamma-ray astronomy comprises the \gls{cta}, whose construction recently started. \gls{cta} will be the first observatory in this energy range to be open to the community. It will be deployed on two sites, one in the northern hemisphere, on the island of La Palma, Spain, and the other in the southern hemisphere, in the Atacama desert, Chile. The southern site will give access, among other major targets, to the \gls{lmc}, which other recent experiments such as the \gls{hawc} or the \gls{lhaaso} do not. In its final configuration, \gls{cta} will be an order of magnitude more sensitive than the current generation of \gls{iact} observatories, over a larger energy range from 20\gev to 200\tev, and with enhanced energy and angular resolution \citep{Acharya:2019}. Thanks to a larger field of view, the instrument will have a survey capability that will be exploited in several ambitious \glspl{ksp} led by the \gls{cta} Consortium on proprietary time \citep{Acharya:2019}. One such project is a deep survey of the \gls{lmc}. It will consist of two phases: over the first 4 years, a scan of the whole galaxy for a total of 340h of observations, which corresponds to about 250h of effective exposure; then, over the following 6 years, a long-term monitoring of SN~1987A for about 150h, if it was detected in phase one. 

The scientific objectives are as many as a star-forming galaxy can offer: population studies of different classes of objects, analyses of the interstellar medium and the population of galactic \glspl{cr}, and indirect searches of \gls{dm}. More specifically, the questions that gave rise to the project and their context are as follows:

\textit{\gls{cr} lifecycle}: What are the properties of \glspl{cr} in the \gls{lmc} at the galaxy scale, as revealed by their gamma-ray interstellar emission? The \gls{lmc} is a different galactic setting compared to the \gls{mw} (different geometry, metallicity, star-formation rate density), and thus constitutes an opportunity to test our understanding of the way a \gls{cr} population builds up over long times and large scales, and whether the conditions of \gls{cr} transport differ from those inferred for our Galaxy (e.g., the respective role of diffusion and advection, or the magnitude of the diffusion coefficient). In particular, a deep survey may inform us about the \gls{cr} lifecycle on small/intermediate scales, typically in the vicinity of major particle accelerators. Due to its lower \gls{cr} background compared to the \gls{mw} \citep{Ackermann:2016}, the \gls{lmc} is a good target to search for inhomogeneities in the \gls{cr} distribution, resulting for instance from recent or sustained \gls{cr} injection episodes and/or enhanced confinement near the source. This may be crucial for a proper understanding of the \gls{cr} lifecycle and associated non-thermal emissions \citep{DAngelo:2018}.

\textit{Particle accelerators}: What is the population of particle accelerators in the \gls{lmc}, and does it differ in any way from the different gamma-ray source classes we know of today ? The handful of objects currently known are rare and extreme sources that make up the high-luminosity end of the population of gamma-ray emitters in the \gls{lmc}. While fine spectral studies of this small number of extreme objects may be instrumental in solving some puzzles in our current understanding of particle acceleration (e.g., the electron-to-proton ratio, or the maximum attainable energy), \gls{cta} will push the census beyond out-of-the-norm sources and may usefully complement the survey of the \gls{mw}. The favourable viewpoint of the \gls{lmc} can make it easier to relate particle accelerators to their environment, owing to a reduced line-of-sight confusion and accurate distance estimate. The increase in the number of known gamma-ray sources in the \gls{lmc} is also interesting as \gls{cta} deep observations will occur in the wake of other major surveys of the \gls{lmc} in the X-ray \citep[with eRosita, e.g.][]{Sasaki:2022} and radio \citep[with the \gls{askap}, e.g.][]{Pennock:2021} bands, providing an exquisite multi-wavelength coverage of sources like \glspl{snr} or \glspl{pwn}.

\textit{Nature of \gls{dm}}: What information can the \gls{cta} survey of the \gls{lmc} bring on the nature of \gls{dm}? The \gls{lmc} has a mass of the order of $10^9$ M$_{\odot}$ enclosed in 8.9 kpc and more than a half is due to a dark halo \citep{vanderMarel:2002}. Study of the rotational curves of the \gls{lmc} revealed that it must contain a dark compact bulge with an anomalously high mass-to-luminosity ratio as large as $M/L \sim 20-50$ \citep{Sofue:1999} compared to that calculated for the \gls{mw} $M/L \sim$ 7 \citep{Sofue:2013}. With these characteristics, the \gls{lmc} is one more potentially suitable source for indirect searches of \gls{dm} signal in our neighborhood. In addition, such an investigation will take place in a specific global context, with different contamination of the hypothetical dark matter signal and various possible biases in the analyses compared to studies of the \gls{gc} or dwarf spheroidals.

In this paper, we provide a quantitative assessment of the detection prospects for the planned survey of the \gls{lmc}. We developed a model for the entire galaxy emission at very high energies, from populations of discrete sources to interstellar emission on various scales and a possible \gls{dm} annihilation component. Based on this model, we simulated \gls{cta} observations of the \gls{lmc} using the latest instrument response functions estimates, and we analysed these data using existing prototypes for the \gls{cta} science tools. In addressing the above questions, we investigated the conditions for the survey realization under which its scientific potential would be maximized, especially the distribution of the exposure. Our goal is to go beyond what is already known and evaluate the prospects for detecting new sources and opening new avenues for high-energy astrophysics in the \gls{lmc}.

The structure of the paper is as follows:
In section \ref{sec:model}, we introduce the gamma-ray emission model used for the \gls{lmc}, including a possible additional emission component produced from \gls{dm} annihilation in the \gls{lmc}. Section \ref{sec:simana} is dedicated to the description of the simulation and analysis methods of \gls{cta} observations. In section \ref{sec:res}, results on detectability of the various classical emission components in our model are presented, as well as sensitivity curves for \gls{cta} detection of a \gls{dm}-related signal. Finally, section \ref{sec:conclu} is dedicated to conclusions.

Throughout the paper, the distance of the \gls{lmc} is assumed to be $d=50.1$\,kpc \citep{Pietrzynski:2013}. Sky positions are given in equatorial coordinates corresponding to the J2000.0 epoch. We will refer to objects such as LHA 120-N~157B as N~157B for short but emphasise that the full denomination should be used when searching for these objects in the CDS/Simbad database.

\section{Emission model} \label{sec:model}

In this section, we describe the model that was developed for the gamma-ray emission of the \gls{lmc} galaxy and used as input to the survey simulations. Since the \gls{vhe} emission of the \gls{lmc} is still largely unexplored, and only a handful of extreme objects have been detected so far, this model is based for the most part on simulated components, inspired by the knowledge of \gls{vhe} source populations in the Galaxy and informed by observations of the \gls{lmc} at other wavelengths (e.g., X-ray SNRs).

We considered a baseline model consisting of classical emission components that can be seen as guaranteed, in the sense that their contribution should exist even if some of their properties may differ from the assumed ones (e.g., the number of \glspl{pwn} or the exact level or spectrum of interstellar emission): (i) the four already known \gls{vhe} sources; (ii) population of \glspl{snr}, \glspl{pwn}, and pulsar halos; (iii) interstellar emission from the galactic population of CRs.

We also envisioned possible emission from the 30 Doradus star-forming region but left it out of the baseline model as such a process cannot be considered to be sufficiently under control theoretically or observationally. We provide in the last subsection a description of the possible spectral and morphological properties of a more speculative component, which is the \gls{vhe} emission from the annihilation of \gls{dm} particles in the mass halo of the \gls{lmc}.

\subsection{Known point sources} \label{sec:model:knw}
    
In the \gls{vhe} domain, there are currently four known sources in the \gls{lmc}: N~157B interpreted as a \gls{pwn}; N~132D interpreted as an \gls{snr}; 30 Doradus C interpreted as a \gls{sb}, although alternative explanations as an \gls{snr} exist; and LMC P3 clearly identified as a gamma-ray binary from the orbital modulation of the signal. Extensive physical context will be given for each known object in Sect. \ref{sec:res:knw}. We left aside other possible sources outside the \gls{lmc} boundaries but within the survey footprint, such as those detected with the Fermi-LAT and whose spectrum could have been extrapolated to the \gls{cta} range (for instance quasar PKS 0601-70). 

All known sources were modeled as point-like objects in our work, although depending on the actual nature of the emission from 30 Doradus C, it might be at the limit of being extended for \gls{cta}. The spectral models for the first three objects were taken from the physical model fits to the H.E.S.S. measurements in \citet{Abramowski:2015}, retaining the hadronic model for N~132D and leptonic model for 30 Doradus C (see Sect. \ref{sec:res:knw} for a justification). There is currently no published broadband physical model for LMC P3 and we used one that is currently being developed by one of us (N. Komin, private communication). 

The source is modeled in a typical way for gamma-ray binaries \citep{Dubus:2013}. The compact object is assumed to be a pulsar that generates a relativistic magnetized outflow, and electron-positron pairs are accelerated in the interaction of this outflow with the stellar wind of the companion star. Gamma-ray emission from the system arises from inverse-Compton scattering of the  population of energetic pairs in the cosmic microwave background and the stellar photon field of the massive star companion. A power-law distribution of positrons and electrons with index 1.5 and energy of $5 \times 10^{37}$\eunit\ in the 0.5-50\tev range reproduces the H.E.S.S. measurements over the 1-10\tev range \citep{Abdalla:2018c}, without exceeding the Fermi-LAT upper limits above 10\gev \citep{Ackermann:2016}. The dense stellar photon field is assumed to have an effective temperature 40000\,K and an orbit-averaged energy density 291\eunit\vunit. Since the orbital light curve of LMC P3 remains poorly characterized as of now \citep{Abdalla:2018c}, we left aside its modeling and analysis as a variable source.

\subsection{Source populations: PWNe, SNRs, pulsar halos} \label{sec:model:pop}

\begin{table*}
\centering
\caption{Summary of the main parameters used in the modeling of the source population model}
\label{tab:srcpop} 
\begin{tabular}{ c | c | c }
\hline
Parameter & Unit & Value \\
\hline
\multicolumn{3}{ c }{Supernovae} \\
\hline
Supernova rate $r_{\rm SN}$ & \snrate & 0.002 \\
ccSNe / SNe Ia ratio $r_{\rm CC}/r_{\rm Ia}$ & - & 1.3 \\
Pulsar-producing fraction & - & 0.75 \\
\hline
\multicolumn{3}{ c }{Pulsars} \\
\hline
Initial magnetic field $B_0$ & G & $\mathcal{L}(12.65,0.55)$ \\
Initial period $P_0$ & ms & $\mathcal{N}(50,35)$ \\
Braking index $n$ & - & 3.0  \\
Neutron star inertia $I_{\rm NS}$ & g\,cm$^2$ & $10^{45}$ \\
Neutron star radius $R_{\rm NS}$ & km & 12 \\
\hline
\multicolumn{3}{ c }{SNRs} \\
\hline
Ejecta mass $M_{\rm ej}$ & \msol & 1.4 for SNe Ia, 5.0 for ccSNe \\
Ejecta energy $E_0$ & erg & $\mathcal{L}(50.7, 0.5)$ \\
Particle injection distribution $S_{\rm SNR}$ & - & PLEC \\
Particle distribution index $\alpha$ & - & $\mathcal{U}(2.3,0.1)$ \\
Particle injection efficiency $\eta_{\rm SNR}$ & - & $\mathcal{U}(0.2,0.1)$ \\
Electron-to-proton injection ratio $\xi$ & - & $10^{-3}$ \\
Age limit & yr & $6 \times10^{4}$ \\
\hline
\multicolumn{3}{ c }{PWNe} \\
\hline
Nebula magnetic field initial strength & G & $5 \times 10^{-5}$ \\
Nebula magnetic field evolution index & - & 0.6 \\
Particle injection distribution $S_{\rm PWN}$ & - & BPLEC \\
Particle distribution index below break $\alpha_1$ & - & 1.5 \\
Particle distribution index above break $\alpha_2$ & - & $\mathcal{U}(2.4,0.4)$ \\
Particle distribution break energy $E_{\rm brk}$ & GeV & $100$ \\
Particle distribution cutoff energy $E_{\rm cut}$ & TeV & $\mathcal{U}(500,300)$ \\
Particle injection efficiency $\eta_{\rm PWN}$ & - & $\mathcal{U}(0.7,0.3)$ \\
Age limit & yr & $10^{5}$ \\
\hline
\multicolumn{3}{ c }{Pulsar halos} \\
\hline
Suppressed diffusion region size $R_{\textrm{SDR}}$ & pc & 50 \\
Suppressed diffusion normalization at 100\tev $D_0^{\textrm{SDR}}$  & \dunit & $4 \times 10^{27}$ \\
Average interstellar diffusion normalization at 10\gev $D_0^{\textrm{ISM}}$ & \dunit & $10^{29}$ \\
Diffusion rigidity scaling index $\delta_{\textrm{D}}$ & - &$1/3$ \\
Particle injection distribution $S_{\textrm{HALO}}$ & - & $S_{\textrm{HALO}} = S_{\textrm{PWN}}$ \\
Particle injection efficiency $\eta_{\textrm{HALO}}$ & - & $\eta_{\textrm{HALO}} = \eta_{\textrm{PWN}}$ \\
Age limit & yr & $4 \times10^{5}$ \\
\hline
\end{tabular} \\
Notes to the table:\\
$\mathcal{U}(\mu,\sigma)$ means uniform distribution of mean and half-width $\mu$ and $\sigma$.\\
$\mathcal{N}(\mu,\sigma)$ means normal distribution of mean and standard deviation $\mu$ and $\sigma$.\\
$\mathcal{L}(\mu,\sigma)$ means log-normal distribution of mean and standard deviation of the logarithm $\mu$ and $\sigma$.\\
(B)PLEC stands for (broken) power law with exponential cutoff 
\end{table*}

The four known objects listed above are the most extreme members of larger populations that \gls{cta} can be expected to unveil, at least partially, and it is one goal of this paper to quantify what fraction of those populations will be probed with the survey. We developed a population model consisting of four classes of sources: shell-like \glspl{snr}, \glspl{isnr}, and \glspl{pwn}, which are the dominant classes of associated \gls{vhe} sources in the Galaxy \citep{Abdalla:2018a,Abdalla:2018b}, and pulsar halos, which constitute an emerging class that has the potential to account for a significant fraction of currently unidentified \gls{vhe} emitters \citep{Linden:2017,Sudoh:2019,Albert:2020,Martin:2022b}. We did not include in our model population components for gamma-ray binaries or microquasars. 

The population synthesis framework is extensively described in \citet{Martin:2022b} (except for \glspl{isnr}), where it was applied to the \gls{mw}. In what follows, we provide a concise description of the different population components and adaptation of the model to the case of the \gls{lmc}, and we refer the reader to the original paper for more details.

\textit{Supernova explosions and pulsar birth}: The four classes of objects considered for our model result from supernova explosions so the rates for such events set the normalisation of the various populations. The \gls{sn} rate in the \gls{lmc} is uncertain by at least a factor of 2, with published values ranging from 0.002 to 0.005\snrate \citep{vandenBergh:1991,Leahy:2017,Bozzetto:2017,Ridley:2010}. Estimates based on the present-day star formation rate or massive star population are shaky because the star formation history of the \gls{lmc} was not steady over the past $\sim100$\,Myr \citep{Harris:2009}, so the current \gls{sn} rate and \gls{sn} types ratio are partially disconnected from the current star formation rate. Building upon the works and arguments of \citet{vandenBergh:1991}, \citet{Leahy:2017}, and \citet{Maggi:2016}, we considered as baseline an \gls{sn} rate of $r_{\rm SN}=0.002$\snrate, with a ratio of core-collapse to thermonuclear \gls{sne} of $r_{\rm CC}/r_{\rm Ia}=1.3$. After a calibration of the population model to known \gls{vhe} sources in the Milky Way, we assumed that the fraction of core-collapse \gls{sne} producing neutron stars is 0.75 \citep{Lorimer:2006}, such that the rate of pulsar birth in the \gls{lmc} is $r_{\rm PSR}=0.00085$\snrate. The source population model starts with the random generation of supernovae over the last 400\,kyr (the lifetime of the longest-lived objects, pulsar halos), with random generation of a number of events from a Poisson distribution, random generation of an age in a uniform distribution, then random generation of a \gls{sn} type in a binomial distribution and finally, for core-collapse \gls{sne}, random selection of those giving birth to pulsars again from a binomial distribution. 

\textit{Locations of objects}: In a first stage, \gls{sne}, and their pulsars when appropriate, are randomly distributed over the \gls{lmc} according to the following prescription: thermonuclear \gls{sne} are uniformly distributed over the gas disk of the galaxy, as defined in Sect. \ref{sec:model:ism}, while core-collapse \gls{sne} are distributed among the different massive star forming regions of the \gls{lmc} in proportion to their ionizing luminosity, following the list of $\rm H\RomNumCap{2}$ regions and their properties in \citet{Pellegrini:2012} and with an added random scatter in position by 0.05\deg\ to account for the typical extent of the regions. In a second stage, we include in our model the present-day knowledge of more than 60 real \glspl{snr} in the \gls{lmc}, with X-ray and dynamical properties derived in a homogeneous way in \citet{Maggi:2016} and \citet{Leahy:2017}. For each real \gls{snr}, we select among the model \glspl{snr} the one with the same type and the smallest distance in the (age, density, energy) space. That a proper match can be obtained is guaranteed, from the statistical point of view, by the fact that the properties of model \glspl{snr} were sampled from distributions inferred from observations of real \glspl{snr} in the \gls{lmc} (see below). For those model objects for which an association is made, the initial random location is replaced by the location of the actual X-ray \gls{snr}, and this affects not only the model \gls{snr} but also the model \gls{pwn}, if any, for objects resulting from core-collapse \gls{sne}.

\textit{Interstellar conditions}: The evolution of all systems and their non-thermal radiation are influenced by the surrounding interstellar conditions, directly or indirectly. For each system, the surrounding magnetic field strength is randomly sampled from a uniform distribution with mean 4\,$\mu$G and half-width 3\,$\mu$G (see Sect. \ref{sec:model:ism} for more details). Similarly, the interstellar radiation field is taken to vary from one object to the other and it was randomly drawn from a uniform distribution between two extreme field models (see Sect. \ref{sec:model:ism} for more details). In both cases, this is meant to incorporate in the model the fact that some objects will arise in star-forming regions with intense radiation fields, while others will be born in more isolated and quiescent environments. Last, the interstellar gas density was also taken to vary from one object to the other, and its value was randomly sampled from a log-normal distribution, inspired by those inferred in \citet{Leahy:2017} for the upstream medium of \glspl{snr} detected in X-rays, that we approximated as a single distribution with mean $\mu_{log(n_{\rm H})}=-1.0$ and standard deviation $\sigma_{log(n_{\rm H})}=0.9$ for $n_{\rm H}$ in units of \nunit.

\textit{\glspl{snr}}: The modeling of the population of \glspl{snr} is based on the framework presented in \citet{Cristofari:2013}. It implements analytical prescriptions for the dynamics of the forward shock in the remnant and computes the evolution of a parameterized distribution of non-thermal particles energized at the shock and trapped in the remnant upon downstream advection. Different treatments are used depending on whether the \gls{snr} results from a thermonuclear or core-collapse explosion: in the former case, the expansion occurs in a uniform circumstellar medium, while in the latter case it occurs in a layered wind-blown cavity shaped by the progenitor massive star. The model is valid over the free expansion and Sedov-Taylor stages and breaks down as the forward shock becomes radiative. We assumed a lifetime $\tau_{\rm SNR}=60$\,kyr for model \glspl{snr} but some do not even reach that limit as they become radiative before.

\textit{\glspl{isnr}}: The modeling of the population of \glspl{isnr} was inspired from a similar work performed in the context of the anticipation of the planned Galactic Plane Survey with \gls{cta} \citep{Remy:2021}. The modeling starts with the generation of a synthetic population of molecular clouds, based on the inferred mass spectrum and cloud density in the \gls{lmc} \citep{Fukui:2008}, and extrapolating it in the range of masses where the catalog is not complete. The probability for a cloud to be interacting with an \gls{snr} is parameterized as a power-law in cloud mass and calibrated on the basis of what is observed in our Galaxy (for a molecular cloud population relevant to the Galaxy). Ultimately, a flat probability distribution seems to be appropriate. For those clouds in interaction, the proton spectrum in each remnant is randomly sampled from parameter distributions derived from the study of such systems in the MW. It is typically a broken power-law spectrum with relatively soft indices. A synthetic population is generated by computing the pion decay spectrum associated to the interacting system\footnote{The gamma-ray production cross section used for these calculations is taken from \citet{Kafexhiu:2014}, as implemented in the Naima python package \citep{Zabalza:2015}.}, given the random-sampled particle spectrum and cloud density. 
These mock \glspl{isnr} are then assigned to the mock core-collapse \glspl{snr} not associated to existing X-ray remnants, after removing the brightest object in the population to account for the fact that we already have a prominent interacting system in our emission model, N~132D.

\textit{\glspl{pwn}}: The modeling of the population of \glspl{pwn} is based on the model presented in \citet{Mayer:2012} and updated in \citet{Abdalla:2018b}. It starts with the random generation of the pulsar population with initial spin periods and magnetic fields sampled from typical distributions for young pulsars, which determines the spin-down history of the pulsars and sets the power available for the production of non-thermal particles in each system. The development of a model \gls{pwn} until its randomly selected age is described as the expansion of a spherical nebula over three dynamical stages, with its content of non-thermal particles evolving as a result of injection, energy losses, and escape. We assumed a lifetime $\tau_{\rm PWN}=100$\,kyr for \glspl{pwn}, a limit consistent with most of the observed population \citep{Abdalla:2018b}, after which they transition to the halo stage (see below). 

\textit{Pulsar halos}: The modeling of the population of halos is based on the diffusion-loss model implementation presented in \citet{Martin:2022}, in which electron-positron pairs injected at a central point diffuse away spherically in a medium characterized by a two-zone concentric structure for diffusion properties, with an outer region typical of the average \gls{ism} and an inner region where diffusion is suppressed down to values inferred for the Geminga pulsar halo \citep{Abeysekara:2017b}. Particle injection into the halo is assumed to start at the end of the \gls{pwn} phase, when the pulsar exits it original nebula as a result of its natal kick, with a spectrum that is similar in shape and normalization to that fed into the \gls{pwn}. Particles experience radiative losses in the randomly sampled magnetic field and radiation fields for the system (see above). The different scalings of the diffusion and loss processes with particle energy result in a characteristic energy-dependent morphology for halos. We assumed a lifetime $\tau_{\rm HALO}=400$\,kyr for the mock halos, which is dictated by the characteristic age of the Geminga pulsar.

\textit{Model calibration}: The population synthesis model features a number of free parameters that should be set to provide a representative emission distribution at a population level. It is not possible to calibrate it on \gls{lmc} observations owing to the small number of sources detected so far, and especially because the latter are most likely extreme objects. Instead, the model was calibrated against the population of known Galactic sources in the \gls{vhe} range, as described in \citet{Martin:2022b}, which resulted in a selection of possible values and statistical distributions for the parameters governing the evolution of the different object classes. Once calibrated, the population synthesis could be run for the specific conditions of the \gls{lmc}. The parameters eventually adopted are summarized in Table \ref{tab:srcpop}.

\begin{figure}
\begin{center}
\includegraphics[width=\columnwidth]{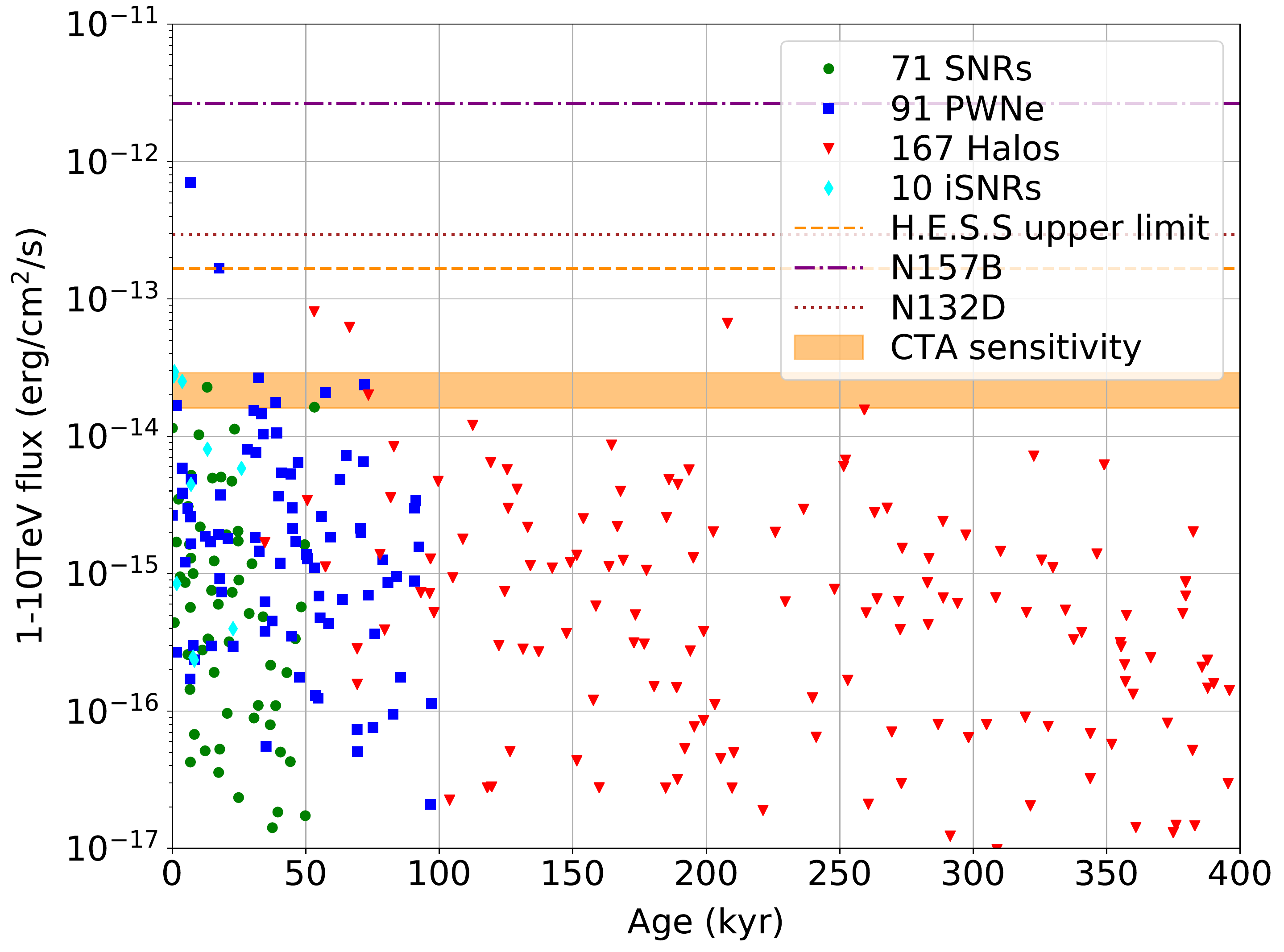}
\caption{Luminosity as a function of age for the random realization of our \glspl{pwn}, \glspl{snr}, \glspl{isnr}, and pulsar halos population model for the \gls{lmc}. Overlaid is the upper limit on point-like emission from H.E.S.S. \citep[using the limit on SN~1987A from][]{Abramowski:2015}, and the range of threshold luminosity for a detection with significance above 5$\sigma$ with \gls{cta} (see Sect. \ref{sec:res:sens}). Also shown are the levels of emission for the strongest (N~157B) and weakest (N~132D) of the currently known sources.}
\label{fig:model:pop:lum}
\end{center}
\end{figure}

The random realization of the source population model that we used in our simulations and analyses of the survey contains 71 \glspl{snr}, 10 \glspl{isnr}, 91 \glspl{pwn}, and 167 pulsar halos within the prescribed age or dynamical limits. Figure \ref{fig:model:pop:lum} displays the 1-10\tev energy flux of mock sources as a function of their age, compared to the H.E.S.S. 99\% confidence level upper limit on SN~1987A \citep[for observations done over 2003-2012; see][]{Komin:2019} and the foreseen \gls{cta} $5 \sigma$ detection threshold as determined in Sect. \ref{sec:res:sens}. The model population, calibrated on Galactic objects, extends nicely up to the H.E.S.S. sensitivity upper limit. In this realization, only two \glspl{pwn} exceed it, which is consistent with the currently detected population which comprises two pulsar-powered sources. This confirms that the population is well normalized and that currently detected objects are the most extreme members of their class. We kept these high luminosity mock objects in the population model as they could well be there and have escaped detection with H.E.S.S. simply because the H.E.S.S. survey did not uniformly cover the full extent of the \gls{lmc}. The discussion on the fraction of the population that could be accessible to \gls{cta} is presented in Sect. \ref{sec:res}.

\begin{figure*}
\begin{center}
\includegraphics[width=\columnwidth]{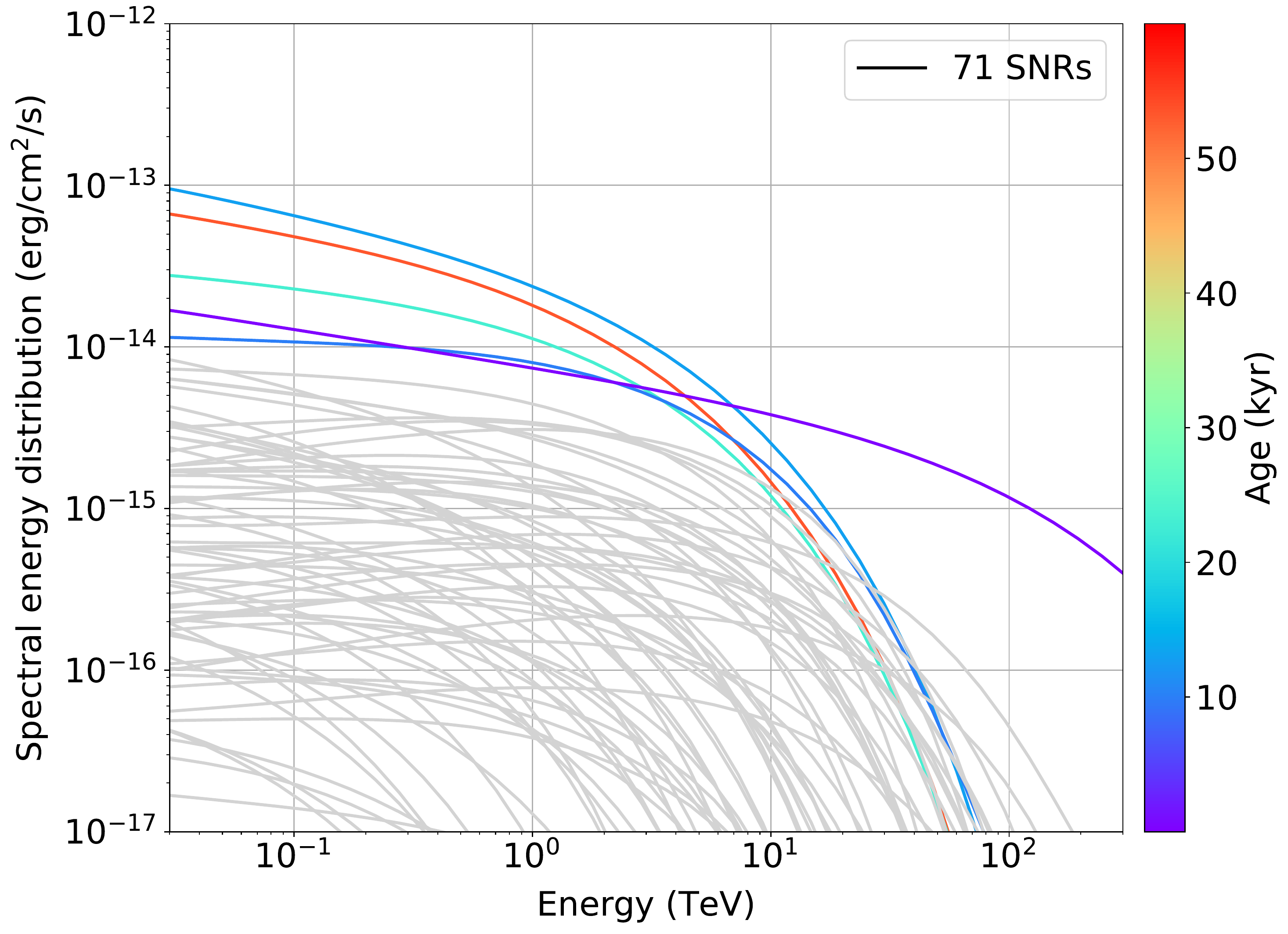}
\includegraphics[width=\columnwidth]{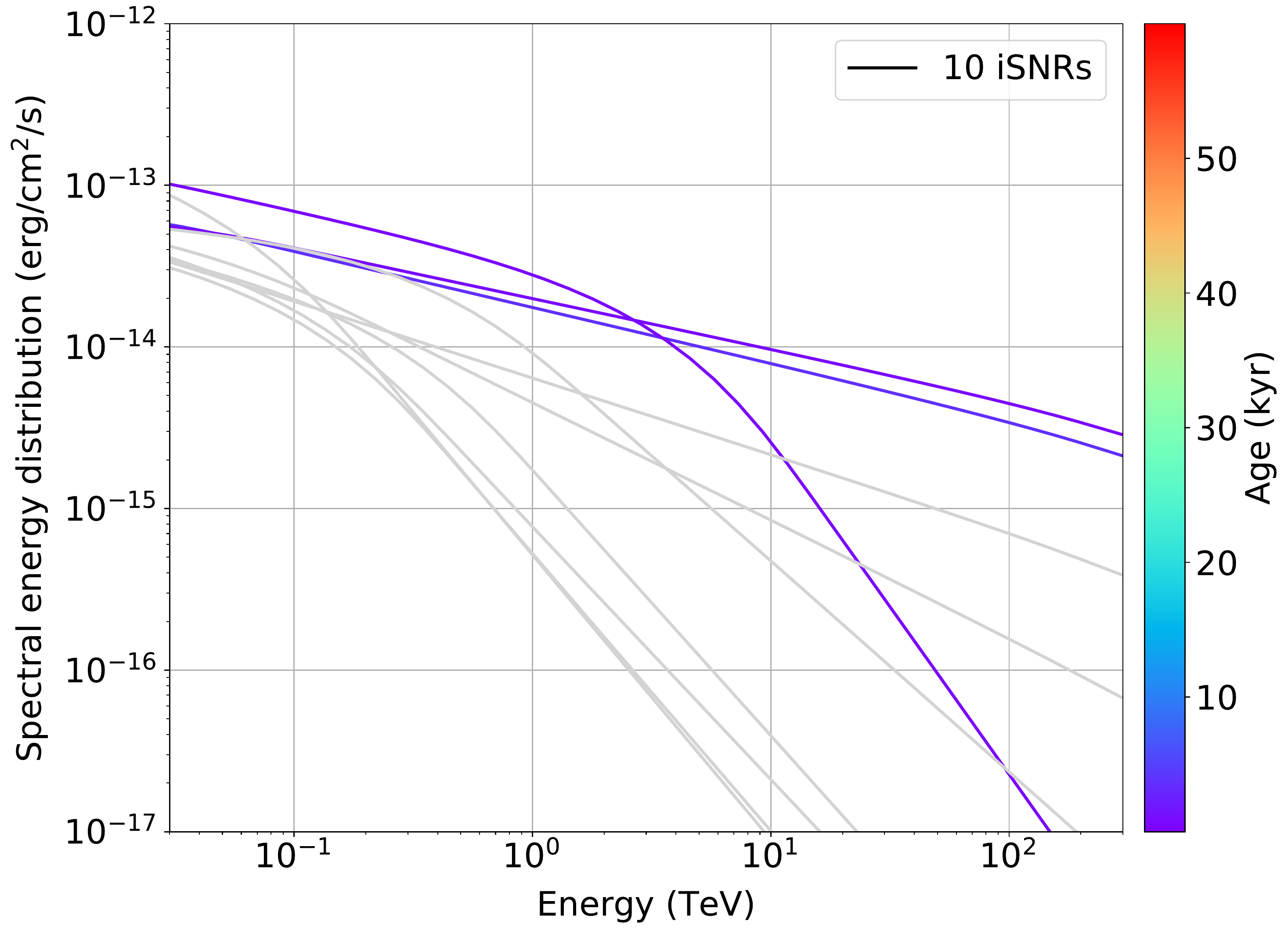}
\includegraphics[width=\columnwidth]{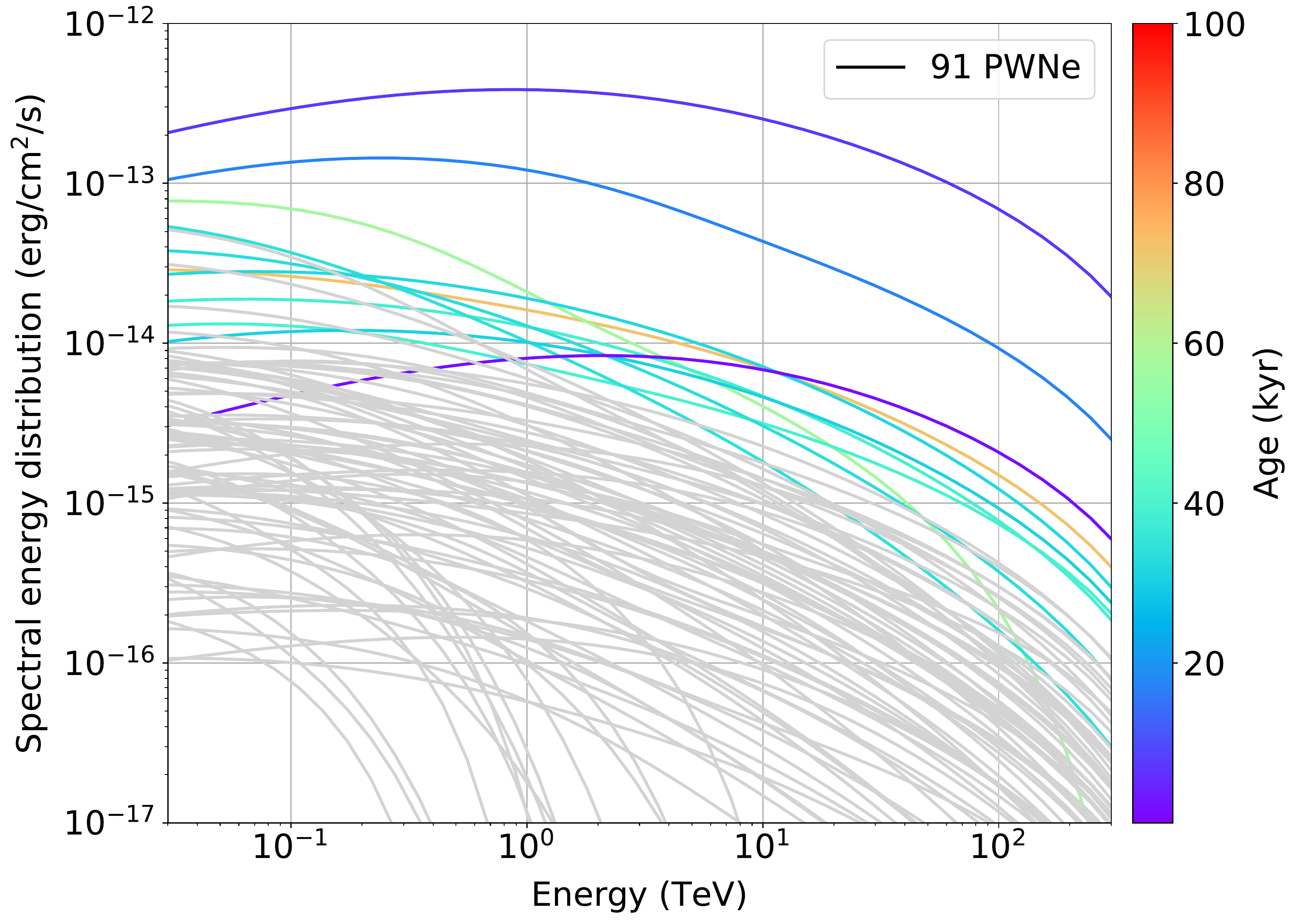}
\includegraphics[width=\columnwidth]{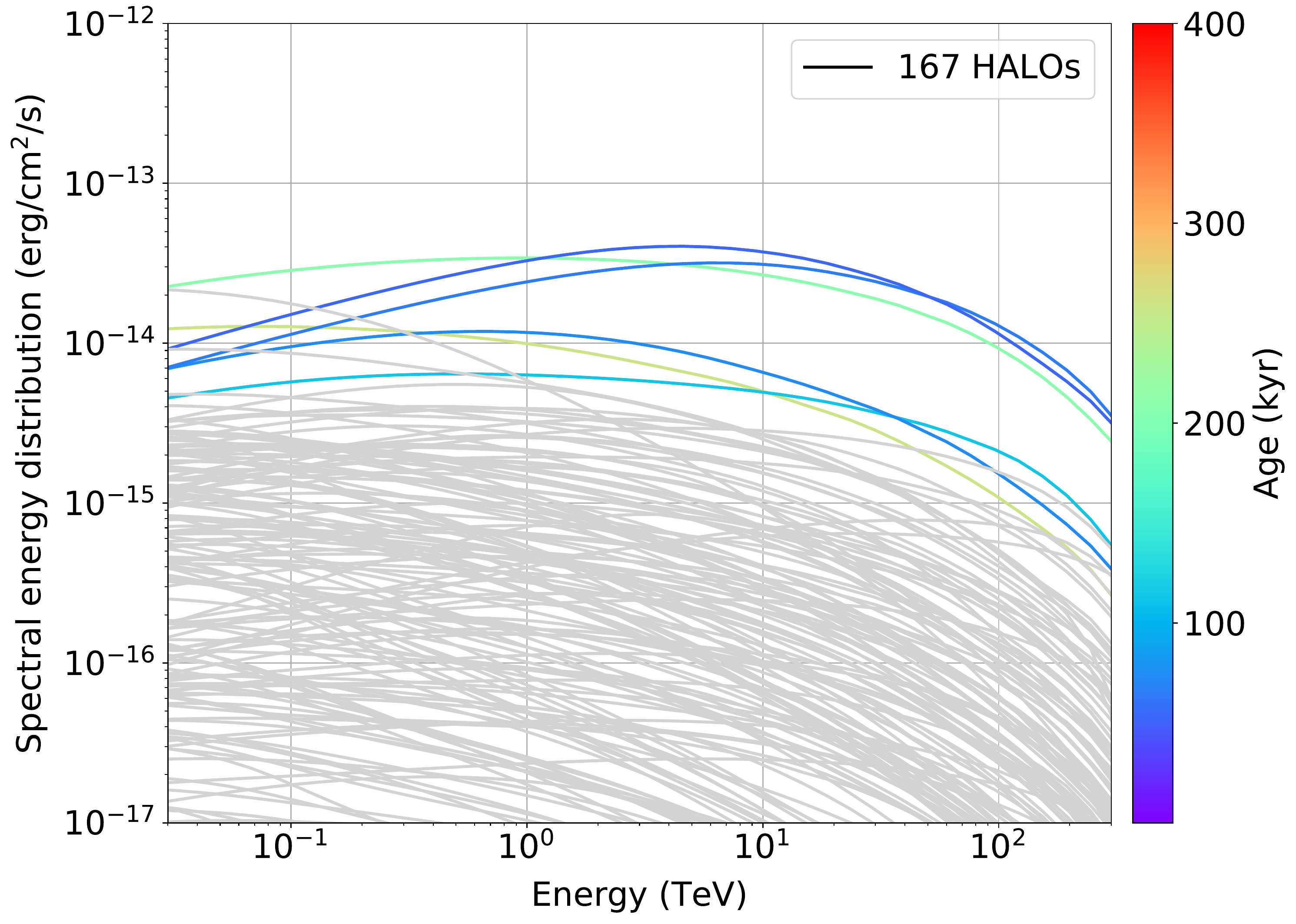}
\caption{Spectra of all individual objects in the realization of our source population model. Shown in light gray are objects below the approximate sensitivity level of the survey (see Sect. \ref{sec:res:sens}).  Curve colors encode the age of the members in each source class (with a scaling specific to each class).}
\label{fig:model:pop:specs}
\end{center}
\end{figure*}

\begin{figure}
\begin{center}
\includegraphics[width=\columnwidth]{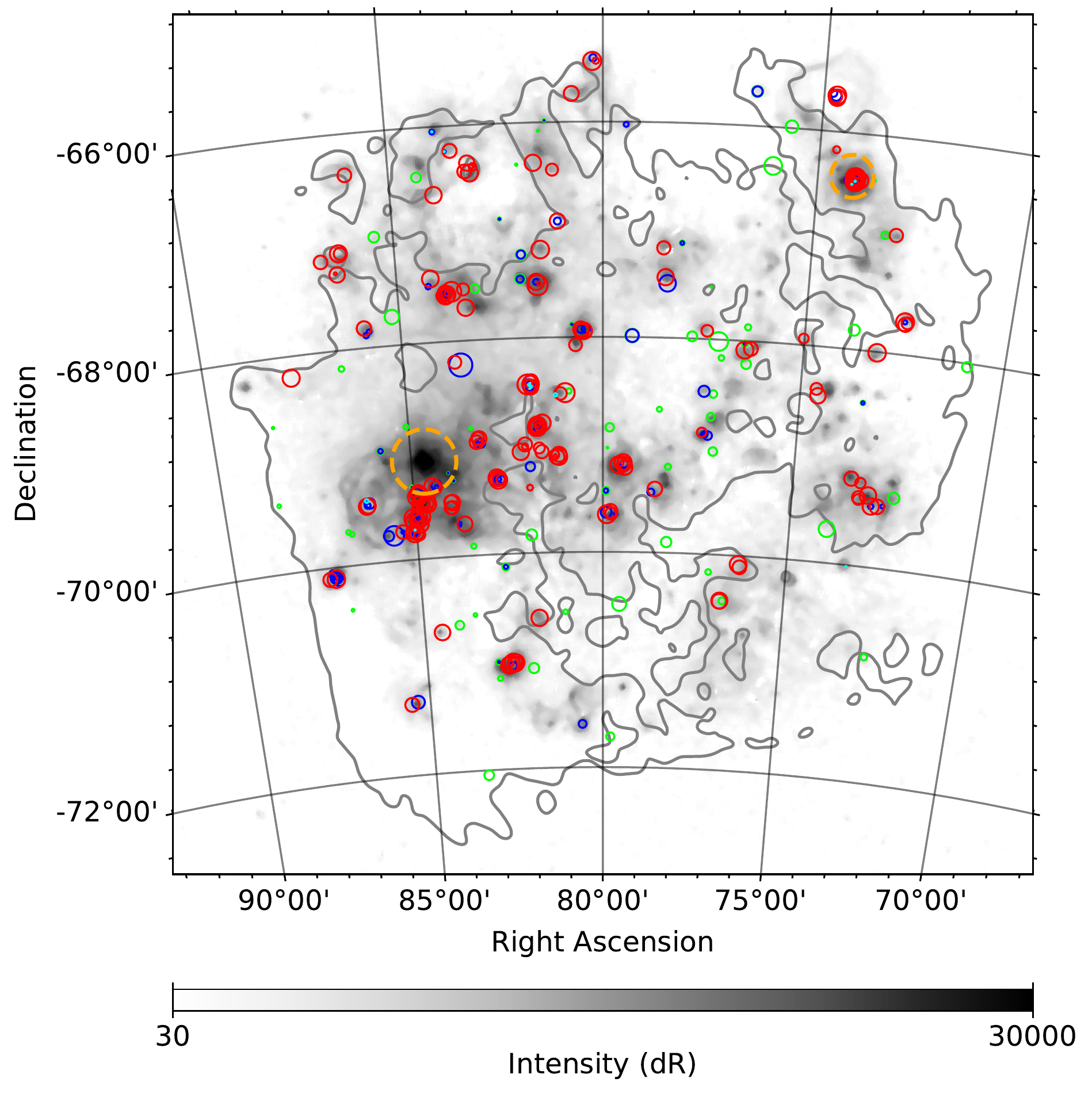}
\caption{Spatial distribution of source population mock objects over the \gls{lmc}. The background image is from the Southern H-Alpha Sky Survey Atlas \citep{Gaustad:2001} and displays H$\alpha$ emission intensity in dR units on a logarithmic scale, thus providing a view on ionized gas distribution in the galaxy. The contours trace the typical extent of the atomic gas disk \citep{Kim:2003}. The small and large dashed orange circles indicate the locations of regions N11 and 30 Doradus, respectively. Model \glspl{snr}, \glspl{isnr}, \glspl{pwn}, and pulsar halos are overlaid as green, cyan, blue, and red circles, respectively. For \glspl{snr} or \glspl{isnr} and \glspl{pwn}, the size correspond to the forward shock and nebula outer radius, respectively, while the sizes of halos correspond to the 95\% containment radius of the 3\tev emission.}
\label{fig:model:pop:dist}
\end{center}
\end{figure}

Figure \ref{fig:model:pop:specs} shows the spectra of all individual objects in the realization of our source population model that we used for simulation and analysis. Figure \ref{fig:model:pop:dist} shows the layout and sizes of the source population model objects over the \gls{lmc}, on top of an H$\alpha$ image of the galaxy. The built-in correlation of most sources with $\rm H\RomNumCap{2}$ regions is clearly apparent (except with 30 Doradus, which we decided to treat separately) and the figure makes it clear that this could result in some degree of source confusion. In some places especially, for instance south of 30 Doradus towards $\rm H\RomNumCap{2}$ regions N158, N159, and N160 (DEM L269, L271, L284) or west of the \gls{lmc} towards $\rm H\RomNumCap{2}$ region N82 (DEM L22), the crowding is quite high. 

Most objects have radial sizes below 0.05\deg, with a handful of rare \glspl{pwn} and \glspl{snr} reaching up to 0.1\deg, which means that the majority of the population will be detected as point-like objects for \gls{cta}. In practice, in the survey simulations described in Sect. \ref{sec:simana}, the morphological information from the source population models was simplified. All \glspl{pwn} and \glspl{snr} were treated as uniform brightness disks if their projected radii is above 3\,arcmin, and as point sources otherwise. For pulsar halos, although the model does include the full energy-dependent morphology, they were modeled as projected two-dimensional Gaussian intensity distribution with a size characteristic of that obtained at 3\tev, except if their 95\% containment radius is smaller than 3\,arcmin, in which case they were treated as point sources \citep[see the discussion on halo size estimate in][]{Martin:2022b}.

\subsection{Interstellar emission from the LMC's population of CRs} \label{sec:model:ism}

Interstellar emission was computed under the assumptions of steady \gls{cr} injection from an ensemble of point sources, followed by diffusive transport in the \gls{ism} and interaction with model distributions for interstellar components (gas, photon and magnetic fields). The final templates for interstellar emissions result from convolving a model distribution of sources with average emission kernels for pion-decay and inverse-Compton processes, plus a correction by the actual gas distribution for the pion decay component. Some of the  assumptions introduced below are inspired by studies of our Galaxy but there is no solid observational evidence that \gls{cr} transport in the \gls{lmc} and the \gls{mw} behaves the same, especially in the \gls{vhe} regime. So when assessing the detectability of interstellar emission from the \gls{lmc}, we will also envision the possibility that some features of our models depart from their baseline values. In what follows, we provide a concise description of the preparation of the large-scale interstellar emission components.

\textit{Cosmic-ray source distribution}: In star-forming galaxies, \glspl{cr} are mainly energized by the mechanical power provided by massive stars in the form of winds and outflows, core-collapse \gls{sn} explosions, and compact objects, and an additional contribution comes from thermonuclear \gls{sn} explosions. Lacking a solid understanding of the relative contribution of each class of \gls{cr} accelerators to the overall \gls{cr} injection luminosity, we simplified the problem by considering that injection occurs in massive-star-forming regions without specifying the objects actually involved or their respective particle acceleration properties. We did not include a source distribution model for injection by thermonuclear \gls{sne}, which can be expected to be more uniformly spread over the galaxy. Instead, we considered the alternative scenario of a distribution of CR injection sites that is less clustered and confirmed that this has no effect on the detection prospects for this component. As a tracer for CR injection sites related to the massive star population, we used a selection of $\rm H\RomNumCap{2}$ regions from the catalog of \citet{Pellegrini:2012}, retaining the most luminous ones, that are populated enough for consistency with our steady-state injection assumption, but excluding the most powerful 30 Doradus, that we will handle separately owing to its extraordinary status. For the 138 regions in our sample, we converted H$\alpha$ luminosities into ionizing luminosities, based on the morphological classification and escape fraction determined by \citet{Pellegrini:2012}, and we took ionizing luminosity as a measure of the richness of each star cluster, to which we assumed \gls{cr} injection power is proportional. Eventually, the \gls{cr} source distribution is of the form:
\begin{equation}
\label{eq:srcdist}
M_{\rm CR}(\mathbf{r})= \sum\limits_{k} L_{k} \delta(\mathbf{r}-\mathbf{r_k}) 
\end{equation}
with a total of 138 injection sites located at the positions $\{\mathbf{r_k} \}$ of selected $\rm H\RomNumCap{2}$ regions and having relative injection luminosities $\{L_{k}\}$ proportional to ionizing luminosities. More details about the selection of $\rm H\RomNumCap{2}$ regions and derivation of their properties can be found in appendix \ref{app:crsrc}.

\textit{Cosmic-ray injection spectrum}: We restricted ourselves to \gls{cr} protons and electrons, treating nuclei via a nuclear enhancement factor when computing hadronic emission (thereby neglecting differences in source spectra for the different species). We assumed that \glspl{cr} at injection in the \gls{ism} follow a power-law distribution in momentum $p$ starting at 1\gev/{\it c} and exponentially cutting off at $p_{\rm cut,p}=1$\pev/{\it c} for protons and $p_{\rm cut,e}=100$\tev/{\it c} for electrons. The latter value is in agreement with the highest electron energies inferred in \glspl{snr} in the \gls{lmc} \citep{Hendrick:2001}. The \gls{cr} power spectral density for species X among protons or electrons (respectively specified with subscripts p or e) reads:
\begin{align}
\label{eq:srcspec}
\frac{\dop{P_{\rm X,inj}(p)}}{\dop{p}} &= P_{\rm X,0} \left( \frac{p}{p_0} \right)^{-\alpha_{\rm X}} e^{-\frac{p}{p_{\rm X,cut}}} \\
&= \beta \frac{\dop{P_{\rm X,inj}(E)}}{\dop{E}} = \beta Q_{\rm X}(E)
\end{align}
where $E$ is kinetic energy and $\beta=v/c$. For our baseline scenario, we started from assumptions inspired by our knowledge of the \gls{cr} population of the \gls{mw} and adopted $\alpha_p= 2.45$ and $\alpha_e= 2.65$ (we neglected the possibility of breaks in the injection spectra). These values are representative of the higher-energy part of the injection spectra in the widely used diffusion+reacceleration propagation models tested against a variety of observables \citep{Trotta:2011,Orlando:2018}. This assumption is however considered a minimal baseline model and the impact of a \gls{cr} population with a harder spectrum will be discussed below.

\textit{Cosmic-ray injection power}: The total \gls{cr} injection power is assumed to be constant in time, at a level corresponding to the long-term average of \gls{cr} injection by \gls{sne} exploding at a rate $r_{\rm SN}$, each event releasing $E_{\rm SN}=10^{51}$\eunit\ of mechanical energy, a fraction $\eta$ of which is tapped by \gls{cr} acceleration:
\begin{align}
\label{eq:srcpow}
&\int{\frac{\dop{P_{\rm X,inj}(E)}}{\dop{E}} E dE} = \eta_{\rm X} r_{\rm SN} E_{\rm SN}
\end{align}
We adopted $\eta_p=10^{-1}$, $\eta_e=10^{-3}$, and $r_{\rm SN}=0.002$\snrate as previously. This translates into a total $6.5 \times 10^{39}$\punit\ for the whole galaxy, to be distributed among the different massive star forming regions in proportion to their ionizing luminosity and then shared into \gls{cr} electrons and protons in a 1:100 ratio.

\textit{Cosmic-ray propagation}: \gls{cr} transport away from each injection site into the \gls{ism} is assumed to occur as a result of spatial diffusion limited by energy losses. Both terms are taken as constant and isotropic over the volume of the galaxy. Diffusion is controlled by a momentum-dependent coefficient of the form:
\begin{align}
\label{eq:diffcoef}
&D(p) = \beta D_0 \left( \frac{p}{p_0} \right)^\delta \\
&\delta=1/3\\
&D_0 = 10^{29} \, {\rm cm}^{2}\,{\rm s}^{-1} \, {\rm at} \,\, p_0 = 10\,{\rm GeV/{\it c}}
\end{align}
The normalisation and index adopted here are typical of the fitted values obtained in the diffusion+reacceleration propagation models from which we borrowed the injection spectra \citep{Trotta:2011,Orlando:2018}. Smaller values of a few $10^{28}$\dunit, as frequently found in the literature \citep[e.g.,][]{Evoli:2019}, would lead to interstellar emission in excess of the Fermi-LAT constraint at 10\gev (see Sect. \ref{sec:model:val}, but also a comment in appendix \ref{app:crprop}). Energy loss processes include hadronic interactions for \gls{cr} protons, synchrotron, inverse-Compton scattering, and Bremsstrahlung radiation for \gls{cr} electrons, plus Coulomb and ionisation losses for both species. They occur in homogeneous gas, photons, and magnetic field distribution models that will be introduced below. Protons and electrons spatial distributions are obtained by solving the diffusion-loss equation for a point-like and stationary source (see appendix \ref{app:crprop} for the details)

\textit{Emission kernels}: Projected particle angular distribution around a source are computed by integrating the particle spatial distributions (defined in appendix \ref{app:crprop}) along the line of sight over a thickness $2H$ and for the assumed distance $d$ to the \gls{lmc}: 
\begin{align}
\label{eq:srcproj}
\frac{\dop{N}}{\dop{E}\dop{\Omega}}(\theta,E) = 2  d^2 \times \int_{0}^{H}{\frac{\dop{N}}{\dop{E}\dop{V}} \left( (\theta^2 d^2+\ell^2)^{1/2},E \right) \dop{\ell}} 
\end{align}
The half-thickness $H$ is taken as representative of the target distribution: a 180\,pc gas disk scale height for \gls{cr} protons \citep{Kim:1999}, and a 1\,kpc magnetic and radiation field halo for \gls{cr} electrons. Inverse-Compton and pion decay angular profiles around a source are computed as\footnote{The calculations were performed with the Naima package \citep{Zabalza:2015}.}:
\begin{align}
\label{eq:srcrad}
&\frac{\dop{\Phi_{\rm IC}}}{\dop{E_{\gamma}}\dop{\Omega}}(\theta,E_{\gamma}) = \iint \frac{\dop{N_e}}{\dop{E_e} \dop{\Omega}} \frac{\dop{F_{\rm IC}}}{\dop{E_{\gamma}}}(E_e,E_{\gamma},\nu) \frac{U(\nu)}{h\nu} \dop{\nu} \dop{E_e} \\
&\frac{\dop{\Phi_{\rm PD}}}{\dop{E_{\gamma}}\dop{\Omega}}(\theta,E_{\gamma}) = \int \frac{dN_p}{\dop{E_p} \dop{\Omega}} \frac{\dop{F_{\rm PD}}}{\dop{E_{\gamma}}}(E_p,E_{\gamma}) n_{\rm H} \dop{E_p}
\end{align}
where $\dop{F_{\rm IC}}/{\dop{E_{\gamma}}}$ is the scattered photon spectrum per electron of energy $E_e$ and target photon of energy $h\nu$, while $\dop{F_{\rm PD}}/{\dop{E_{\gamma}}}$ is the decay photon spectrum per relativistic proton of energy $E_p$ and target proton. Quantities $U$ and $n_{\rm H}$ are the photon and gas target densities and we use for all injection sites the same values that are averages over the galaxy (more details are provided below). The resulting emission kernels are convolved with the \gls{cr} source distribution defined above:
\begin{align}
\label{eq:srcconv}
\mathcal{S}_{\rm PD,IC}(\bold{r},E_{\gamma}) &= M_{\rm CR}(\bold{r}) \otimes \frac{\dop{\Phi_{\rm PD,IC}}}{\dop{E_{\gamma}} \dop{\Omega}} (\theta,E_{\gamma}) \\
&= \sum\limits_{k} \frac{\dop{\Phi^k_{\rm PD,IC}}}{\dop{E_{\gamma}} \dop{\Omega}} (\| \bold{r}-\bold{r_k} \|,E_{\gamma})
\end{align}
where the sum runs over $k$ injection sites. In the equations above, gamma-ray photon energy was denoted as $E_{\gamma}$, to distinguish it from proton or electron kinetic energy $E_p$ or $E_e$, but in the following we will denote it simply as $E$ for convenience. For the pion decay component, a nuclear enhancement factor of 1.753 is introduced to account for the contribution of helium and heavier nuclei in \glspl{cr} and the \gls{ism} \citep[computed from][under the assumption of a 0.4 solar metallicity medium\footnote{We used the median metallicity found in \citet{Cole:2005} from intermediate-age and old field stars in the central regions of the \gls{lmc}. This is however a simplification since the metallicity in the LMC appears to be strongly position-dependent \citep[see][and references therein]{Lapenna:2012}. Interestingly, the 0.4 solar metallicity is consistent with the value obtained for the Fe element from X-ray spectroscopy of \glspl{snr} in the LMC \citep{Maggi:2016}, although the latter study also shows element-wise variations.}]{Mori:2009}. The resulting emission cube is rescaled by a gas column density map of the \gls{lmc} to recover the actual gas distribution structure of the galaxy, using the following formula:
\begin{equation}
\label{eq:gascorr}
\mathcal{S}_{\rm PD}^{\rm corr}(\bold{r},E_{\gamma}) = \mathcal{S}_{\rm PD}(\bold{r},E_{\gamma}) \times \frac{N^{\rm obs}_{\rm H}(\bold{r})}{2 n_{\rm H} z_{\rm gas}}
\end{equation}
where the denominator of the fraction on the right-hand side is the average gas column density assumed in this work, computed from parameters defined in the next paragraph, while the numerator refers to the gas column density map derived from observations of the atomic and molecular gas phases (plus a correction for the dark gas), as introduced in \citet{Abdo:2010d}.

\textit{Interstellar gas}: We define the gas disk of the \gls{lmc} as having a radius $R_{\rm gas}=3.5$\,kpc and a scale height $z_{\rm gas}=0.18$\,kpc \citep{Staveley-Smith:2003,Kim:1999}. Within this radius, the total interstellar atomic hydrogen mass of the \gls{lmc} is $M_{\rm H\RomNumCap{1}} = 3.8 \times 10^8$\msol \citep{Staveley-Smith:2003}, and the molecular hydrogen mass is $M_{\rm H_2} = 5.0 \times 10^7$\msol \citep{Fukui:2008}. Following \citet{Abdo:2010d}, building upon the results of \citet{Bernard:2008}, we increased these amounts by 50\% to account for the presence of dark neutral gas that could be cold atomic gas with optically thick 21\,cm line emission and/or pure molecular hydrogen gas with no CO emission. The content of ionised hydrogen gas is computed following \citet{Paradis:2011}, using electron density $n_e=1.52$\vunit\ and mean H$\alpha$ intensity of 26.3 Rayleigh corresponding to the regime defined as ``typical $\rm H\RomNumCap{2}$ regions" in the article. This yields a total ionised hydrogen mass of $M_{\rm H\RomNumCap{2}} = 1.7 \times 10^7$\msol, and thus a total interstellar hydrogen mass of $M_{\rm H} = 6.6 \times 10^8$\msol, with an estimated uncertainty of $2.1 \times 10^8$\msol that stems mostly from the uncertain amount of dark neutral gas \citep{Bernard:2008}. Assuming a typical volume for the gas disk of the galaxy of $V = 2 \pi R_{\rm gas}^2 z_{\rm gas}$, the average hydrogen density of the \gls{lmc} is $n_{\rm H}=1.93$\nunit.

\textit{Interstellar magnetic field}: The interstellar magnetic field can be expected to vary across the extent of the galaxy and fluctuate on $\sim50-100$\,pc scales, for instance because of the stirring by \glspl{snr} and \gls{sb}s. In the context of source populations, the magnetic field in different locations of the galaxy was randomly sampled from a uniform distribution with mean 4\,$\mu$G and deviation 3\,$\mu$G. The minimum 1\,$\mu$G value corresponds to the strength of the ordered component of the magnetic field only, while the 4\,$\mu$G mean value corresponds to the average strength for the ordered plus random components \citep{Gaensler:2005}. What the maximum strength could be is unclear, as are the frequency and scales at which it is encountered, so we adopted a uniform and symmetric distribution extending up to 7\,$\mu$G by lack of any better prescription. Yet, in the context of interstellar emission on large scales, the diffusion framework used here cannot handle inhomogeneous conditions so the magnetic field strength considered in electron diffusion is uniform at a value of 4\,$\mu$G. The magnetic field topology can have an influence on particle transport, for instance the specific orientation of the regular component of the field or the spatially-dependent ratio of turbulent to regular components \citep[see, e.g.,][in the context of the gamma-ray interstellar emission from the \gls{mw}]{Gaggero:2015b}. Exploring these effects is however beyond the capabilities of the diffusion model framework used here, which cannot handle anisotropic diffusion.

\textit{Interstellar radiation field}: The model for the \gls{isrf} was developed from the work of \citet{Paradis:2011}, in which the broadband infrared dust emission of the \gls{lmc} was linearly decomposed into gas phases and fitted to dust emission models, eventually yielding dust emissivity spectra $Q_Y(\nu)$ per unit column density for each gas phase $Y$ and levels of stellar radiation heating the dust in each phase. From these and the average gas column densities corresponding to the adopted gas disk model, we could construct a complete \gls{isrf} average model that, for simplicity, we approximated as a sum of five Planck distributions: 
\begin{align}
&U(\nu) \sim \sum_k \frac{u_k}{a T_k^4} B(\nu,T_k)  \\
&T_k \, {\rm in} \, \{2.73,35,330,3800,35000\} \, {\rm in \, K} \\
&u_k \, {\rm in} \, \{0.26,0.12, 0.025, 0.30, 1.20\} \, {\rm in \, eV/cm}^3
\end{align}
The \gls{isrf} model is uniform and has no spatial dependence. More details about the derivation of the \gls{isrf} model can be found in appendix \ref{app:isrf}.

\textit{Alternative interstellar medium model}: The model defined above for interstellar gas, magnetic and radiation fields is intended as a set of average conditions applicable to the large scales over which most \glspl{cr} will evolve and will be referred to as the ``average \gls{ism} model". We also used a second set of interstellar conditions that may be more relevant to small scales and the vicinity of some \gls{cr} sources, where large amounts of gas that fed massive star formation are still present. We assumed that, in such regions, the average neutral gas density is 10 times the galactic average density computed above, while the ionized gas column density becomes $6.18 \times 10^{20}$\cunit, as computed following \citet{Paradis:2011}, using electron density $n_e=3.98$\vunit, and an H$\alpha$ intensity of 113.3 Rayleigh corresponding to the limit between ``typical $\rm H\RomNumCap{2}$ regions" and ``very bright $\rm H\RomNumCap{2}$ regions'' in the article. The radiation field model is stronger as a result of infrared radiation components scaling linearly with gas column densities. In the absence of any solid estimate, the magnetic field strength in these gas-rich regions is kept at its large-scale average value. This second model will be referred to as the ``gas-rich \gls{ism} model" and may be more appropriate for \glspl{cr} confined to the vicinity of their sources (either because of suppressed diffusion or because of efficient energy losses as in the case of very-high-energy electrons), or to \glspl{snr} or \glspl{pwn} located in rich star-forming regions.

\begin{figure}
\begin{center}
\includegraphics[width=\columnwidth]{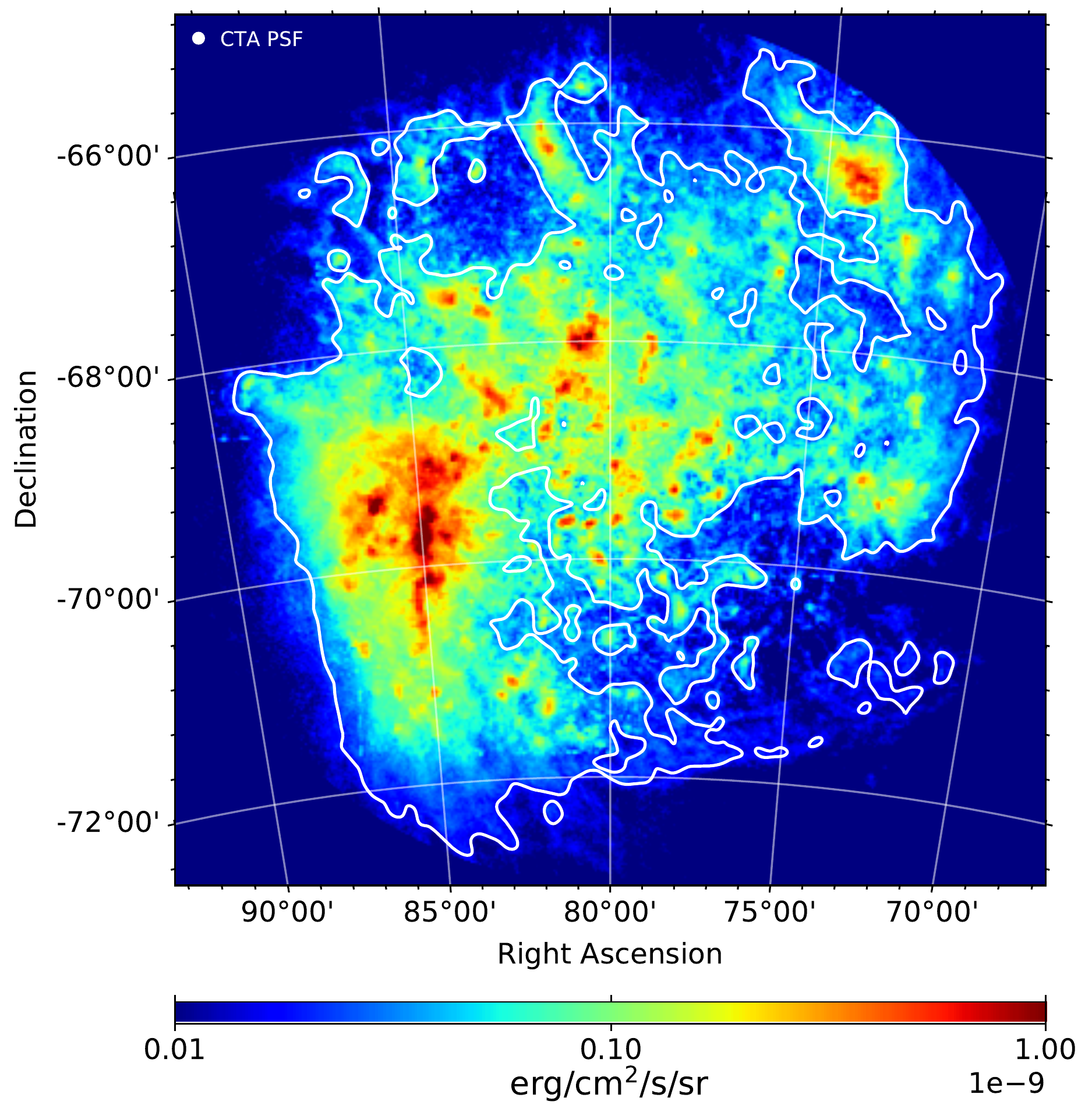} \\
\includegraphics[width=\columnwidth]{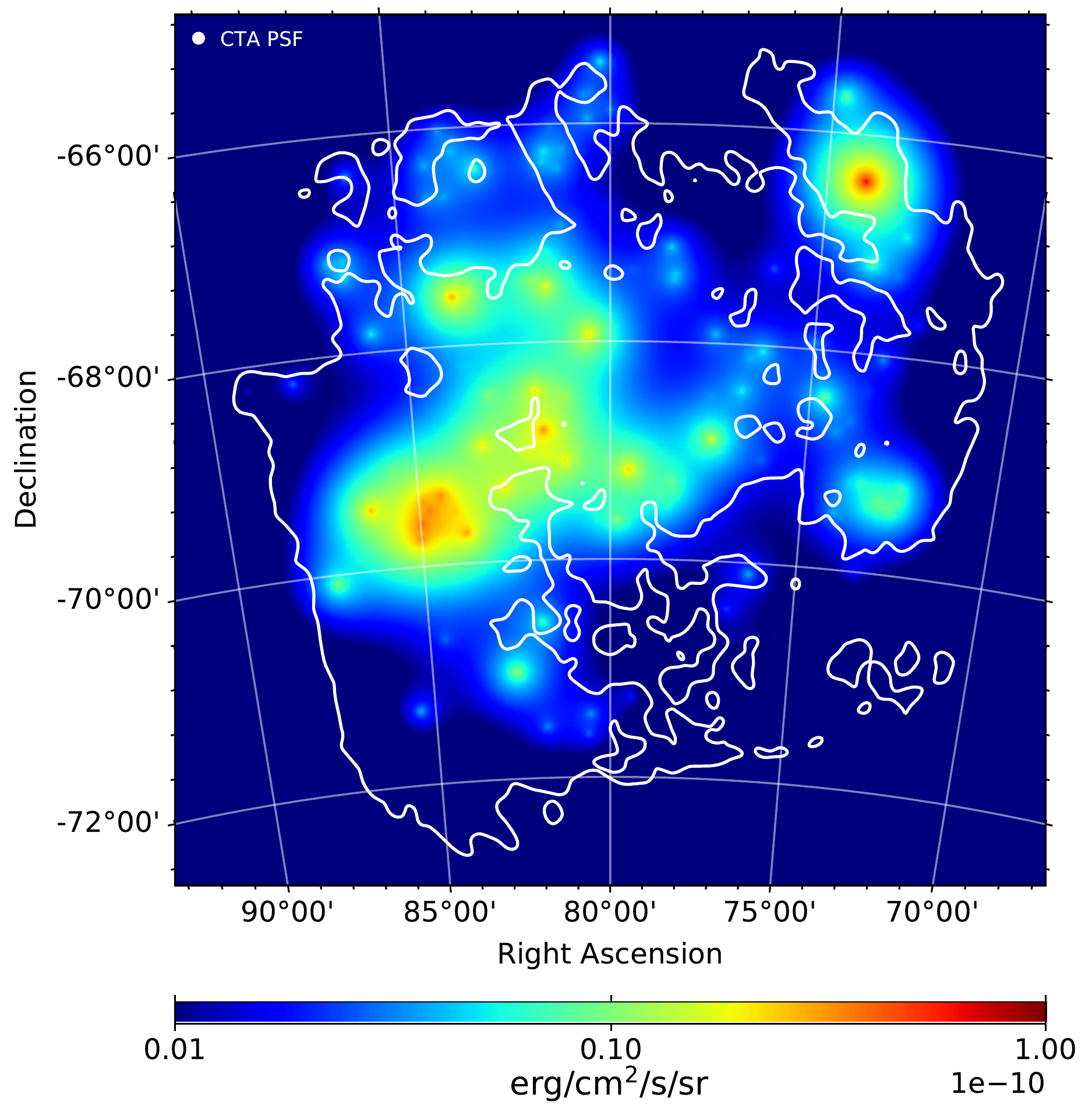}
\caption{Emission maps in $E^2 \times \mathcal{S}(E)$ at $E = 1$\tev for large-scale interstellar emission from pion-decay (top panel) and inverse-Compton scattering (bottom panel). Note the different color scale between the two maps.}
\label{fig:model:ism}
\end{center}
\end{figure}

The layout of pion-decay and inverse-Compton interstellar emission over the galaxy is illustrated in Fig. \ref{fig:model:ism}, at a reference photon energy of 1\tev. Hadronic emission is strongly correlated with the distribution of interstellar gas, owing to the long propagation range of \gls{cr} protons that can fill the entire galactic volume, while leptonic emission is strongly correlated with the assumed distribution of injection sites, because the reach of \gls{cr} electrons is limited by inverse-Compton and synchrotron losses.

\subsection{Emission from the 30 Doradus star-forming region} \label{sec:model:sfr}

Over recent years, the question of the behaviour of \glspl{cr} during the very early stage of interstellar propagation, in the vicinity of their parent sources, has been the focus of numerous theoretical and observational analyses. A recent review of the current status of observations of this stage in the \gls{cr} life cycle can be found in \citet{Tibaldo:2021}.

A rationale behind that interest is that this stage can have consequences on several key observables of the CR phenomenon, for instance the isotopic and spectral properties of the local flux of CRs, or the morphology and spectrum of the large-scale interstellar emission \citep{DAngelo:2016,DAngelo:2018}. The vicinity of sources is also where evidence for acceleration of galactic \glspl{cr} to PeV energies and beyond may more likely be found, if the latter are produced and confined only for a short phase in the evolution of (a subset of) the accelerators.

On the theoretical side, \gls{crs} freshly released from their accelerator can be expected to influence the transport conditions around it by the same kinetic processes that governed their confinement into the source during acceleration, i.e. self-generation of magnetic turbulence from resonant and non-resonant instabilities \citep{Malkov:2013,Bell:2013}. This may lead to enhanced local confinement over several 10\,pc scales and 10-100\, kyr durations, depending on particle energy and surrounding gas conditions \citep{Nava:2016,Nava:2019,Brahimi:2020}. On the observational side, there is growing evidence that specific transport conditions occur in the vicinity of some \gls{cr} sources, from individual isolated objects such as \glspl{snr}, pulsars or \glspl{pwn}, up to more extended sites such as \glspl{sfr} and \glspl{sb}.

The interpretation of Galactic observations is challenging because of the need for careful modeling and subtraction of foreground and background interstellar emission along the line of sight to a given source, to properly isolate interstellar emission on small/intermediate scales around it. In that respect, the external viewpoint on the nearby \gls{lmc} can be a valuable complementary source of information. The distance to the galaxy, however, restricts our probing of the vicinity of sources to physical scales of the order of $50-100$\,pc and above (or $0.06-0.11$\deg, compared to the $0.05-0.06$\deg anticipated angular resolution of the southern array at 1\tev), not to mention the need for sufficient \gls{cr} injection power to produce a detectable signal. For that reason, we investigated the possibility for the survey to constrain \gls{cr} transport in the vicinity of the most prominent \gls{sfr} in the \gls{lmc}, 30 Doradus. N11 may also constitute an interesting target, although the lack of detectable nonthermal X-ray emission suggests it may be different in nature \citep{Yamaguchi:2010}. A region like 30 Doradus hosts massive stars by the hundreds \citep{Walborn:2014}, and is thus potentially able to produce \gls{crs} in large quantities; combined with the vast amounts of gas and intense photon fields found in such a location, conditions are ideal for the study of young \gls{crs}. In addition to allowing the investigation of how \glspl{cr} behave close to their sources, major \glspl{sfr} are also well-motivated candidates for the acceleration of particles to the highest energies, in the PeV range or even beyond \citep{Bykov:2001,Parizot:2004,Aharonian:2019,Bykov:2020}.

\begin{figure}
\begin{center}
\includegraphics[width=\columnwidth]{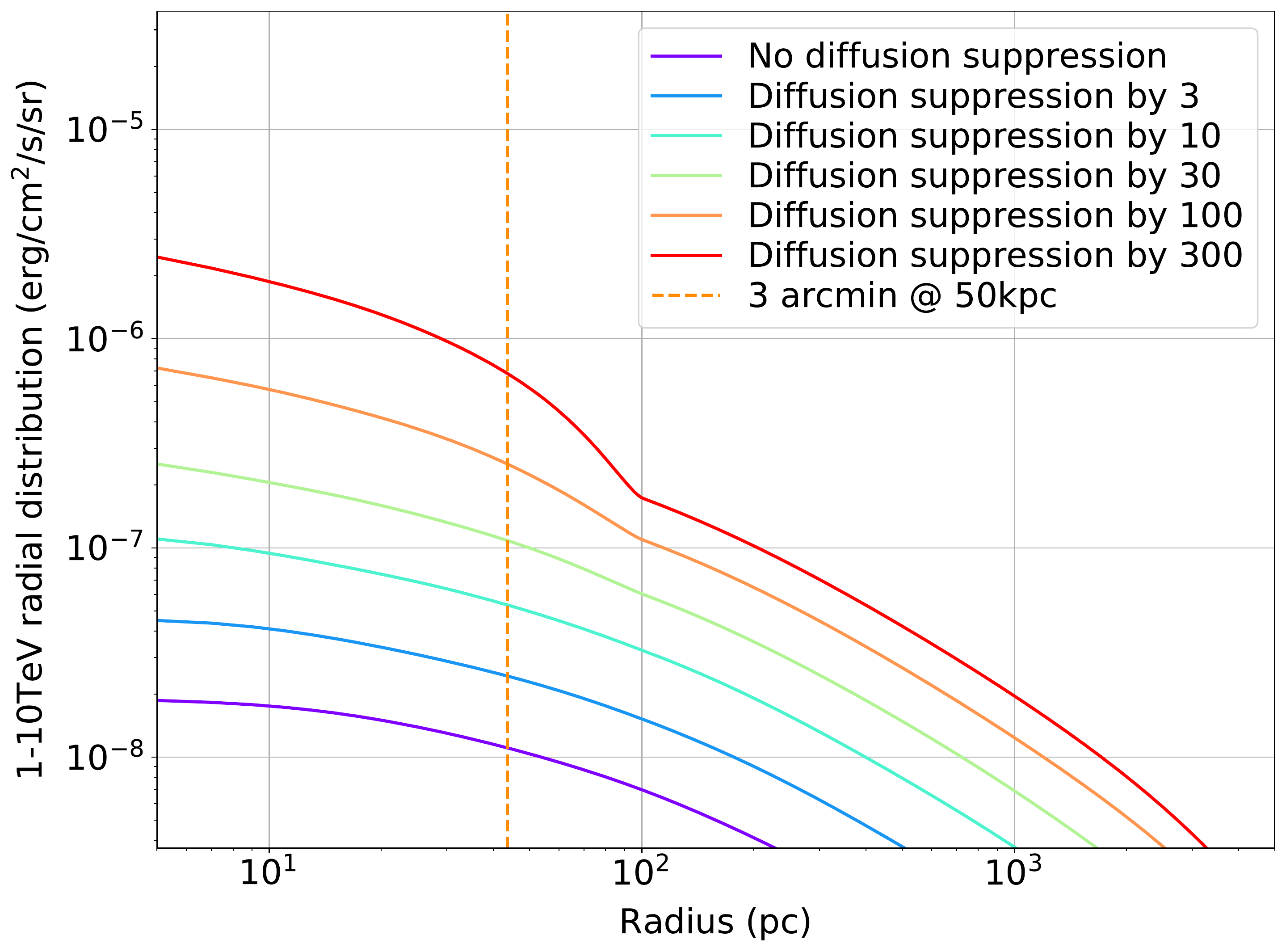}
\caption{Radial 1-10\tev intensity profiles for pion-decay emission resulting from \gls{cr} protons continuously injected by a point-source over 5\,Myr and diffusing away in a medium characterized by two zones: an outer region $>100$\,pc typical of the average ISM and an inner region $\leq100$\,pc where diffusion is suppressed relative to the ISM. The profiles shown here correspond to emission from a region filled with a homogenous gas density $n_{\rm H}=1.93$\nunit, and the injection luminosity is arbitrarily set at $10^{40}$\punit. When used in specific cases, for instance the 30 Doradus region, the intensity distribution is corrected for the actual gas content of the region (see text).}
\label{fig:model:sfr:emiss}
\end{center}
\end{figure}

\begin{figure}
\begin{center}
\includegraphics[width=0.8\columnwidth]{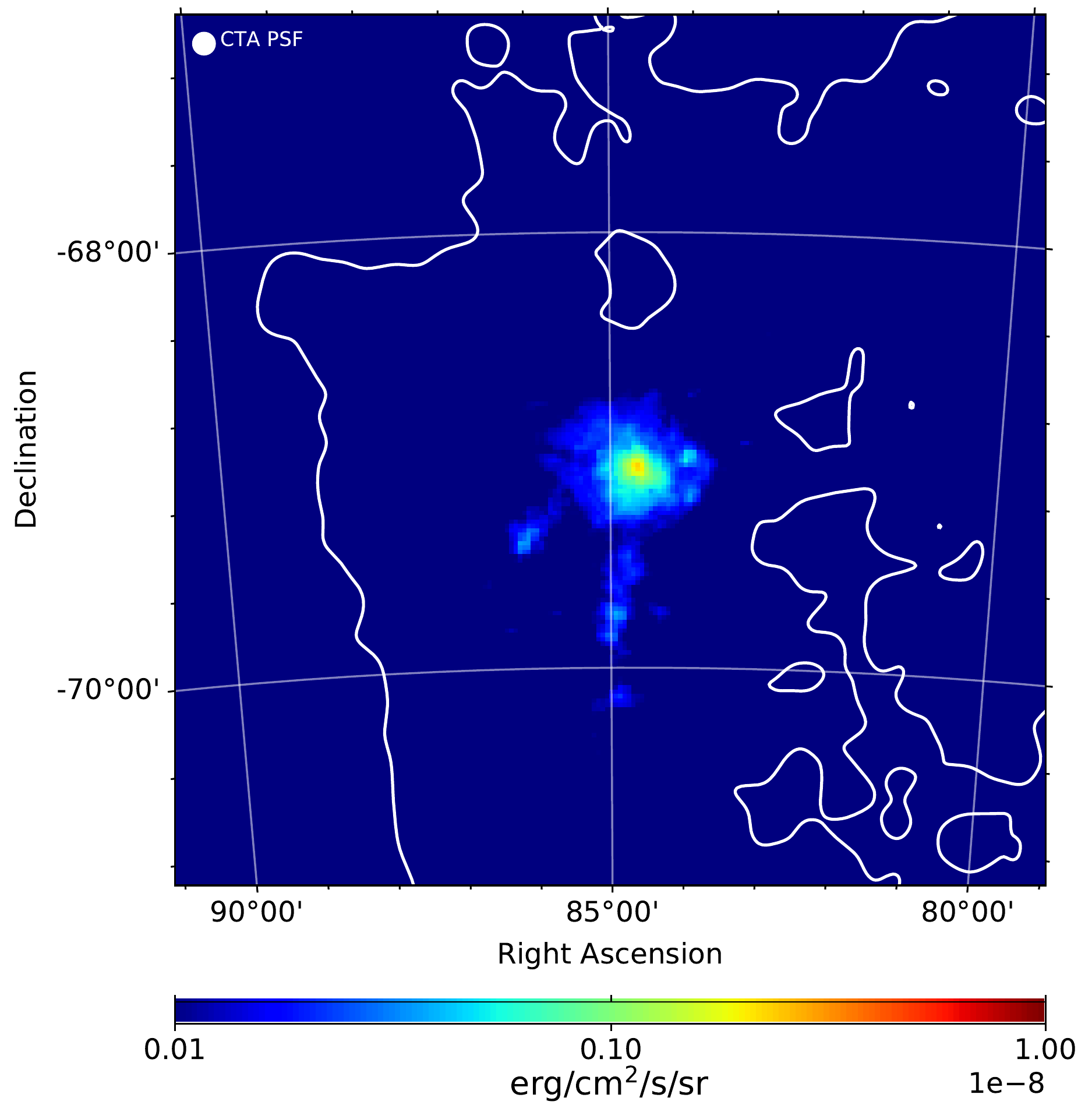}
\includegraphics[width=0.8\columnwidth]{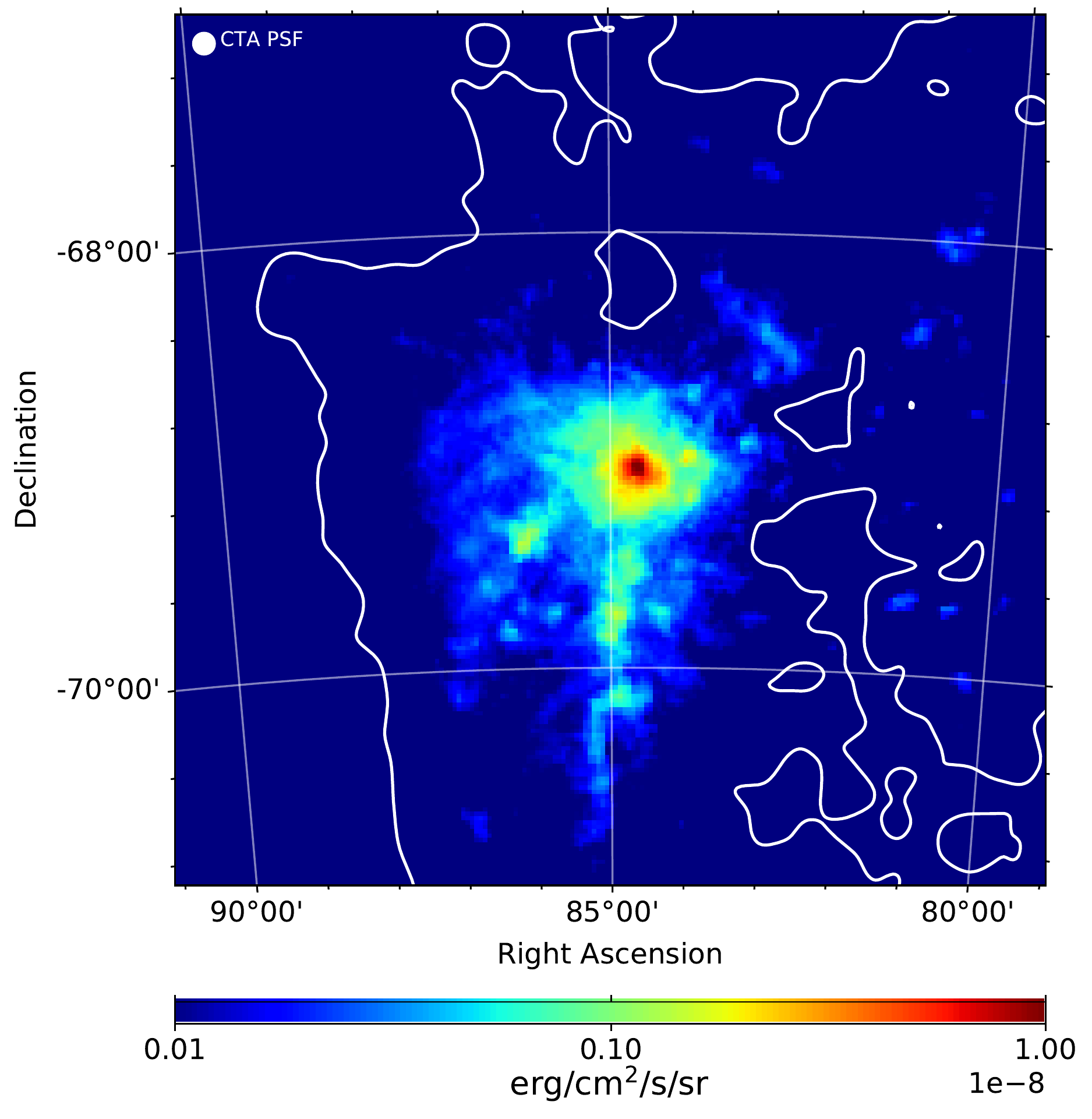}
\includegraphics[width=0.8\columnwidth]{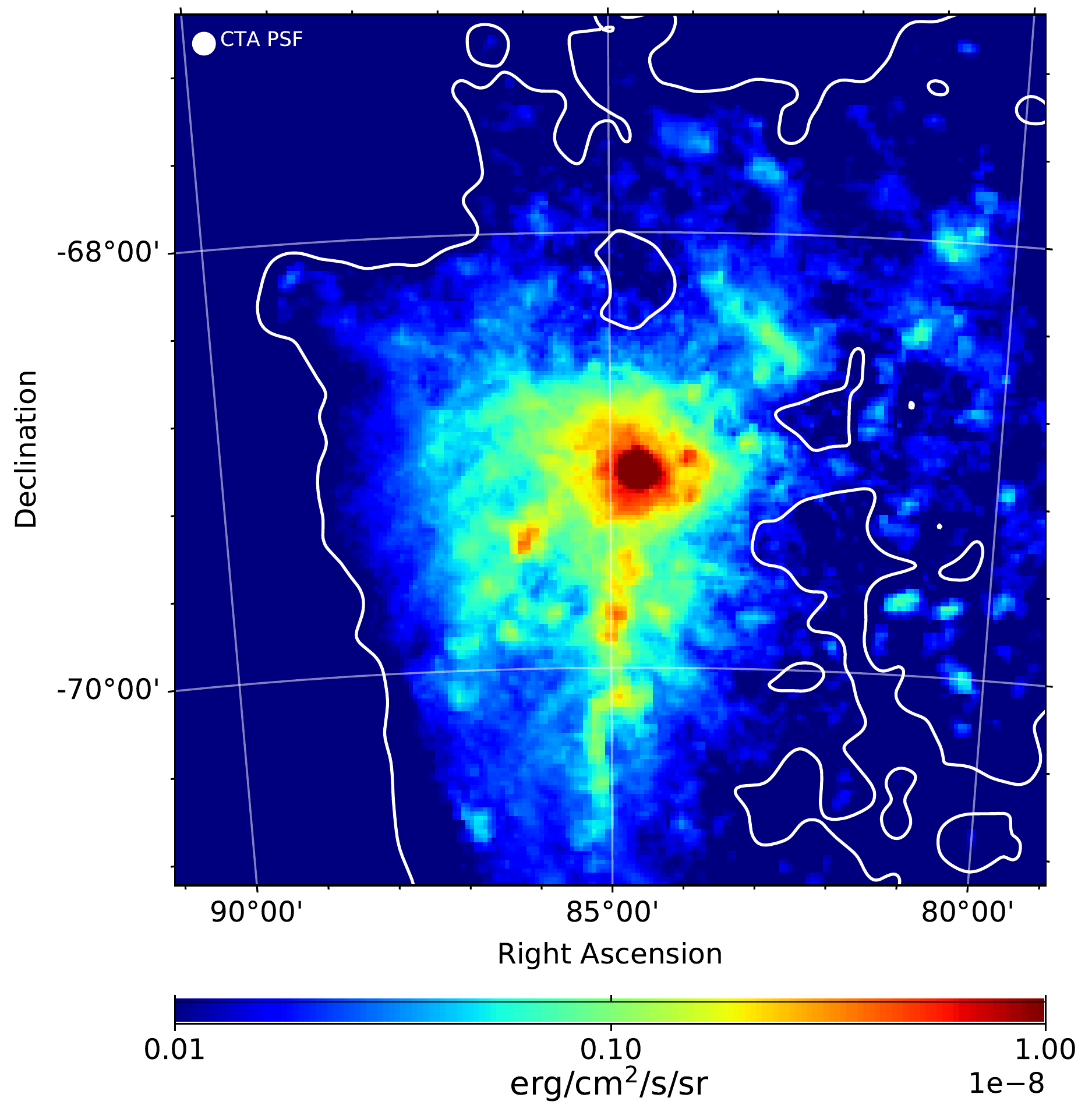}
\caption{Intensity maps at 1\tev for pion-decay emission from the 30 Doradus star-forming region, under three assumptions for the suppression of spatial diffusion within 100\,pc of the central source: reduction by a factor of 3, 30, and 300 (from top to bottom panel, respectively).}
\label{fig:model:sfr:30dor}
\end{center}
\end{figure}

We adopted a generic approach to the problem, independent of any specific scenario for \gls{cr} acceleration in \glspl{sfr} \citep[e.g., acceleration by individual stars in the cluster, or via repeated shocks, or at the cluster's wind termination shock; see][]{Parizot:2004,Ferrand:2010,Bykov:2018,Morlino:2021}. We restricted the physical description of the phenomenon to the following:
\begin{enumerate}
\item continuous injection of accelerated particles from a point source, with constant power and constant spectral shape assumed to be a power-law in momentum with exponential cutoff; in practice, we considered the injection of protons over a duration of 5\,Myr, with a hard spectrum with power-law index 2.25 and a cutoff at 1\pev; 
\item spatial diffusion in a medium characterized by a two-zone concentric structure for diffusion properties, with an outer region typical of the average ISM and an inner region where diffusion is suppressed relative to the ISM; the ISM diffusion coefficient has the form introduced in Sect. \ref{sec:model:ism}, and we considered diffusion suppression as an overall reduction by factors ranging from a few to a few hundred, within a distance of 100\,pc from the source;
\item alongside with spatial diffusion, particles experience homogenous energy losses over the entire volume explored, both the inner and outer diffusion regions; for protons, these consist of hadronic interactions losses for which we adopted the average gas density introduced in Sect. \ref{sec:model:ism} (limited variations around that value have little influence on the final outcome as the emission model is eventually corrected for the actual gas distribution around a given source; see below).
\end{enumerate}
The above assumptions allow to compute a three-dimensional emissivity kernel that we integrate along lines of sight over the typical thickness of the gas disk, and then renormalize in each direction by the actual gas distribution towards a given region (similarly to what was done for the large-scale pion-decay emission model in Sect. \ref{sec:model:ism}), finally yielding an intensity distribution. Figure \ref{fig:model:sfr:emiss} shows radial intensity profiles for different values of the suppression factor, and before correction of the intensity for any actual gas distribution. Figure \ref{fig:model:sfr:30dor} shows the resulting intensity distribution for the 30 Doradus region, after correction for the actual gas distribution and for three cases of diffusion suppression.

With such a description of the problem, we restrict the discussion to that of knowing under which conditions a given \gls{sfr} can be detected and identified as such. Specifically, we want to determine the requirements in terms of injection power and diffusion suppression for the latter two objectives to be fulfilled (the spectral index of the injection spectrum is also a relevant parameter but we already assumed as reference scenario a rather low value). Since such parameters are essentially unknown, it is not possible to incorporate all \glspl{sfr} in our global emission model for the galaxy in a coherent and justified way; instead, we will present below, in the results section, a parametric study of the prospects for the detection of 30 Doradus in the survey.

\subsection{Dark Matter} \label{sec:model:dm}

\begin{table}
\centering
\caption{Benchmark \gls{dm} profiles adopted in this work, using parameters extracted from Table II in \citet{Buckley:2015}, keeping the same nomenclature. The J-factor, in the last column, is integrated over a field of view of 10\deg.}
\label{tab:dmprofiles} 
\begin{tabular}{ c  c  c  c  c  c  c }
\hline
Profile & $\alpha$ & $\beta$ & $\gamma$ & $r_s$ & $\rho_0$ & J-factor \\
 &  &  &  & (kpc) & ($\rm M_{\odot}/kpc^3$) & ($\rm GeV^2/cm^5$) \\
\hline
{\tt iso-min} & 2 & 2 & 0 & 2.4 & $2.9\times10^7$ & $5.96\times 10^{20}$ \\
{\tt iso-mean} & 2 & 2 & 0 & 2.4 & $3.7\times10^7$ & $9.71\times 10^{20}$\\
{\tt iso-max} & 2 & 2 & 0 & 2.0 & $6.2\times10^7$ & $1.67\times 10^{21}$ \\
{\tt nfw-min} & 1 & 3 & 1 & 12.6 & $1.8\times10^6$ & $6.52 \times 10^{20}$ \\
{\tt nfw-mean} & 1 & 3 & 1 & 12.6 & $2.6\times10^6$ & $1.36 \times 10^{21}$ \\
{\tt nfw-max} & 1 & 3 & 1 & 17.6 & $2.5\times10^6$ & $2.85 \times 10^{21}$ \\
\hline
\end{tabular}
\end{table}

We assume that \gls{dm} is made of stable particles, which may however annihilate with each other, producing a shower of standard model particles. This in turn would lead to either direct or secondary production of gamma--rays, at energies of a few GeV and above, thus making them potentially detectable with \gls{cta} (and other gamma--ray telescopes). We address the reader to the vast literature existing on the \gls{dm} candidates and models complying with the many requirements and characteristics \citep[e.g.,][and references therein]{Bertone:2010, Boyarsky:2009, Bohm:2004, Hu:2000, Blais:2002nd}, adopting here for our purposes the generic definition of \gls{wimp}.

In the \gls{wimp} \gls{dm} scenario, the gamma--ray flux produced by the interaction follows:
\begin{equation}
\frac{\dop{\Phi}}{\dop{E}}=\frac{1}{8 \pi} \frac{\sigv}{m_{\chi}^2} \frac{\dop{N_{\gamma}}}{\dop{E}} \int_{\Delta\Omega}\int_{\rm l.o.s} \rho^2(\ell) \dop{\ell} \label{eq:dmflux}                                     
\end{equation}
where $\frac{\dop{\Phi}}{\dop{E}}$ is the gamma--ray flux produced, $\sigv$ is the \gls{dm} annihilation velocity-averaged cross section, $m_{\chi}$ is the mass of the \gls{dm} candidate, $\frac{\dop{N_{\gamma}}}{\dop{E}}$ is the gamma--ray spectrum produced by one single annihilation event (two \gls{dm} particles annihilating into a shower of  standard model particles), and $\rho(l)$ is the \gls{dm} density distribution within the target, with $l$ being a generic variable representing position along the line of sight. The latter integral term is also known as ``J-Factor'', and that is how it will be referred to from now on. Our goal is to test different \gls{dm} models according to the parameters of Eq. \ref{eq:dmflux}. Each \gls{dm} model will be included in the \gls{lmc} emission model as a new diffuse source, and the potential of \gls{cta} to detect a source of this nature will be assessed.

It is important to stress that, according to the custom in high-energy \gls{dm} searches with gamma--rays, we will treat both $\sigv$ and $\mdm$ as free parameters, and adopt ``single annihilation'' spectra assuming at each time the branching ratio of the interaction is one, namely that the entire annihilation takes place in the specific channel, then showing the results for different channels in order to bracket the possible outcome. Model--specific analyses relating $\sigv$ and $\mdm$ to the parameters of the particle theory (Lagrangian) can be performed separately, and are outside the scope of this study.

\begin{figure}
\begin{center}
\includegraphics[width=0.45\textwidth]{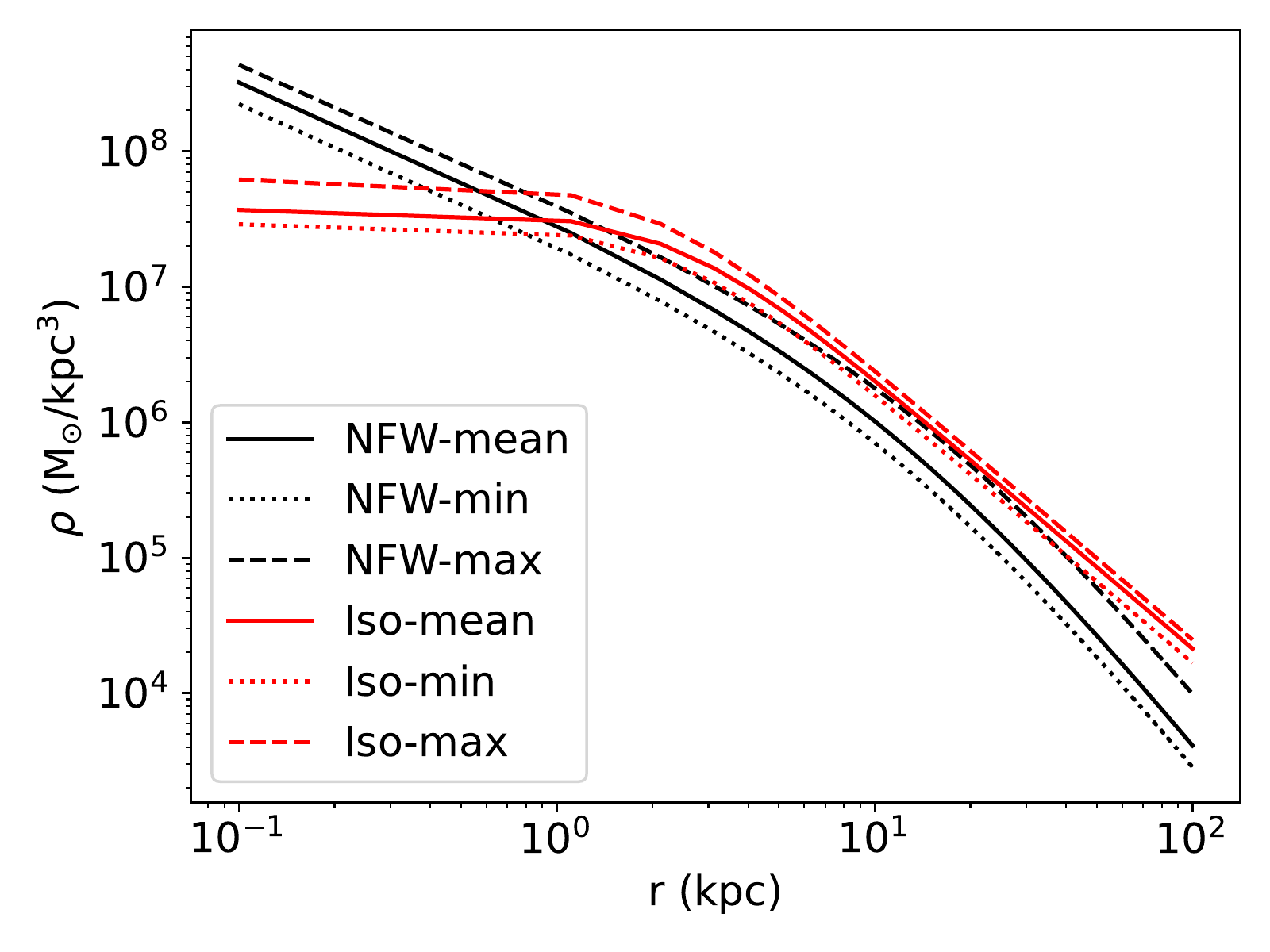}
\includegraphics[width=0.45\textwidth]{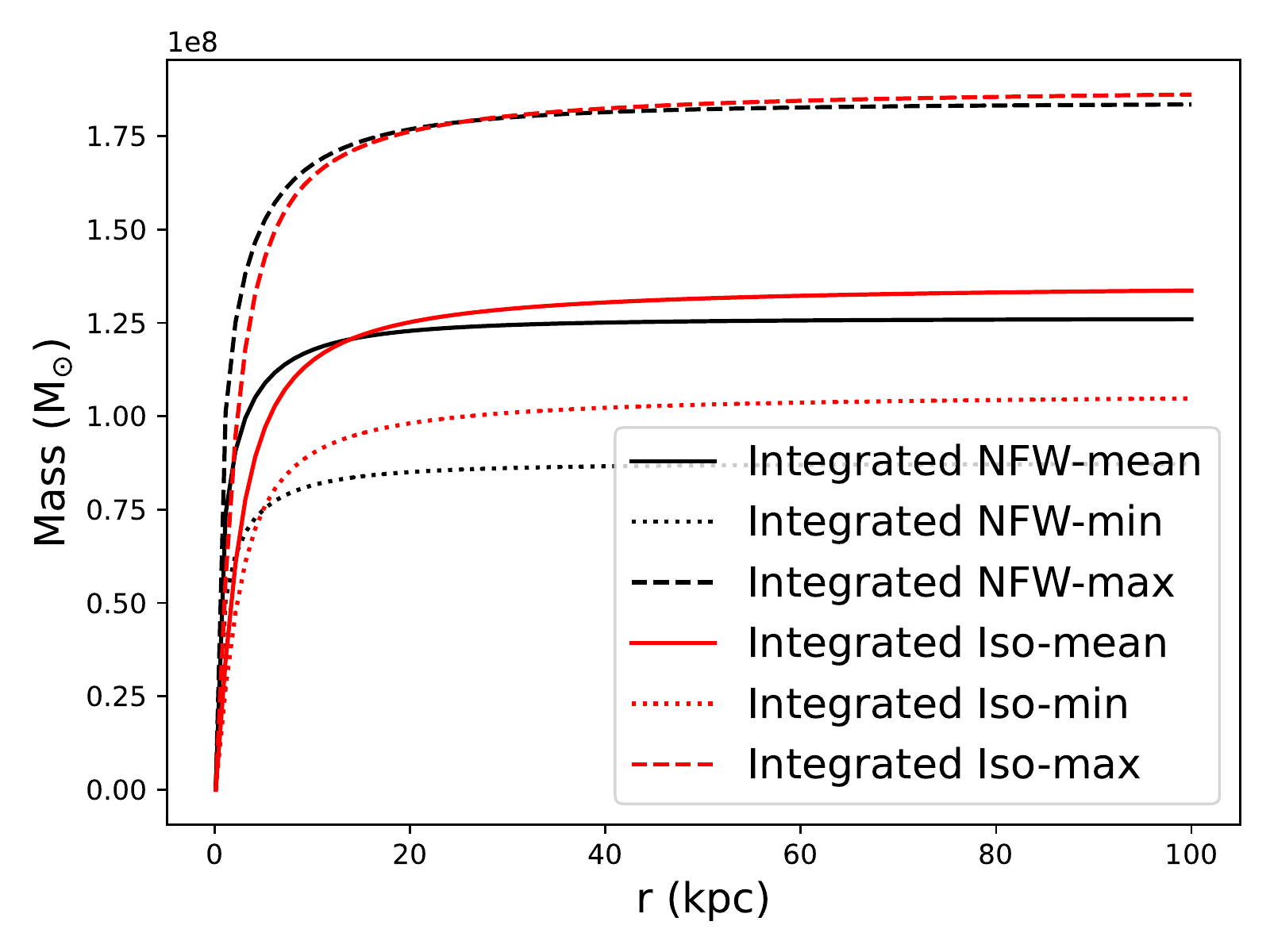}
\caption{\gls{dm} benchmark density profiles and integrated mass as a function of radius (top and bottom panel, respectively), computed using the parameters in Table \ref{tab:dmprofiles}.}
\label{fig:dmdensity} 
\end{center}
\end{figure}

The \gls{dm} distribution of the \gls{lmc} can be inferred by the gravitational structure of its disk, following the well-known ``rotation curve method''. This allows to infer the DM component of the gravitational potential for disk galaxies in an extended mass range, once a suitable set of tracers for the circular motion of the disk -- at different galactocentric distances -- and a good understanding of the visible component are available. DM is usually assumed to be spherically distributed, as there is little evidence for sizable departure from symmetry in hydrodynamical cosmological numerical simulations of galaxy formation and evolution, and we kept that assumption here. In order to be consistent with previous literature and allow direct comparison, while at the same time performing an independent analysis, we have closely followed the results of \citet{Buckley:2015}, which in turn adopt the data available in the literature and presented in \citet{Kim:1998, Luks:1992, vanderMarel:2013}.
We have adopted a Hernquist-Zhao six-parameter profile \citep{Zhao:1996}:
\begin{equation}
\rho(r) = \frac{\rho_{0}}{\left(\frac{r}{r_{S}}\right)^{\gamma}\left[ 1+\left(\frac{r}{r_{S}} \right)^{\alpha}\right]^{\frac{\beta-\gamma}{\alpha}}}
\label{gNFW}
\end{equation}
centered at $(\alpha_{\rm J2000},\delta_{\rm J2000})=(80.0\deg,-69.0\deg)$, where $r_{S}$ is the scale radius and $\rho_{0}$ is the characteristic density, both of which can be derived from the rotation curves of the \gls{lmc}. These last two parameters are the ones that most affect the total mass of the specific \gls{dm} halo, and therefore are most constrained by the observations of the \gls{lmc} baryonic mass and dynamics mentioned above. 
When $\alpha$ = 1 and $\beta$ = 3, the Hernquist-Zhao profile is called a generalised NFW profile (gNFW) with flexible inner \gls{dm} density slope $\gamma$. Setting $(\alpha,\beta,\gamma) = (1,3,1)$, we retrieve the \gls{nfw} profile \citep{Navarro:1996}, while an isothermal profile is obtained setting $(\alpha,\beta,\gamma) = (2,2,0)$. Variations of these two profiles have been tested, with their parameters shown in Table \ref{tab:dmprofiles} and plotted in Fig. \ref{fig:dmdensity}. These variations maximise and minimise the \gls{dm} density, but are still compatible with the rotation curves.

For the computation of the density profiles and their corresponding J-factors, we have used the public code {\tt CLUMPY}, a code for gamma--ray and neutrino signals from \gls{dm} structures \citep{Charbonnier:2012, Bonnivard:2016, Hutten:2019}. We have generated two-dimensional sky maps of the J-factor in Eq.~\ref{eq:dmflux}, with the parameters listed in Table \ref{tab:dmprofiles}, in a field of view of 10$^{\circ}$. The J-Factor integrated in the 10$^{\circ}$ field of view, given in the last column of the table, was also calculated with {\tt CLUMPY}. These sky maps correspond to the spatial part of the model and are combined with the gamma--ray spectra of different annihilation channels in the final \gls{dm} emission model.
 
For the spectral part of the \gls{dm} emission model (the $\dop{N_{\gamma}}/\dop{E}$ term in Eq. \ref{eq:dmflux}), the recipes from \citet{Cirelli:2011} were used, where the energy spectra of gamma--rays produced by different \gls{dm} annihilation channels are provided. We study the $b \overline b$, $W^+ W^-$, $\tau^+\tau^-$, and $\mu^+ \mu^-$ channels, including electro-weak corrections as computed in \citet{Ciafaloni:2011}.

\subsection{Emission model validation} \label{sec:model:val}

The consistency of our emission model with our present-day knowledge of the \gls{lmc} is checked against the following criteria :
\begin{enumerate}
\item The total predicted interstellar gamma-ray emission should not exceed the integrated flux measured at 10\gev. In \citet{Ackermann:2015}, 0.1-100\gev extended emission was decomposed into a large-scale emission seemingly correlated with the gas disk and a handful of additional components of unclear nature. We therefore assumed that the interstellar emission at 10\gev predicted by our model should not exceed the sum of all extended emission components found in \citet{Ackermann:2015}, which corresponds to an upper limit in flux density $F(E)$ at 10\gev of $E^2 \times F(E) = 2 \times 10^{-11}$\feunit.
\item Gamma-ray emission on small scales $\leq 50$\,pc, either from individual sources or fine structures in interstellar emission, should not exceed upper limits on point-like emission in the 1-10\gev and 1-10\tev bands. As typical values, we used upper limits on SN 1987A derived in \citet{Ackermann:2015} and \citet{Abramowski:2015} and corresponding to $5.4 \times 10^{-13}$\feunit\ at 10\gev and $1.2 \times 10^{-13}$\feunit\ at 1\tev. Constraints have certainly improved since these studies due to increased exposure, but by a factor likely $\leq 2$.
\item The total predicted interstellar radio synchrotron emission should not exceed the integrated flux measured at 1.4\,GHz. Synchrotron emission at frequency 1.4\,GHz arises mostly from 5\gev \gls{cr} leptons in the assumed mean 4\,$\mu$G interstellar magnetic field \citep{Blumenthal:1970}, which are not those contributing to the gamma-ray signal in the \gls{cta} band, but such a check guarantees some continuity and consistency in leptonic emission over a wide range of energies. The total radio flux at 1.4\,GHz measured from ATCA+Parkes observations is 426\,Jy \citep{Hughes:2007}. This includes an estimated 50\,Jy from background point sources and $\geq$20\% from thermal bremsstrahlung from ionized gas. The total synchrotron emission therefore has an intensity $\leq$291\,Jy at most. We checked that the assumptions made in computing the large-scale interstellar emission of leptonic origin lead to a total interstellar synchrotron intensity below this limit. 
\end{enumerate}
In practice, with the assumptions introduced in Sect. \ref{sec:model}, the three criteria listed above are fulfilled and the comparison confirms that the various components of our model are well calibrated.

The criterion on the total interstellar gamma-ray emission is the most constraining since our baseline model predicts a 10\gev flux that nearly saturates the maximum acceptable level. The Fermi-LAT measurement is thus very informative already and will restrict the allowed space for some parameters: for instance, it is not possible to strongly increase the \gls{cr} proton injection rate while keeping all other parameters untouched. Actually, for given \gls{cr} injection and spatial diffusion indices, the luminosity of the large-scale pion-decay component is set to first order by the product of injection rate, gas mass, and the inverse of spatial diffusion normalization, and none of these parameters is known to high accuracy (the most constrained being the gas mass, with 30\% uncertainty, and the least constrained the diffusion coefficient). 

The criterion on small-scale, almost point-like, gamma-ray emission is also fulfilled. Small-scale emission peaks in the pion-decay model are about 4 times below the 10\gev limit, which again shows that Fermi-LAT measurements are already constraining, and about 50 times below the 1\tev limit. Small-scale emission peaks in the inverse-Compton model are more than two orders of magnitude below the 10\gev and 1\tev limits, in the baseline model relying on the average \gls{ism} model. Using the gas-rich \gls{ism} model instead, which comes with a higher \gls{isrf} and may be more appropriate for regions harbouring rich stellar populations, leads to higher emission maxima by a factor 2-3, while using a harder injection spectrum for \gls{cr} electrons, with a power-law index of 2.25, increases the small-scale emission peaks at 1\tev by a factor 20, which still remains largely below the current constraints. Last, in our realization of the source populations model, apart from a couple of extreme objects that would already be detectable with H.E.S.S., which nicely matches the current census of gamma-ray sources in the \gls{lmc}, the populations of \glspl{snr}, \glspl{isnr}, \glspl{pwn}, and pulsar halos reach emission levels that are at most 2-3 times below the 10\gev and 1\tev limits. 

The predicted 1.4\,GHz synchrotron intensity in our model is 53\,Jy, which is far below the limit defined above and may suggest our model would significantly underpredict the actual level of synchrotron emission. Our model assumes that 1\% of the total \gls{cr} injection power is in the form of primary electrons, in agreement with estimates for the Galaxy \citep{Strong:2010}, so increasing the predicted interstellar synchrotron flux would have to be done by acting on other parameters. Reducing the diffusion coefficient normalization or increasing the injection power are not options because of constraints on the pion decay component, which saturates the allowed level at 10\gev (although a smaller diffusion coefficient would be allowed if the diffusion region has a finite size; see the comment in appendix \ref{app:crprop}). Instead, a slightly higher interstellar magnetic field would alleviate the discrepancy. Taking into account the contribution from secondary electrons would also reduce the gap, although by no more than 30\% according to the estimate of the contribution of secondaries presented below. On the other hand, the measured flux includes more than purely interstellar emission, for instance contribution from a population of unresolved discrete \glspl{snr}, or thermal emission from ionized gas, if it contributes more than 20\% of the total radio flux. 

\begin{figure}
\begin{center}
\includegraphics[width=\columnwidth]{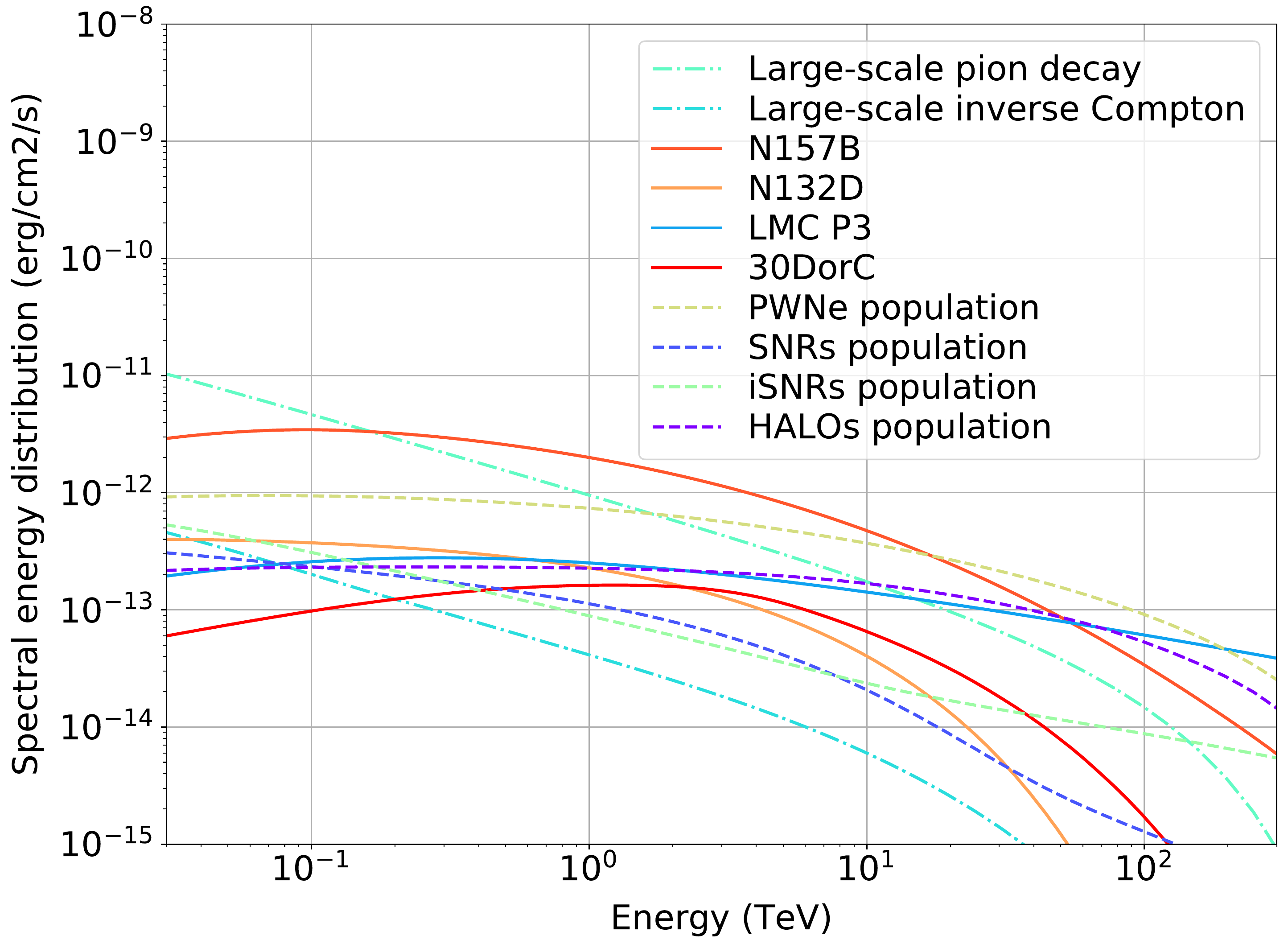}
\caption{Spectral energy distribution of all emission components in our model (except for possible emission from dark matter and \gls{crs} in the vicinity of \glspl{sfr}): the four currently known TeV sources (SNR N~132D, PWN N157D, 30 Doradus C, and binary LMC P3), large-scale interstellar emission from pion decay and inverse-Compton scattering, and total emission from the \glspl{snr}, \glspl{isnr}, \glspl{pwn}, and pulsar halos populations.}
\label{fig:model:seds}
\end{center}
\end{figure}

The integrated emission spectra for all components discussed above are shown in Fig.\ref{fig:model:seds}, except for possible emission from 30 Doradus. PWN N~157B dominates the galaxy's emission over most of the $0.1-100$\tev range; as a confirmation of its outlier nature, it is two to three times more luminous at 1\tev than all \glspl{pwn} in our synthetic population taken together. Similarly, N~132D is as bright at 1\tev as the rest of the \glspl{snr} population, including interacting ones. The second most luminous component overall is large-scale interstellar pion-decay emission up to about 1\tev, and the mock \glspl{pwn} population at higher energies. Large-scale inverse-Compton interstellar emission appears as a comparatively subdominant component, in agreement with \citet{Persic:2022}.

Secondary leptons from charged pions are not included in our model. The magnitude of their contribution can be estimated from the luminosity of the pion-decay gamma-ray component, which is $3.7 \times 10^{37}$\punit above 1\gev. In that range, the spectrum of secondary leptons is very similar to that of gamma rays, albeit at least two times lower \citep{Kelner:2006}. The spectrum of secondary leptons at injection would thus be close to a power law with index 2.7 and luminosity above 1\gev of  $<1.9 \times 10^{37}$\punit. Compared to the injection spectrum for primary electrons, a power law with index 2.65 and luminosity above 1\gev of $6.5 \times 10^{37}$\punit, this suggests that secondaries would be a correction to our model at the level of <30\% in the energy range of interest.

\section{Survey simulation and analysis} \label{sec:simana}

\subsection{Observation simulations} \label{sec:simana:obs}

Observation simulation in this work means the generation of high-level data, ready for scientific analysis. In practice, it produces lists of events such as those that passed Cherenkov light detection, shower reconstruction, and gamma-hadron discrimination. Photon and background events are randomly generated from a model for celestial emission in the region of interest, a description of the instrument's performances, and a definition of the observations. The latter is addressed in the next section and sets the number, positions, and durations of all pointings in the survey. Due to the availability of instrument responses for a limited subset of observing conditions and in the absence of realistic scheduling constraints, this is done under simplifying assumptions.

The performances of the \gls{cta} observatory are defined in instrument response functions and background rates. The former describes how an incident gamma ray is converted into a measured event and is factorized into three terms for effective area, point spread function, and energy dispersion. The latter defines how events that are not gamma rays in origin are generated over the data space as a function of observing conditions. In this work, we used response \textit{South\_z40\_50h} of the \textit{prod5-v0.1} release\footnote{\url{https://zenodo.org/record/5499840\#.Y9D4nvGZMbY}}, valid over 50\gev-200\tev and appropriate for observations at 40\deg\ zenith angle averaged over azimuth angles (the \gls{lmc} will be seen at best at $\sim 46$\deg\ elevation from the southern site). This is a description of the southern array that will be built during an initial construction phase of the project and will consist of 14 medium-sized telescopes and 37 small-sized telescopes. The project may later evolve towards a final full-scope configuration comprising 4 large-sized telescopes, 25 medium-sized telescopes and 70 small-sized telescopes on the southern site, but we did not investigate the prospects for such a configuration.

The emission model $\mathcal{S}$ for the \gls{lmc} was introduced in previous sections and will be convolved with the instrument response functions $\mathcal{R}$ for given observing conditions. In the particular case of this work, we consider mainly sources that are steady (on human time scales). The only exception to this would be the gamma-ray binary LMC P3, which has its emission modulated by orbital motion, but we do not focus on that particular aspect and assumed its phase-averaged emission to be constant. We therefore leave aside the general time dependence of the signals and the biases introduced by the instrument in photon arrival time measurements.

In the field of a pointing defined by parameters $\bold{p}$, the expected event measurement rates at a given position in the sky $\bold{r}$ and reconstructed energy $E$ can be split into background events $\mathcal{M}^B$ and gamma-ray events $\mathcal{M}^S$:
\begin{align}
&\mathcal{M}^B_{\bold{p}} (\bold{r},E,\bold{\theta_B}) = \mathcal{B}(\bold{r},E | \bold{\theta_B}, \bold{p}) \\
&\mathcal{M}^S_{\bold{p}} (\bold{r},E,\bold{\theta_S}) = \iint  \mathcal{S}(\bold{r_0},E_0 | \bold{\theta_S}) \mathcal{R}(\bold{r},E | \bold{r_0},E_0,\bold{p}) \dop{E_0} \dop{\bold{r_0}} \\
& \mathcal{R}(\bold{r},E | \bold{r_0},E_0,\bold{p}) = \mathcal{A}(\bold{r_0},E_0,\bold{p}) \mathcal{P}(\bold{r} | \bold{r_0},E_0,\bold{p}) \mathcal{D}(E | \bold{r_0},E_0,\bold{p}) 
\end{align}
Lists of events with reconstructed energy, direction, and arrival time are randomly generated for each pointing from the above expected measurement rates.

The dependence of background rate $\mathcal{B}$, effective area $\mathcal{A}$, point spread function $\mathcal{P}$ and energy dispersion $\mathcal{D}$ on vector $\bold{p}$ encapsulates the general dependence of the instrument response on observation conditions (e.g., detector center and orientation on the sky, pointing zenith and azimuth). In this work, however, we will neglect energy dispersion. Vectors $\bold{\theta_S}$ and $\bold{\theta_B}$ hold the various spectral and spatial parameters on which celestial and background models $\mathcal{S}$ and $\mathcal{B}$ depend. In the following, we will denote $\bold{\theta_S^T}$ and $\bold{\theta_B^T}$ the true values of these parameters, and $\bold{\tilde{\theta}_S}$ and $\bold{\tilde{\theta}_B}$ their estimated values (from the maximum likelihood estimator, see below).

\subsection{Pointing strategy} \label{sec:simana:pnt}

The \gls{lmc} is slightly larger than the field-of-view of \gls{cta} so the survey will involve a number of overlapping observations to encompass the full galaxy. Because of the diversity of targets in the \gls{lmc}, the optimal pointing layout is not obvious: concentrating the exposure over a smaller patch of the sky will maximize sensitivity to point-like sources in the innermost regions (e.g., \glspl{pwn}, \glspl{snr} and pulsar halos); conversely, spreading the exposure well beyond the outskirts of the \gls{lmc} will include nearly empty fields and provide more contrast for the detection of very extended sources with a size comparable to the field-of-view of the instrument (e.g., interstellar emission on galactic scales). 

To ensure uniformity of the exposure at all energies, we aimed at a pointing pattern with a large number of short-duration pointings equally-spaced from one another. We considered a layout in which pointings are distributed along concentric hexagons and equidistant from their closest neighbours. We searched for the optimal pointing spacing by evaluating its impact on the sensitivity to several representative source morphologies and positions in the survey field: (i) a point source at the position of SN~1987A, i.e., in central regions; (ii) a point source at the position of star-forming region N11, i.e., on the edge of the galaxy; (iii) interstellar pion-decay emission with the morphology computed in our emission model.

\begin{figure}
\begin{center}
\includegraphics[width=\columnwidth]{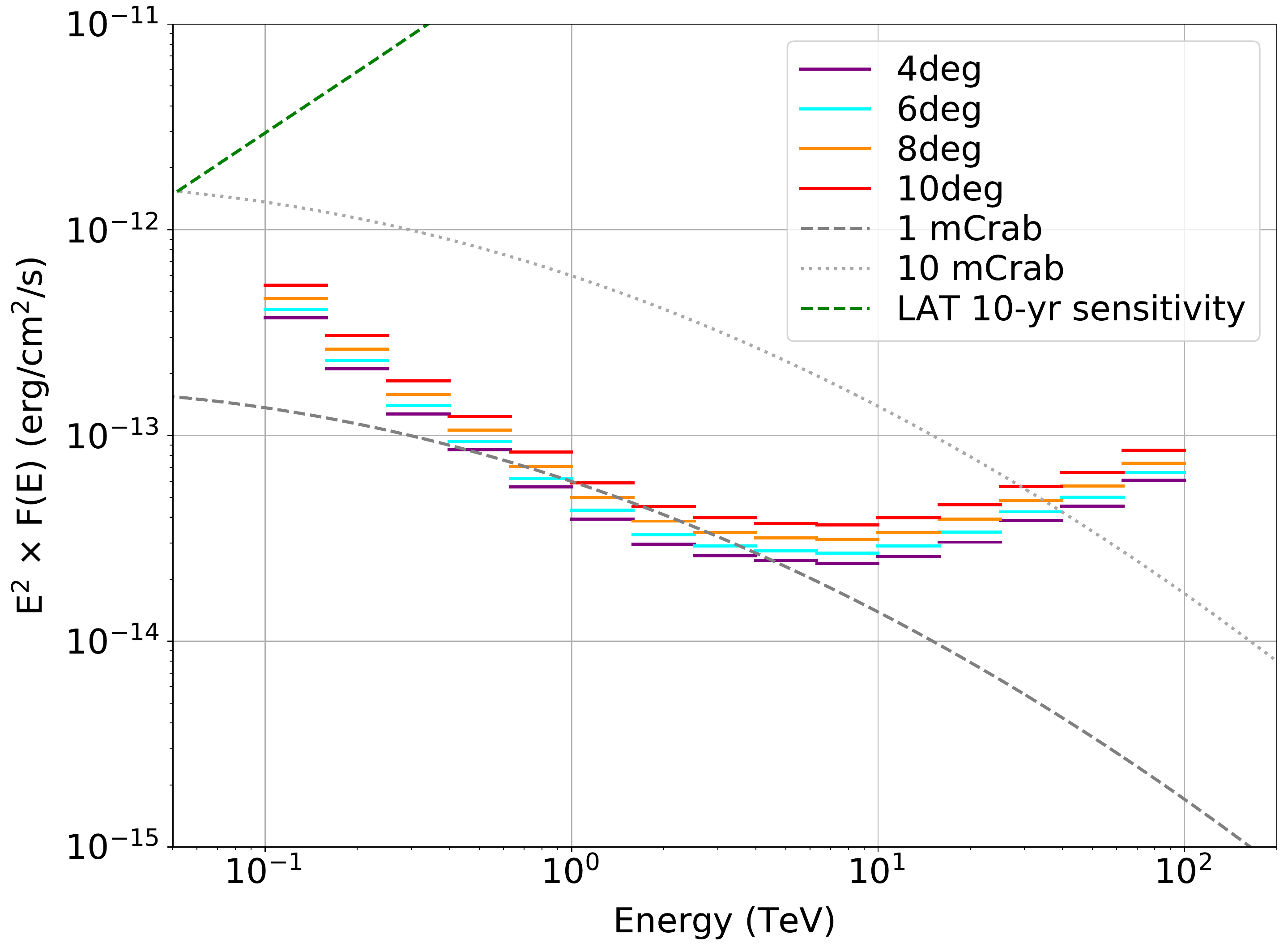}
\includegraphics[width=\columnwidth]{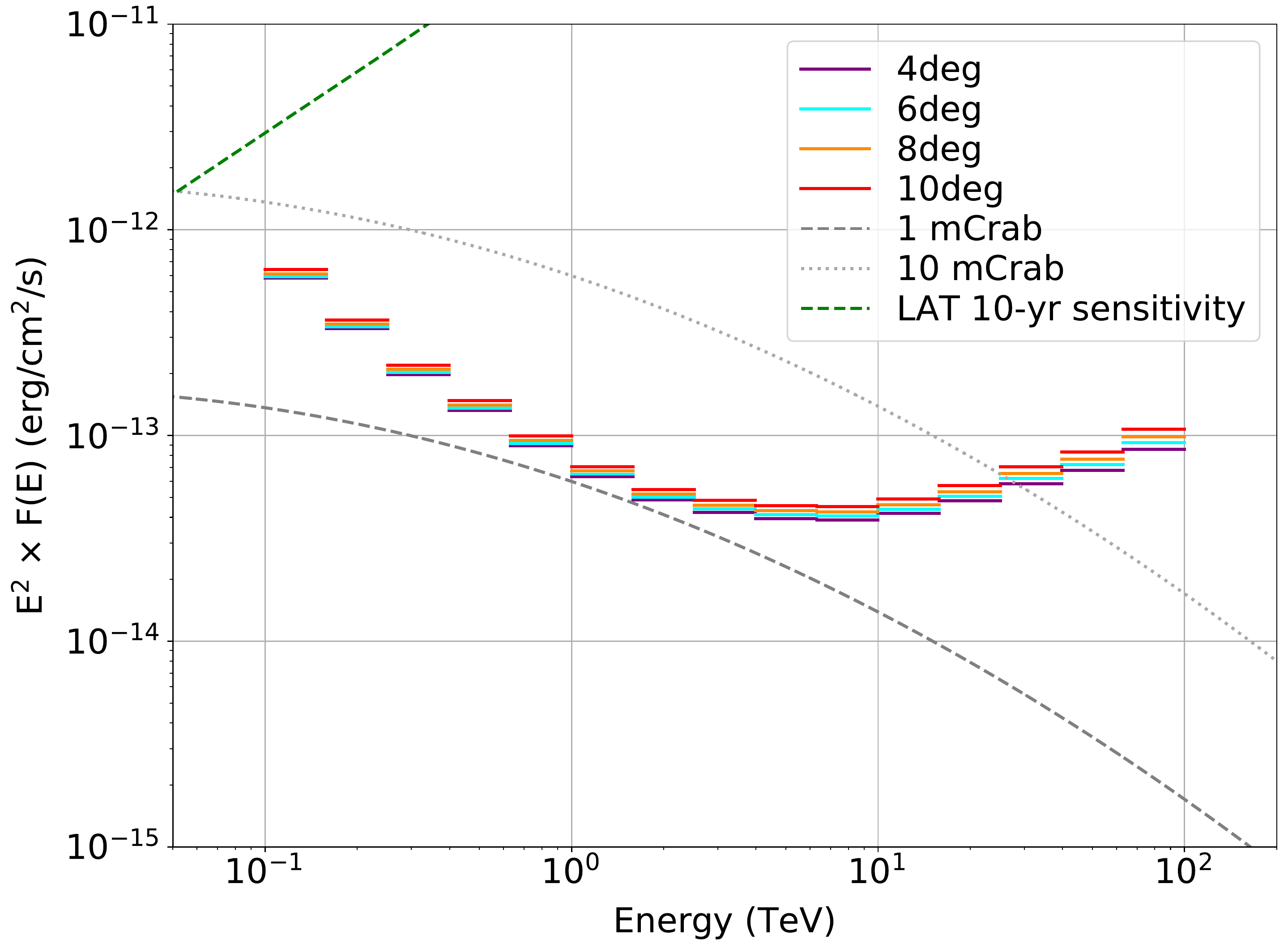}
\includegraphics[width=\columnwidth]{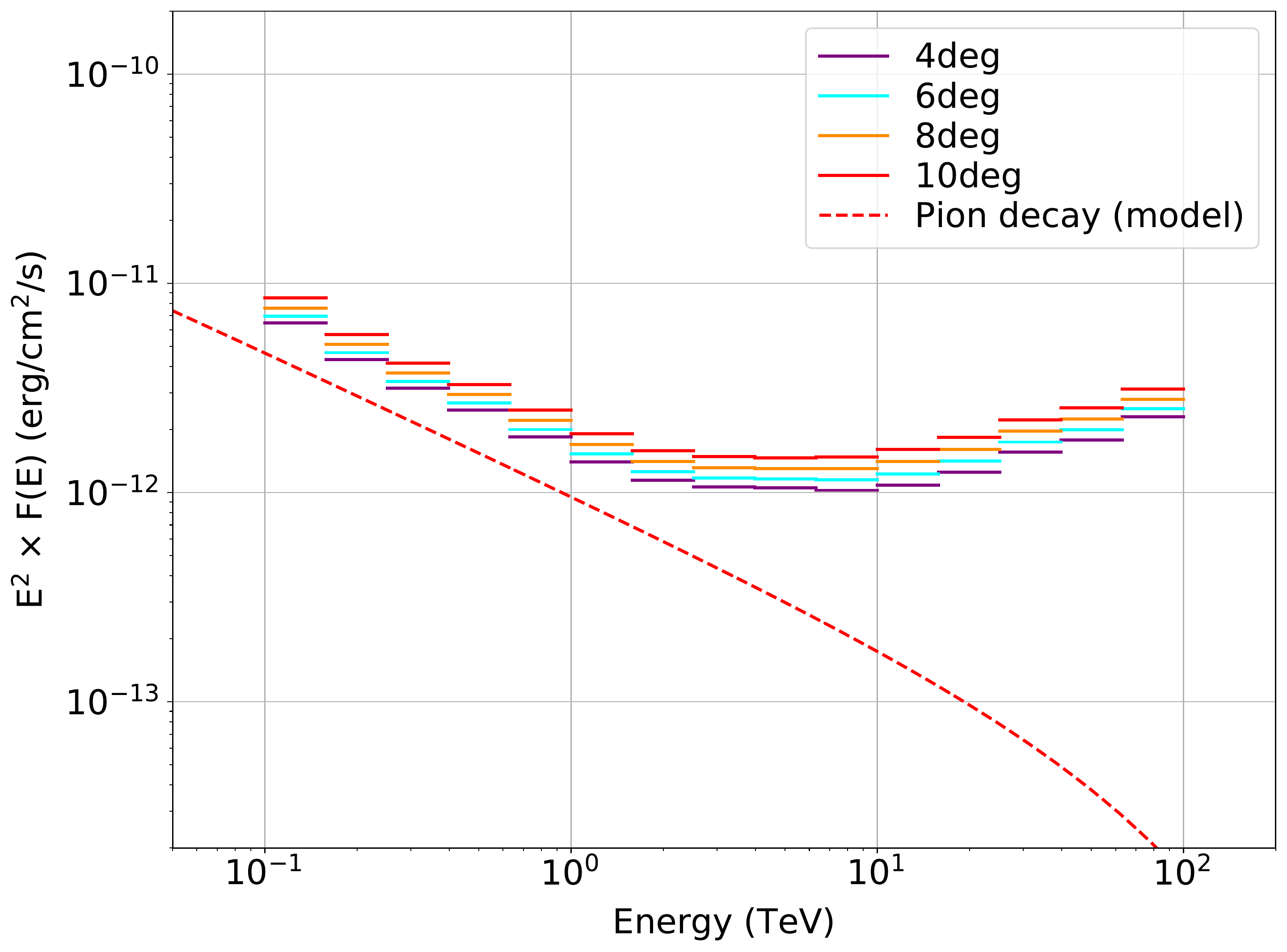}
\caption{Sensitivity curves for point sources at the positions of SN~1987A and N11 and for our large-scale pion-decay emission model (from top to bottom panel, respectively), as a function of the pointing pattern. The latter consists of a large number of equally-spaced pointings arranged along concentric hexagons, and we compared different spreads of the exposure, from 4\deg\ to 10\deg. Overlaid as reference for the point sources are spectra of the Crab nebula rescaled by factors of 0.01 and 0.001, and the LAT P8R3 10-yr sensitivity for Galactic coordinates (l,b)=(120\deg,45\deg).}
\label{fig:simana:sens}
\end{center}
\end{figure}

We compared different spreads of the exposure, parameterized as the maximum extent of the pattern (i.e. full width of outermost hexagon) and varied from 4\deg\ to 10\deg. Sensitivity curves for the three test sources listed above are presented in Fig. \ref{fig:simana:sens} and, in the case of point sources, compared to spectra of the Crab nebula rescaled by factors of 0.01 and 0.001 \citep{Abeysekara:2019}, and to the LAT P8R3 10-yr sensitivity\footnote{\url{https://www.slac.stanford.edu/exp/glast/groups/canda/lat\_Performance.htm}} for Galactic coordinates (l,b)=(120\deg,45\deg). The meaning and computation of sensitivity curves will be defined below but we emphasize that we checked that some parameters of the data analysis have no impact on the conclusions reported here (in particular the size of the region of interest used in the binned analysis).

As could be anticipated, sensitivity to a centrally located point-like source improves as exposure becomes more concentrated, by a factor $<2$ that is approximately constant over the energy band. The sensitivity gain seems however to flatten as the pattern size drops below 6\deg. Sensitivity to a diffuse source such as our large-scale pion-decay emission model shows a similar behaviour, although less pronounced at low energies $<1$\tev. There does not seem to be any benefit of adding nearly empty fields to the survey, probably because interstellar emission as modeled here has sufficient structure on small angular scales that it can be easily disentangled from instrumental background. In contrast to the two previous sources, sensitivity to a point source located in the outskirts of the galaxy is nearly insensitive to the exposure spread, with a maximum effect at the level of 30\% at 100\tev. This likely results from exposure spread being compensated by a higher number of pointings having their centers close to the boundaries of the galaxy.

\begin{figure}
\begin{center}
\includegraphics[width=\columnwidth]{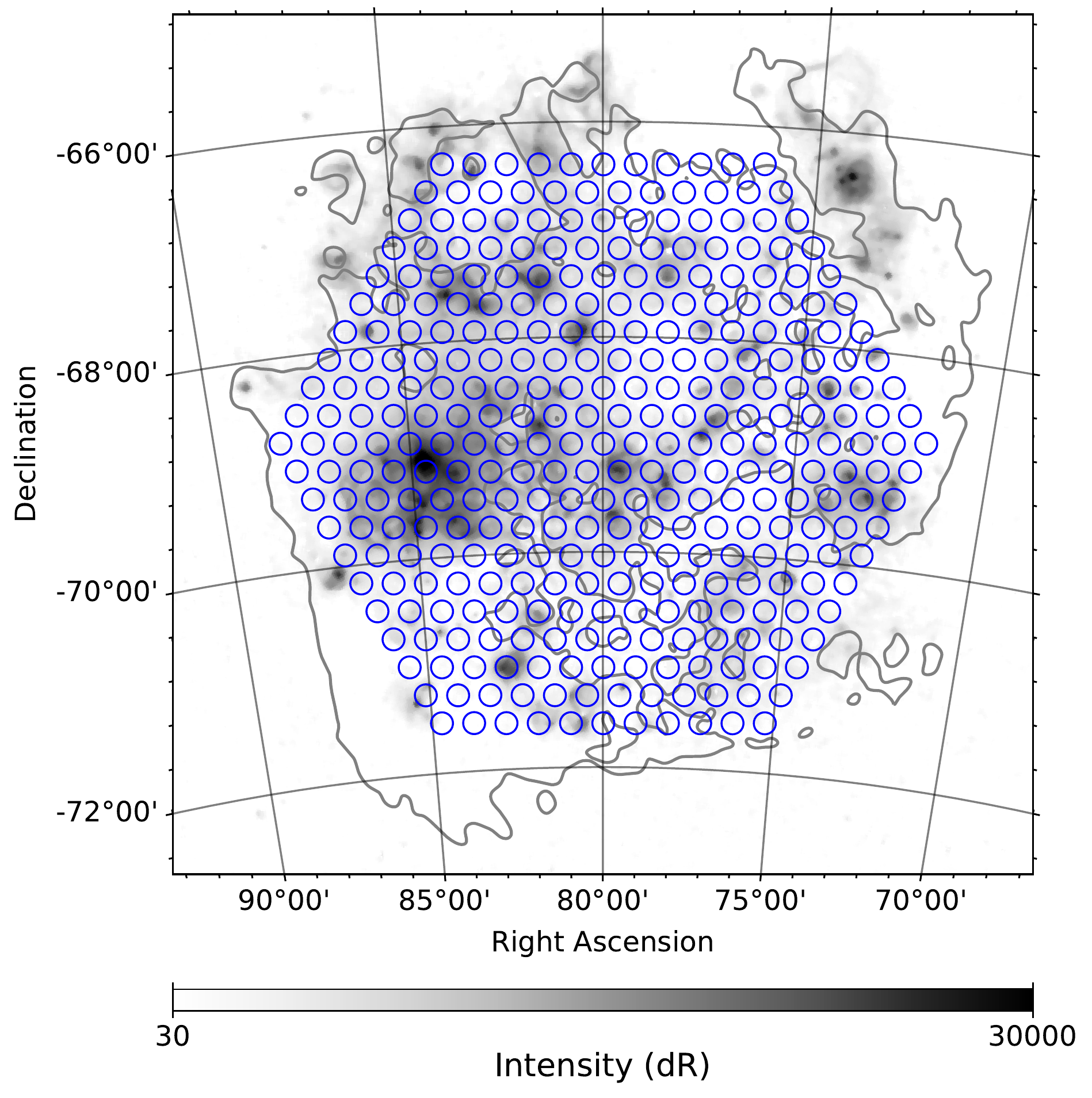}
\caption{Pointing pattern adopted in the study. Each blue circle corresponds to the center of one among 331 equally spaced pointings of nearly 1h each, arranged along 10 concentric hexagons centered on $(\alpha_{\rm J2000},\delta_{\rm J2000})=(80.0\deg,-69.0\deg)$. The whole pattern spans 6\deg, and each pointing has an effective gamma-ray field-of-view of $3-4\deg$ in off-axis radius in the $0.1-10$\tev range. For comparison, the anticipated angular resolution of the southern array at 1\tev is about $0.05-0.06$\deg.}
\label{fig:simana:pntpattern}
\end{center}
\end{figure}

We eventually adopted a pointing pattern consisting of 331 pointings of 3698\,s each, equally spaced along 10 concentric hexagons centered on $(\alpha_{\rm J2000},\delta_{\rm J2000})=(80.0\deg,-69.0\deg)$. This corresponds to a spacing between adjacent pointings of 0.3\deg\ and to a maximum extent of 6\deg, as illustrated in Fig. \ref{fig:simana:pntpattern}. In a given pointing, sensitivity typically drops beyond an off-axis angle of $3-4\deg$, depending on energy in the $0.1-10$\tev range, which ensures a broad enough coverage of the galaxy and its outskirts. Although a smaller pointing spacing would have provided a slightly better sensitivity to all emission components tested here, keeping a wide enough survey footprint covering the galaxy at large is key for making discoveries.

\subsection{Simulated data analysis} \label{sec:simana:fit}

Source characterization is achieved by maximum likelihood estimation of the parameters of a model for some region of interest in the simulated observations. In this work, we used a likelihood analysis for binned data and Poisson statistics, as implemented in the ctools package \citep{Knodlseder:2016}, and we stacked data such that events from all pointings are added and instrument response functions are averaged over all observations (see ``Combining observations" in ctools user manual). The applicability of such an approach was demonstrated in \citet{Knodlseder:2019} on real data from the H.E.S.S. experiment.

The region of interest is typically a $8\deg \times 8\deg$ square centered on $(\alpha_{\rm J2000},\delta_{\rm J2000})=(80.0\deg,-69.0\deg)$ and aligned on equatorial coordinates, except for  DM analyses where a $10\deg \times 10\deg$ region was used to fully capture the very extended signals considered. Within this area, events are binned in $0.02\deg \times 0.02\deg$ spatial pixels and 0.1\,dex spectral intervals spanning 100\gev to 100\tev. The high lower energy bound compared to the full range that should be accessible to \gls{cta} is warranted by the rapid degradation of performance $<100$\gev for zenith angles $>40$\deg\ at which the \gls{lmc} will be observed.

The logarithm of the likelihood is computed from measured number of counts in the data cube $D$ and predicted number of counts in the model cube $M$:
\begin{align}
&D=\left\{n_{i,j}\right\}, M=\left\{\mu_{i,j}\right\} \\
&\ln\mathcal{L} (\mathcal{M}^B,\mathcal{M}^S | D) = \sum\limits_{i,j} n_{i,j} \ln{\mu_{i,j}}  - \mu_{i,j}
\end{align}
In the above equations, $i$ is the index on spectral intervals and $j$ the index on spatial pixels. The dependence of the likelihood on the parameters and functional form of the models for instrumental background and celestial emission is expressed as a dependence on model functions $\mathcal{M}^B$ and $\mathcal{M}^S$. 

For a given set of observations, predicted model counts are obtained by sampling $\mathcal{M}^B$ and $\mathcal{M}^S$ at bin centers $(\bold{r}_j,E_i)$, multiplying by bin volume $\Delta \Omega_j  \Delta E_i$ and pointing livetime $\Delta T_{\bold{p}}$, and finally summing over all pointings.
\begin{align}
&\mu_{i,j} = \mu_{i,j}^B + \mu_{i,j}^S \\
&\mu_{i,j} = \sum\limits_{\bold{p}}( \mathcal{M}^B_{\bold{p}}(\bold{r}_j,E_i,\bold{\theta_B}) + \mathcal{M}^S_{\bold{p}}(\bold{r}_j,E_i,\bold{\theta_S}) )  \times \Delta \Omega_j \times \Delta E_i \times \Delta T_{\bold{p}}
\end{align}
In the framework of this analysis, models are factorized into two terms $\mathcal{M} = \mathcal{H} \times \mathcal{F}$, with $\mathcal{H}$ describing the (possibly energy-dependent) morphology and $\mathcal{F}$ defining the spectral shape.

Optimum parameters $\bold{\tilde{\theta}_S}$ and $\bold{\tilde{\theta}_B}$ are searched for iteratively such that the likelihood is maximized:
\begin{align}
\ln\mathcal{L} (\tilde{\mathcal{M}}^B,\tilde{\mathcal{M}}^S | D) = \ln\mathcal{L} (\mathcal{M}^B(\bold{\tilde{\theta}_B}),\mathcal{M}^S(\bold{\tilde{\theta}_S}) | D) 
\end{align}
The significance of a source component or source parameter in the model is assessed in terms of the \gls{ts}:
\begin{align}
TS = 2 \ln \frac{\mathcal{L}(\tilde{\mathcal{M}}^B,\tilde{\mathcal{M}}^S_{\rm test}|D)}{\mathcal{L}(\tilde{\mathcal{M}}^B,\tilde{\mathcal{M}}^S_{\rm null}|D)} 
\end{align}
where $\tilde{\mathcal{M}}^S_{\rm test}$ is the optimum model including the additional tested source component or parameter, for instance an additional source component with non-zero normalization or a cutoff parameter in the spectrum of a component, and $\tilde{\mathcal{M}}^S_{\rm null}$ is the optimum model without it \citep{Cash:1979}. A value $TS>25$ is adopted as a criterion for significant detection of a source (either over the full energy range or within narrower intervals as in the case of sensitivity curves). In practice, the fitting of model parameters and calculation of the significance of sources was done using the \textit{ctlike} function from ctools.

For low-significance source components, we calculated flux upper limits, usually in narrow energy bins. Keeping the spatial and spectral shape parameters of the component of interest fixed, and varying only its normalization, Wilks' theorem \citep{Wilks:1938} states that the $\gls{ts}$ function asymptotically approaches a $\chi^2$-distribution with one degree of freedom under the null hypothesis. We therefore adopted as upper limit the flux normalization such that $\gls{ts} = 2.71$, which corresponds to a 95\% \gls{cl} upper limit. The calculation of flux upper limits was performed with function $\textit{ctulimit}$ from ctools and used in particular to set constraints on the \gls{dm} annihilation cross section $\sigv$, as described in Sect. \ref{sec:model:dm}. 

In the analyses presented below, we frequently made use of so-called \textit{Asimov} data sets. An \textit{Asimov} data set \citep{Cowan:2011} is a representative data set in which the number of counts in a given bin in the data space corresponds exactly to the model expectation, without any statistical fluctuation. When fitting a model to such a data set, the true values of the model parameters are perfectly recovered, if the model used for simulation and fitting is the same. The main interest of such an approach is to get mean results for source significance and detection upper limits, without the need for a large number of realizations of simulated data (which in the present case is quite computer-intensive as it would require simulating the full 340h of observations about 1000 times or more for each analysis setup).

\section{Detection prospects} \label{sec:res}

\subsection{Sensitivities} \label{sec:res:sens}

We begin by presenting the survey sensitivity to some of the components in our emission model. Sensitivity was computed in independent energy bins, typically 5 per decade, as the source flux yielding on average a detection with $\gls{ts}=25$ in each bin. It depends strongly on source morphology but also on position in the field, first because the exposure is slightly uneven and second because other neighbouring or overlapping emission components may increase the detection threshold. Yet, the sensitivity curves presented below were computed for each source independently, considering only the instrumental background as other source component and not the full emission model. In most cases, this is partly justified by the fact that diffuse interstellar components in our baseline emission model are too weak to seriously alter sensitivity. In specific regions, however, source confusion may be a problem and limit our ability to detect and/or separate weak source components. The data points for the sensitivities to the emission components discussed below are provided in Table \ref{tab:sens} for convenience and may be used in future assessments of the detectability of some sources for specific models (e.g., SN~1987A).

\begin{figure}
\begin{center}
\includegraphics[width=\columnwidth]{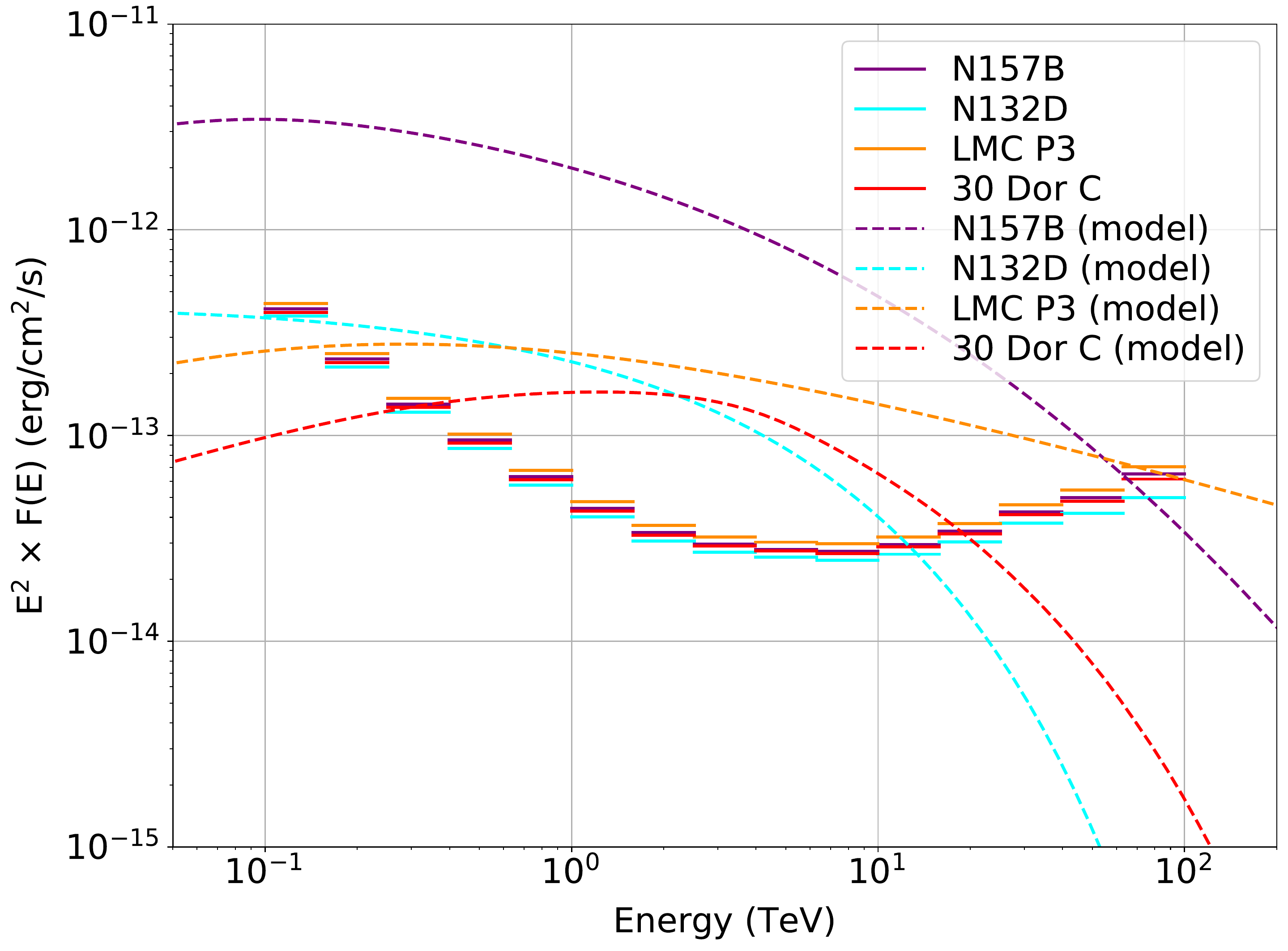}
\caption{Sensitivity of the survey to point sources at the positions of the four \gls{vhe} objects currently known in the \gls{lmc}. Overlaid in dashed lines are the true spectra of the components in the emission model.}
\label{fig:res:sens:knw}
\end{center}
\end{figure}

Figure \ref{fig:res:sens:knw} presents sensitivity curves for point sources at the positions of the four \gls{vhe} objects currently known in the \gls{lmc}, together with the original true spectra used for these components in our emission model. Obviously, these objects will be detected with high significance in small individual energy bins over most of the band, thus allowing fine spectral studies as will be discussed below in Sect. \ref{sec:res:knw}. Also apparent in this plot is the fact that sensitivity slightly depends on position in the field, with sensitivity loss of the order of $\sim$20\% over most of the range, peaking at $\sim$50\% at the very highest energies $\sim100$\tev, as we go from central (N~132D, cyan sensitivity curve) to more peripheral (LMC P3, orange sensitivity curve) positions within the galaxy. We checked how the sensitivity to 30 Doradus C is affected by the proximity of the very strong N~157B source, and found that it degrades only by $\sim$10\% at the lowest energies $\sim100$\gev.

\begin{figure}
\begin{center}
\includegraphics[width=\columnwidth]{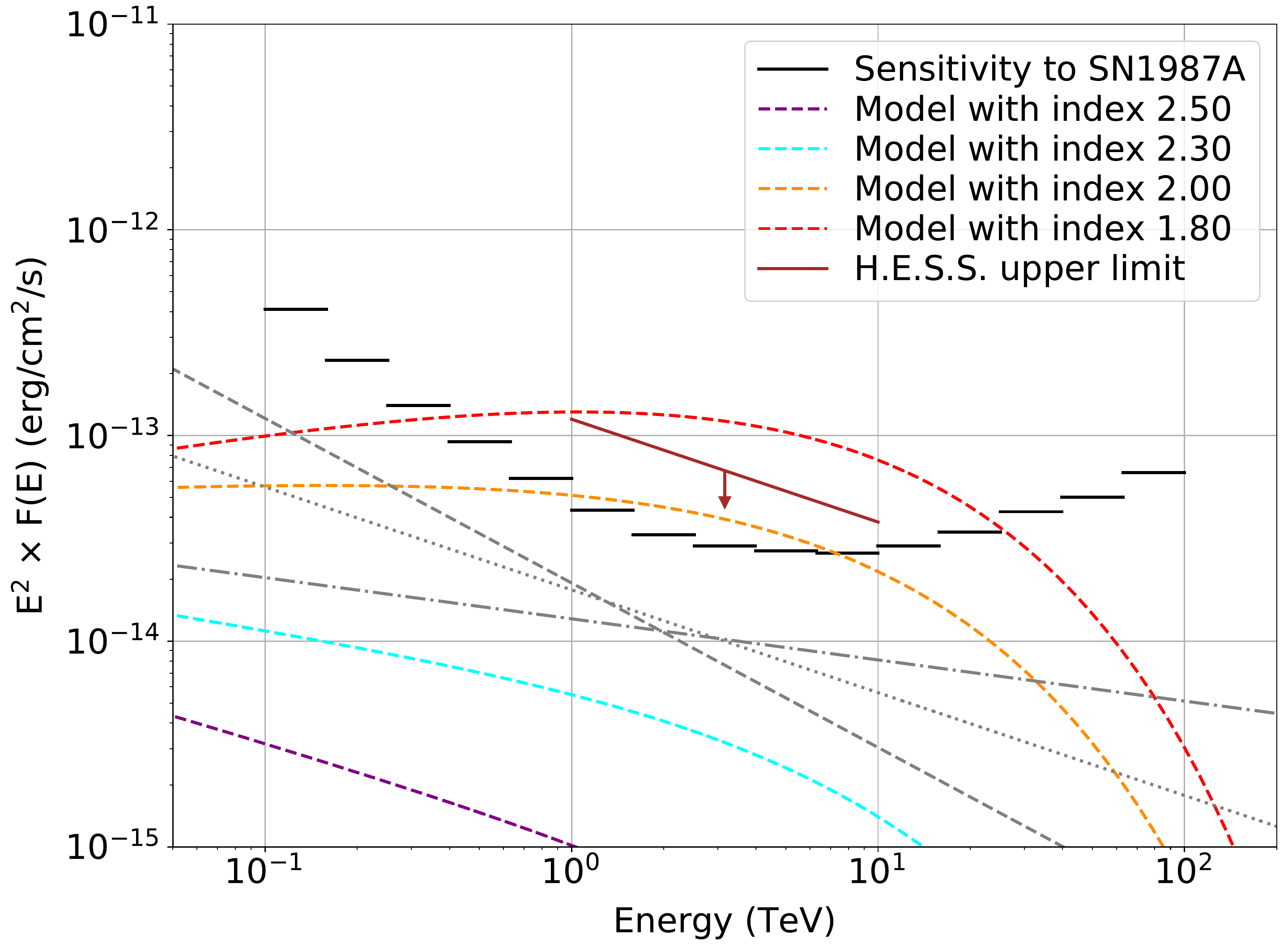}
\caption{Sensitivity of the survey to a point source at the position of SN~1987A, compared to simple models for pion decay emission from the remnant (see Sect. \ref{sec:res:87a} for details). Also shown as gray lines are power-law spectra with photon indices 2.2, 2.5, and 2.8 normalized such that they yield a broadband detection with $TS=25$ for a point source at the position of SN~1987A.}
\label{fig:res:sens:sn87a}
\end{center}
\end{figure}

Figure \ref{fig:res:sens:sn87a} displays the sensitivity of the survey to a point source at the position of SN~1987A, in order to help assessing the detectability of the object as function of different models of particle acceleration in the remnant. A more quantitative discussion of the prospects is provided below, in Sect. \ref{sec:res:87a}. Also shown as gray lines are power-law spectra normalized such that they yield a global detection with $TS=25$ for a point source at the position of SN~1987A. Those were determined iteratively from a series of simulated observations and model fits (using only instrumental background and SN~1987A as model components), adjusting the normalization of the SN~1987A source until convergence to a global $TS=25$. As a reference for the discussion to follow, these detection thresholds correspond to 1-10\tev luminosities in the range $2.0-2.4 \times 10^{-14}$\feunit. Taking into account variations at the 20\% level on sensitivity, depending on position in the field, we will henceforth assume that the typical 1-10\tev sensitivity of the survey to point sources is in the range $1.6-2.9 \times 10^{-14}$\feunit.

\begin{figure}
\begin{center}
\includegraphics[width=\columnwidth]{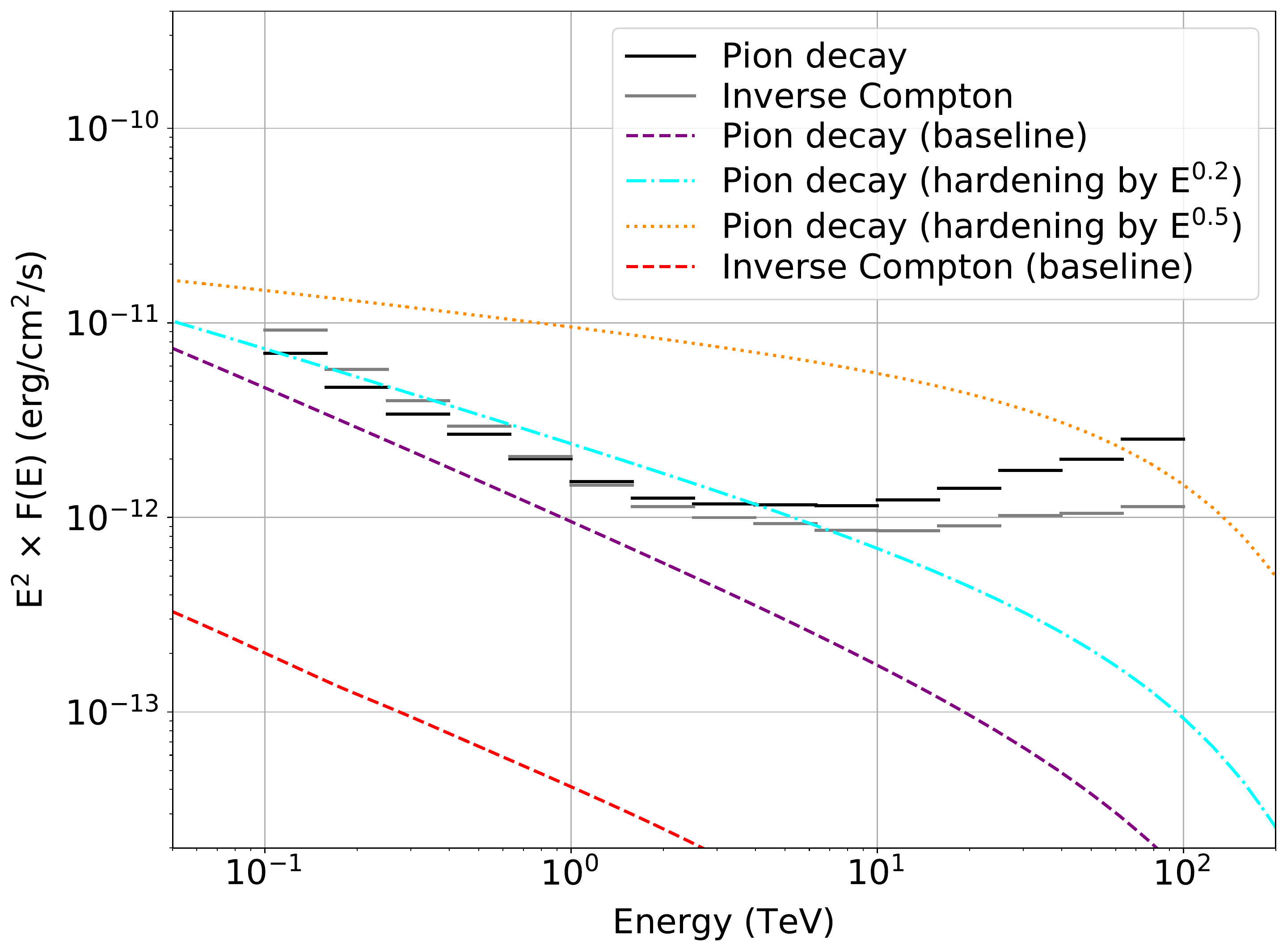}
\caption{Sensitivity of the survey to large-scale interstellar emission from pion decay and inverse-Compton scattering, with the specific morphology resulting from our emission model assumptions. Overlaid in dashed lines are the baseline spectra of the components in the emission model. In the case of pion decay, two variants are also shown in which the spectrum was hardened by multiplication with a power-law of index +0.2 and +0.5 and a pivot energy at 10\,GeV.}
\label{fig:res:sens:pionic}
\end{center}
\end{figure}

Figure \ref{fig:res:sens:pionic} shows the sensitivity of the survey to large-scale interstellar emission from pion decay and inverse Compton scattering. In contrast to the sensitivities introduced before for point sources, which were mostly influenced by the position in the survey footprint, the survey to such extended components depends on the specific morphology resulting from assumptions made when building the emission model (e.g., diffusion coefficient, distribution of \gls{cr} injection sites; see Sect. \ref{sec:model:ism}). In addition, we emphasise that the sensitivity is computed in very optimistic conditions, using the true source energy-dependent morphology and instrumental background properties in a full spatial-spectral likelihood fit. Nevertheless, this provides a useful reference as the best situation one can hope for and Fig. \ref{fig:res:sens:pionic} shows that, even if this ideal setup, large-scale inverse-Compton emission remains out of reach while pion decay could be detected with modest significance ($TS \sim 60$). The plot however illustrates the potential of the survey to detect or constrain large-scale pion-decay emission if it happens to be harder than assumed in our model. Such a prospect will be addressed more extensively below, in Sect. \ref{sec:res:ism}.

\begin{table*}
\centering
\caption{Sensitivities to several emission components considered in this work, expressed in $E^2 \times F(E)$ with the threshold flux density $F(E)$ defined in Sect. \ref{sec:res:sens}: in columns four to seven, sensitivities to point sources at the positions of currently known \gls{vhe} sources in the \gls{lmc}; in the eighth column, sensitivity to a point source at the position of SN~1987A; in the last two columns, sensitivities to extended emission templates for large-scale pion-decay and inverse-Compton radiation from the \gls{ism}. The sensitivities were computed in each energy bin as the flux yielding a detection with $\gls{ts}=25$ on average, for a binned and stacked analysis over a $8\deg \times 8\deg$ region of interest centered on  $(\alpha_{\rm J2000},\delta_{\rm J2000})=(80.0,-69.0)$.}
\label{tab:sens}
\begin{tabular}{cccccccccc}
\hline
Bin & Lower bound & Upper bound & N~157B & N~132D & 30 Dor C & LMC-P3 & SN~1987A & LMC-IC & LMC-Pion \\
 & (TeV) & (TeV) & & & & \feunit & & & \\
\hline
1 & 0.100 & 0.158 & 4.137e-13 & 3.803e-13 & 3.968e-13 & 4.387e-13 & 4.103e-13 & 9.193e-12 & 6.962e-12 \\
2 & 0.158 & 0.251 & 2.356e-13 & 2.152e-13 & 2.261e-13 & 2.505e-13 & 2.322e-13 & 5.772e-12 & 4.664e-12 \\
3 & 0.251 & 0.398 & 1.421e-13 & 1.297e-13 & 1.370e-13 & 1.515e-13 & 1.397e-13 & 3.986e-12 & 3.397e-12 \\
4 & 0.398 & 0.631 & 9.498e-14 & 8.671e-14 & 9.179e-14 & 1.015e-13 & 9.336e-14 & 2.941e-12 & 2.675e-12 \\
5 & 0.631 & 1.000 & 6.297e-14 & 5.732e-14 & 6.084e-14 & 6.759e-14 & 6.185e-14 & 2.052e-12 & 2.001e-12 \\
6 & 1.000 & 1.585 & 4.419e-14 & 4.022e-14 & 4.290e-14 & 4.769e-14 & 4.329e-14 & 1.468e-12 & 1.527e-12 \\
7 & 1.585 & 2.512 & 3.370e-14 & 3.070e-14 & 3.282e-14 & 3.648e-14 & 3.294e-14 & 1.135e-12 & 1.256e-12 \\
8 & 2.512 & 3.981 & 2.961e-14 & 2.703e-14 & 2.908e-14 & 3.216e-14 & 2.898e-14 & 9.983e-13 & 1.173e-12 \\
9 & 3.981 & 6.310 & 2.796e-14 & 2.563e-14 & 2.744e-14 & 3.030e-14 & 2.743e-14 & 9.277e-13 & 1.161e-12 \\
10 & 6.310 & 10.000 & 2.729e-14 & 2.478e-14 & 2.665e-14 & 2.970e-14 & 2.682e-14 & 8.583e-13 & 1.149e-12 \\
11 & 10.000 & 15.849 & 2.942e-14 & 2.647e-14 & 2.869e-14 & 3.207e-14 & 2.899e-14 & 8.546e-13 & 1.229e-12 \\
12 & 15.849 & 25.119 & 3.418e-14 & 3.034e-14 & 3.326e-14 & 3.726e-14 & 3.396e-14 & 9.046e-13 & 1.413e-12 \\
13 & 25.119 & 39.811 & 4.256e-14 & 3.750e-14 & 4.129e-14 & 4.608e-14 & 4.259e-14 & 1.021e-12 & 1.744e-12 \\
14 & 39.811 & 63.096 & 4.999e-14 & 4.176e-14 & 4.793e-14 & 5.419e-14 & 5.022e-14 & 1.049e-12 & 1.993e-12 \\
15 & 63.096 & 100.000 & 6.492e-14 & 4.992e-14 & 6.139e-14 & 7.047e-14 & 6.610e-14 & 1.138e-12 & 2.521e-12 \\
\hline
\end{tabular}
\end{table*}

\subsection{Known point sources} \label{sec:res:knw}

Figure \ref{fig:res:sens:knw} makes it clear that the sensitivity level reached by the survey will allow detections of the four currently known \gls{vhe} sources with very high significance and fine spectral studies over most of the energy range. Each one of these sources would deserve its own broadband modelling, taking into account a wealth of multi-wavelength data, to establish which particular aspects of particle acceleration can be addressed by the \gls{cta}. This is however left out of the scope of the present paper and, in the following, we illustrate the potential of future spectral studies and tie these prospects to considerations on the physics at play in these objects.

Pulsar wind nebula N~157B belongs to a plerion involving PSR~J0537$-$6910 and the $\sim$5000-year old SNR 0537.8$-$6910. The whole system has a diameter of 24\,pc and presumably results from the explosion of a $\sim$25\msol O8-O9 star and is likely associated with the OB association LH99 \citep{Chen:2006,Micelotta:2009}. The 16-ms pulsar is the most rapidly spinning and most powerful young pulsar known, with a spin-down luminosity of $\dot E = 4.9 \times 10^{38}$\,erg\,s$^{-1}$ \citep{Marshall:1998}. N~157B was detected in H.E.S.S. observations and is the first and only PWN detected outside of the Milky Way in this energy band \citep{Abramowski:2012,Abramowski:2015}. This is accounted for by the high spin-down power of the pulsar combined with an intense infrared photon field for inverse-Compton scattering. In the \gls{cta} survey of the \gls{lmc}, N~157B would be detected significantly from 100\gev up to 100\tev. With the physical model assumed here, which saturates the 2015 upper limit from Fermi-LAT at 100\gev \citep{Ackermann:2016}, detection in the low-energy range $<1$\tev will allow an unambiguous connection of the \gls{he} and \gls{vhe} domains and the exploration of a spectral range that probes a region of the particle spectrum where most of the spin-down power of the pulsar may be channeled \citep[if the particle spectrum peaks around 100\gev as assumed in][]{Zhang:2008}. In particular, it might be useful to figure out why N~157B seems to be a much less efficient particle accelerator than the Crab \citep{Abramowski:2015}. At the other end of the band, a better characterization of the cutoff region around 100\tev will provide key information about the maximum energies that can be reached and retained in young and powerful pulsar wind nebulae, especially in the context of X-ray observations revealing a cometary nebula expanding to large volume into a low-pressure parent \gls{snr} \citep{Chen:2006}. Pulsed gamma-ray emission from PSR~J0537$-$6910, similar to that observed from the Crab pulsar \citep{Ansoldi:2016}, will also most likely be searched for in the 100\gev-1\tev range, although non-detection at GeV energies suggests a low pulsed fraction \citep{Ackermann:2015}, and the strong steady emission from the nebula will make it difficult to extract such a signal.

N~132D is the brightest X-ray (and gamma-ray) \gls{snr} in the galaxy, with an estimated age of $2450 \pm 195$\,yr \citep{Law:2020}. The blast wave has a physical diameter of about 20\,pc and exhibits a quite irregular morphology characterized by a horseshoe shape, with the southern part plowing through dense molecular clouds \citep{Sano:2015}, giving rise to copious X-ray thermal emission \citep{Borkowski:2007}, while the northern part is blowing out in a lower-density medium and is much dimmer in most bands. It was detected with both Fermi-LAT and H.E.S.S. \citep{Abramowski:2015,Ackermann:2016,Abdalla:2021}, with a hard spectrum $<100$\gev in the Fermi-LAT band and a much softer spectrum $>1$\tev in the H.E.S.S. band. N~132D is extremely bright in the GeV range, actually the highest luminosity of all known GeV \glspl{snr} \citep{Bamba:2018,Acero:2016}, and therefore seems to be a prolific accelerator, in transition between young GeV-hard/TeV-bright and middle-aged GeV-bright/TeV-dim \glspl{snr}. In this evolutionary stage, the highest-energy particles $\gtrsim 100$\tev have escaped the remnant and the content of $\sim$10\tev particles may be close to its maximum \citep{Ptuskin:2003,Ptuskin:2005}. Transition objects like N~132D may be key to study how \glspl{cr} progressively enrich the \gls{ism} and propagate in the vicinity of sources. The authors of \citet{Bamba:2018} estimated that the gamma-ray emission cannot be predominantly leptonic in origin because the high GeV luminosity and faint non-thermal X-ray emission impose a small magnetic field strength and exceedingly large energy in accelerated electrons; the gamma-ray emission thus has to be mostly hadronic, and the authors estimated that particles still contained in N~132D have a maximum energy of 30\tev. In this scenario, emission in the 10\tev range is a mix of inverse-Compton and pion decay contributions. An additional contribution to the signal, not considered in \citet{Bamba:2018} but mentioned in \citet{Vink:2021}, may come from the highest-energy particles that escaped the remnant and ought to radiate efficiently in the large amounts of molecular gas located in the close neighbourhood \citep[$10^5$\msol, according to][]{Banas:1997}. Given the young age of N~132D, and if some kind of self-confinement is at work despite the abundance of neutrals in the medium, such particles should still be in the vicinity of the remnant \citep{Nava:2016}. N~132D would be detected with high significance in the low-energy range $<1$\tev, and up to $\sim10$\tev, which will allow testing the above ideas. 

30 Dor C should not be confused with $\rm H\RomNumCap{2}$ region 30 Dor and was actually named so because it appeared as a clearly separated component in the structure of the region in radio wavelengths \citep{LeMarne:1968}. It was identified as a \gls{sb} from radio, H$\alpha$, and X-ray observations \citep{Mathewson:1985,Mills:1984,Dunne:2001}. It has a diameter of about 90\,pc and is powered by OB association LH 90 \citep{Lucke:1970}, composed of several clusters with estimated ages spanning $3-7$\,Myr \citep{Testor:1993}. Its most distinctive feature is strong non-thermal X-ray emission filling most of its volume but being particularly bright along the northwestern part of its bounding shell \citep{Bamba:2004,Kavanagh:2015}. The non-thermal emission spectrum is well accounted for from radio to X-rays under the assumption of an exponentially cut-off synchrotron model with a maximum electron energy of 80\tev for a magnetic field of 10\bunit\ \citep{Kavanagh:2015}. The origin of emitting particles however remains unclear. They could result from acceleration in the \gls{sb} volume, for instance from strong stellar wind collisions in the central star clusters, interior \gls{snr} shocks interacting with high density gas clumps within the bubble, or turbulent acceleration, and later be captured in the magnetized \gls{sb} shell \citep{Kavanagh:2015}; alternatively, they could be produced at the forward shock of an SNR that expanded during $\lesssim 20$\,kyr in the tenuous \gls{sb} interior until reaching and colliding with its shell \citep{Bamba:2004,Yamaguchi:2009}. The latter scenario has received recent support from an observed anticorrelation between optical and X-ray shell morphologies, suggesting that the expanding shock has reached the shell and that parts of it are stalled in the densest regions while others continue with high velocities through gaps in the layer \citep{Kavanagh:2019}. This also provides a convenient explanation for enhanced X-ray emission from the \gls{sb} walls, reheated by the interior \gls{snr} shock collision, an idea well motivated by both observations \citep{Jaskot:2011} and simulations \citep{Krause:2014}. 30 Doradus C was detected at TeV energies in H.E.S.S. observations but the origin of the gamma-ray emission associated remains unclear. In the scenario promoted in \citet{Kavanagh:2019}, the TeV emission comes from inverse-Compton scattering of the same population of energetic electrons that powers X-ray synchrotron emission, and any hadronic contribution would be secondary. The survey should enable us to test this idea via significant detection in the $<1$\tev region, where hadronic and leptonic contribution would markedly differ \citep{Abramowski:2015}. The search for a hadronic contribution can also be carried out at the other end, above 10\tev, where subdominant pion decay emission could extend much beyond the inverse-Compton spectrum downturn because protons and nuclei are not loss-limited, contrary to electrons.

LMC P3 is a binary system comprising a compact object of unknown nature, a neutron star being preferred \citep{Corbet:2016,vanSoelen:2019}, and an O5-type stellar companion. This source was first discovered in the X-ray band as a hard point source in very energetic \gls{snr} \citep{Bamba:2006}, and later identified as a high-mass X-ray binary in the X-ray band again \citep{Seward:2012}. A GeV source was later detected in coincidence with \gls{snr} DEM L241 \citep{Ackermann:2016}, and later confirmed as a gamma-ray binary through a 10.3\,d orbital modulation of the signal in the GeV \citep{Corbet:2016} and TeV range \citep{Abdalla:2018c}. The orbital parameters of the system were subsequently refined in \citet{vanSoelen:2019}. It is an object of a rare kind, with only about half a dozen currently known members of the class \citep{Dubus:2013}, and in any case the first such source detected outside the Milky Way and the most luminous of all, with an orbit-averaged luminosity in the $1-10$\tev range of $1.4 \pm 0.2 \times 10^{35}$\punit. This binary is also very unique since it is in a \gls{snr}, implying a relatively young system, and this may explain why it is the brightest gamma-ray binary. H.E.S.S. observations of LMC P3 (also named HESS J0536--675) led to the detection of significant emission over only 20\% of the orbit, in contrast to similar object LS 5039 but akin to 1FGL J1018.6--5856 \citep{Abdalla:2018c}. High-significance detection of the object over most of the \gls{cta} energy range should allow a more accurate characterization of the orbital light curve, together with phase-resolved spectral analyses hardly accessible with the current H.E.S.S. sensitivity. Combined with ever-improving data from Fermi-LAT until \gls{cta} becomes operational, this will allow thorough investigation of the origin of the GeV and TeV emission from LMC P3, especially their different behaviours such as the observed phase offset in the orbital variability \citep{Abdalla:2018c}. Given the recent identification of the system, the questions to be addressed in LMC P3 are still rather generic to gamma-ray binaries: clarifying the various processes shaping gamma-ray emission from binaries and their relative contributions, in that specific object and along its orbit, and in particular, moving forward in the growing consensus that different populations of particles, accelerated in different regions (pulsar magnetosphere, pulsar wind, wind-wind collision layer) and/or by different mechanisms (shock acceleration, reconnection), are involved and manifest themselves through specific spectral and temporal signatures in different (gamma-ray) wavebands \citep{Dubus:2013}.

\subsection{SN 1987A} \label{sec:res:87a}

SN 1987A is a source of major interest in the \gls{lmc}, as a well-studied remnant that could provide insight into the very first stages of particle acceleration following a core-collapse \gls{sn} explosion. Particle acceleration in SN 1987A is already at work, as evidenced by the increasing radio synchrotron emission detected from the remnant since about 1200 days after outburst \citep{Zanardo:2010}. The radio spectrum exhibits a power-law shape and its index over the 843\,MHz to 8.6\,GHz frequency range has evolved from $-0.932 \pm 0.051$ to $-0.758 \pm 0.037$ over year 5 to year 19 after explosion \citep{Zanardo:2010}. As of late 2013 - early 2014, the spectral index over the 0.072 to 8.64\,GHz range is $-0.74 \pm 0.02$ \citep{Callingham:2016}, which implies an emitting electron population having a power-law distribution of index 2.5. This is steeper than the canonical flat distribution with index 2.0 expected for diffusive shock acceleration in the test particle limit. Such a steep distribution could result from acceleration in a cosmic-ray modified shock, with a subshock compression ratio of 3 that affects the acceleration of the lower-energy particles \citep[][and references therein]{Zanardo:2010,Callingham:2016}; alternatively, it can result from acceleration in the presence of subdiffusion transport across the shock front due to trapping of particles in braided magnetic field structures \citep{Kirk:1996}, or from the drift of magnetic structures with respect to the downstream thermal plasma \citep{Caprioli:2020}. Whatever the origin for this steep distribution, radio observations probe particle energies that are quite far from what would be probed with \gls{cta}. For a typical magnetic field with an order-of-magnitude strength of 10\,mG downstream of the forward shock in SN 1987A \citep{Berezhko:2011,Berezhko:2015}, synchrotron emission in the 1-10\,GHz range arises from sub-GeV electrons. How the particle distribution behaves beyond this range, up to what maximum energy, and in which proportions for electrons and nuclei remains currently unknown in the absence of detection at gamma-ray energies, and is precisely a major science case for \gls{cta}. Non-thermal X-ray emission was also detected from SN~1987A and could be a more direct probe of particles with energies relevant to \gls{cta}, but the very origin of the emission is unclear and contamination by an absorbed \gls{pwn} is likely \citep{Greco:2021}.

Several analyses or models of the shock dynamics linked to particle acceleration can be found in the literature \citep{Zhekov:2010,Berezhko:2011,Dwarkadas:2013,Berezhko:2015,Petruk:2017}. The full problem is quite complex. The \gls{sn} ejecta drove a blast wave in a circumstellar medium that was shaped by the progenitor star or system into a highly structured and anisotropic matter distribution summarized in \citet{Potter:2014}. Most of the mass encountered by the blast wave so far lies in an equatorial disk, a density enhancement in the circumstellar medium likely resulting from the interaction of dense wind emitted during the red giant phase of the progenitor and a subsequent fast wind emitted in a blue supergiant phase before explosion. It has a half-opening angle $15 \pm 5$\deg\ and typical density $\sim 100$\vunit, and contains knots or fingers with density tens to hundreds times higher. In the polar directions is a bipolar bubble of hot shocked blue supergiant wind with density of 0.1\vunit. In such an environment, the initial blast wave gave rise to a variety of shocks with different velocities: forward shock propagating in most of the volume of the equatorial ring, reverse shock propagating in the \gls{sn} ejecta, transmitted and reflected shocks triggered from interaction with dense clumps. Each of the models mentioned above relies on specific assumptions for the respective contributions of the various shocks at play in SN~1987A and the conditions in which they evolve. It is beyond the scope of this work to present a complete and up-to-date model or discussion of the non-thermal processes in SN~1987A. Instead, for illustrative purposes mostly, we provide below model spectra from a very simple model that however captures some of the important constraints available today.

Starting with the energetics, the total swept up mass so far is of the order of 0.1\msol \citep{Potter:2014}, by a forward shock that has been expanding over the past years at a velocity of 3890\kms\ estimated from radio observations \citep{Ng:2013}. This corresponds to $1.5 \times 10^{49}$\,erg of kinetic energy that flowed into the forward shock front, about 1\% of the estimated total initial ejecta kinetic energy of the explosion, or 4\% if we restrict this ratio to the solid angle subtended by the equatorial disk \citep{Potter:2014}. We supposed that a canonical 10\% of this kinetic energy went into diffusive shock acceleration of protons. The proton spectrum is assumed to follow a broken power-law distribution in momentum starting from 100\mev/{\it c}, with a change in slope at 1\gev/{\it c} and an exponential cutoff at an arbitrary momentum of 100\tev/{\it c}. The sub-GeV part of the spectrum has a fixed index 2.5, as required by radio observations of synchrotron emission from electrons. We checked that injecting a fraction of order $10^{-4}$ of the energy into accelerated electrons, and assuming they radiate in a 10\,mG downstream magnetic field \citep{Berezhko:2011,Berezhko:2015}, yields a synchrotron intensity at the Jy level at 1\,GHz, as observed \citep{Callingham:2016}. 

We then considered different assumptions for the spectrum slope at higher energies. A minimalist and worst-case model assumes that the spectrum simply extends with the same slope up to the cutoff energy. Alternative more optimistic models rely on the possibility of a hardening of the spectrum above some energy. Such concave shapes are characteristic of the non-linear \gls{dsa} theory, at least in its most basic version, with spectra steeper or harder than the test-particle \gls{dsa} prediction below or above transrelativistic energies, respectively \citep{Drury:1981,Eichler:1984,Berezhko:1999}. In the context of SN~1987A, this possibility was explored in \citet{Berezhko:2011,Berezhko:2015}, whose latest model predicts a hardening up to a power-law index of about 1.8. Yet, the degree of hardening or even its very existence were largely questioned over the past decade. This stems mostly from the non-detection of a corresponding signature in radio or gamma-ray observations of young \glspl{snr}, with inferred indices for the emitting particle populations of $\sim 2.2-3.0$ \citep{Caprioli:2019}. Additional evidence challenging concave distributions from non-linear \gls{dsa} came from analyses of locally measured \gls{cr} spectra in the framework of standard Galactic propagation models \citep[see, e.g.,][]{Trotta:2011,Evoli:2019}; despite coming with different assumptions on source and transport terms, they invariably require power-law indices in the range $\sim$2.3-2.4 for the nuclei injection spectra above $\sim$10\gev (although it remains unclear in this context how exactly such injection spectra are shaped by the processes of acceleration in the source and escape from it). 

For a series of possible proton spectra above 1\gev, from the softest option with index 2.5 to the hardest option with index 1.8, we computed the associated pion-decay signal, under the assumption that the particle population is interacting with compressed gas downstream of the forward shock having a typical density $\sim 400$\vunit\ (this corresponds to the above-mentioned upstream density in the equatorial disk, increased by a compression ratio of 4 appropriate for a strong shock; higher compression ratios of 6-7 are possible in the context of a \gls{cr}-modified shock). A nuclear enhancement factor of 1.753 appropriate for the \gls{lmc} is applied (see previously).

The resulting model spectra are shown in Fig. \ref{fig:res:sens:sn87a}, together with the sensitivity of the \gls{cta} survey to a point source at the position of SN~1987A and the H.E.S.S. upper limit published in 2015 \citep{Abramowski:2015}. Our most optimistic model with a hard spectrum is similar to that presented in \citet{Berezhko:2015} and is already dismissed by the H.E.S.S. observations \citep{Abramowski:2015}. At the other end, the worst-case model with a soft spectrum extending that of sub-GeV electrons is more than a factor 20 below the sensitivity curve, a gap that it would not be easy to close with corrections of order unity on the acceleration efficiency and/or downstream density. Our simple modeling suggests that meaningful spectral analyses would be accessible to \gls{cta} for rather flat spectra, with power-law slopes of $\sim2.0$, which would imply processes flattening or hardening up the spectrum towards high energies. 

While this is already interesting enough, we emphasise that an object like SN~1987A deserves a tailor-made modeling harnessing the growing wealth of multiwavelength information on the shock history and environment. The very simple model used here implicitly relates the radio and gamma-ray emission whereas they could arise from distinct regions owing to different distributions of the \gls{cr} species and targets involved in the respective radiative processes; for instance, radio could be mainly produced in the lower-density ionized regions above and below the equatorial plane, while gamma rays would predominantly come from the higher-density clumps \citep[see Figs. 4 and 5 in][]{Berezhko:2015}. Furthermore, there ought to be more than plain and simple acceleration at the forward shock as presented here, for instance a contribution from reflected shocks reaccelerating freshly accelerated particles while enhancing further the density in which they radiate, such that prospects for detection and study may well be more promising. In addition, the conditions of particle acceleration and non-thermal emission are most likely strongly time-dependent, and may significantly evolve over the next $10-20$ years, a reality that is not captured in the rough model used here where the spectral index or downstream density are fixed to a single value; gamma-ray emission with a soft spectrum with photon index $\sim2.3-2.4$, as observed in many Galactic \glspl{snr}, would become within reach of simple broadband detection pending a flux increase by a factor $3-4$ (compare the cyan curve and gray lines in Fig. \ref{fig:res:sens:sn87a}). At the very least, the above discussion demonstrates that the survey should bring us into the right ballpark, especially if the emission from SN~1987A is on the rise, and tell us something about proton acceleration in SN~1987A, at last.

\subsection{Large-scale interstellar emission} \label{sec:res:ism}

As illustrated in Fig. \ref{fig:res:sens:pionic}, the spectral sensitivity of the survey to the specific intensity distribution of inverse-Compton emission is at best a factor 30 above the expected emission level in the case of the average \gls{ism} model. Using the more intense \gls{isrf} of the gas-rich \gls{ism} model pushes the emission level up by a factor of a few and only a very hard electron injection spectrum with power-law index $\lesssim 2.2$ would lead the predicted inverse-Compton emission to rise close to the survey sensitivity. Including the contribution of secondary leptons from charged pions would increase the level of the emission by no more than 30\%, according to the estimate of the contribution of secondaries presented in Sect. \ref{sec:model:val}, and this additional component would have a more uniform spatial distribution likely making its detection more challenging. Overall, interstellar inverse-Compton emission on galactic scales may be out of reach of the \gls{cta} survey of the \gls{lmc}.

Prospects for detecting galactic pion-decay emission are more encouraging. The spectral sensitivity of the survey to the intensity distribution of our baseline pion decay emission model is a factor $\sim 2$ above the expected total flux level at energies below a few TeV. High-significance detection as a very extended component and fine-binned spectral studies therefore seem to be excluded for the baseline model, but pion-decay emission with a harder spectrum could be detectable in several individual energy bins, at the very least in the regions most rich in gas. We review in appendix \ref{app:hard} some motivations for considering the possibility of harder emission than assumed in our baseline model, and we provide below quantitative prospects for the detectability of pion decay over galactic scales in the \gls{lmc}.

The potential of \gls{vhe} diffuse interstellar emission for the study of the highest-energy \glspl{cr}, up to the knee, has been demonstrated already for the Galaxy \citep{Lipari:2018,Cataldo:2019}. The \gls{cta} survey of the \gls{lmc} may provide valuable complementary information. This is illustrated in Fig. \ref{fig:res:sens:pionic}, where the sensitivity to the pion decay emission template is compared to total emission spectra for the baseline model, and for the baseline model hardened by multiplication with power laws with indices +0.2, to reflect a possible harder interstellar population of \glspl{cr} above $200-300$\,GV as measured locally by AMS-02, and +0.5, to mimic the extreme case of the \gls{lmc} having a spectrum similar to that of starburst galaxies M82 or NGC253 \citep{Acciari:2009,Abdalla:2018e,Ajello:2020}. Henceforth, we will refer to these models as BASE, HARD020, and HARD050, respectively. The hardened variants of the model yield spectra with power-law photon indices ranging approximately from 2.55 to 2.25 (away from the $\sim$100\tev cutoff that arises from the assumed 1\pev maximum proton energy). We emphasize here that these modified models, with emission levels up to $\sim10-20$ times that of our baseline model in the 1-10\tev range, are still consistent with the non-detection with H.E.S.S. of small-scale features in interstellar emission.

\begin{figure}
\begin{center}
\includegraphics[width=0.8\columnwidth]{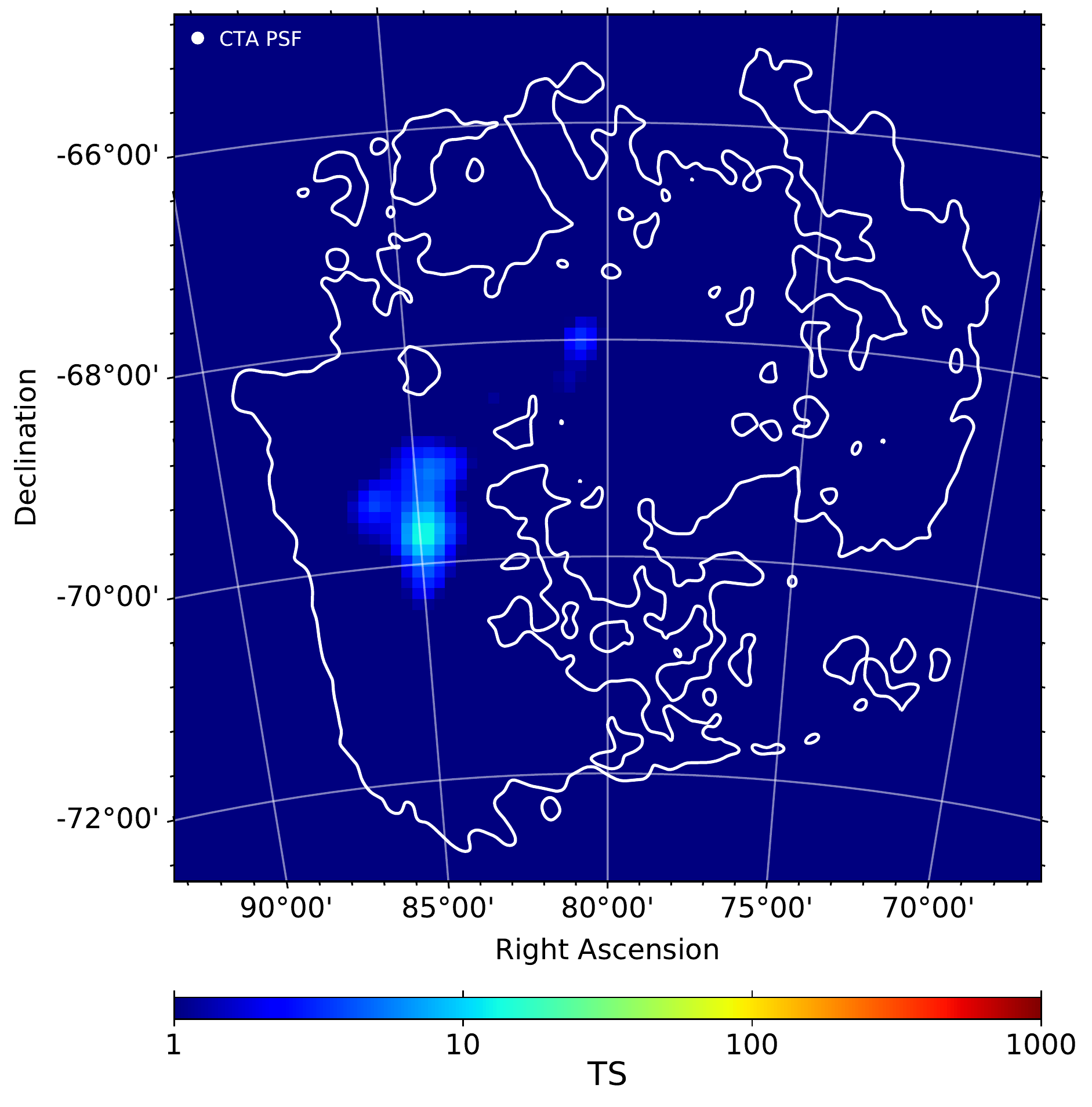}
\includegraphics[width=0.8\columnwidth]{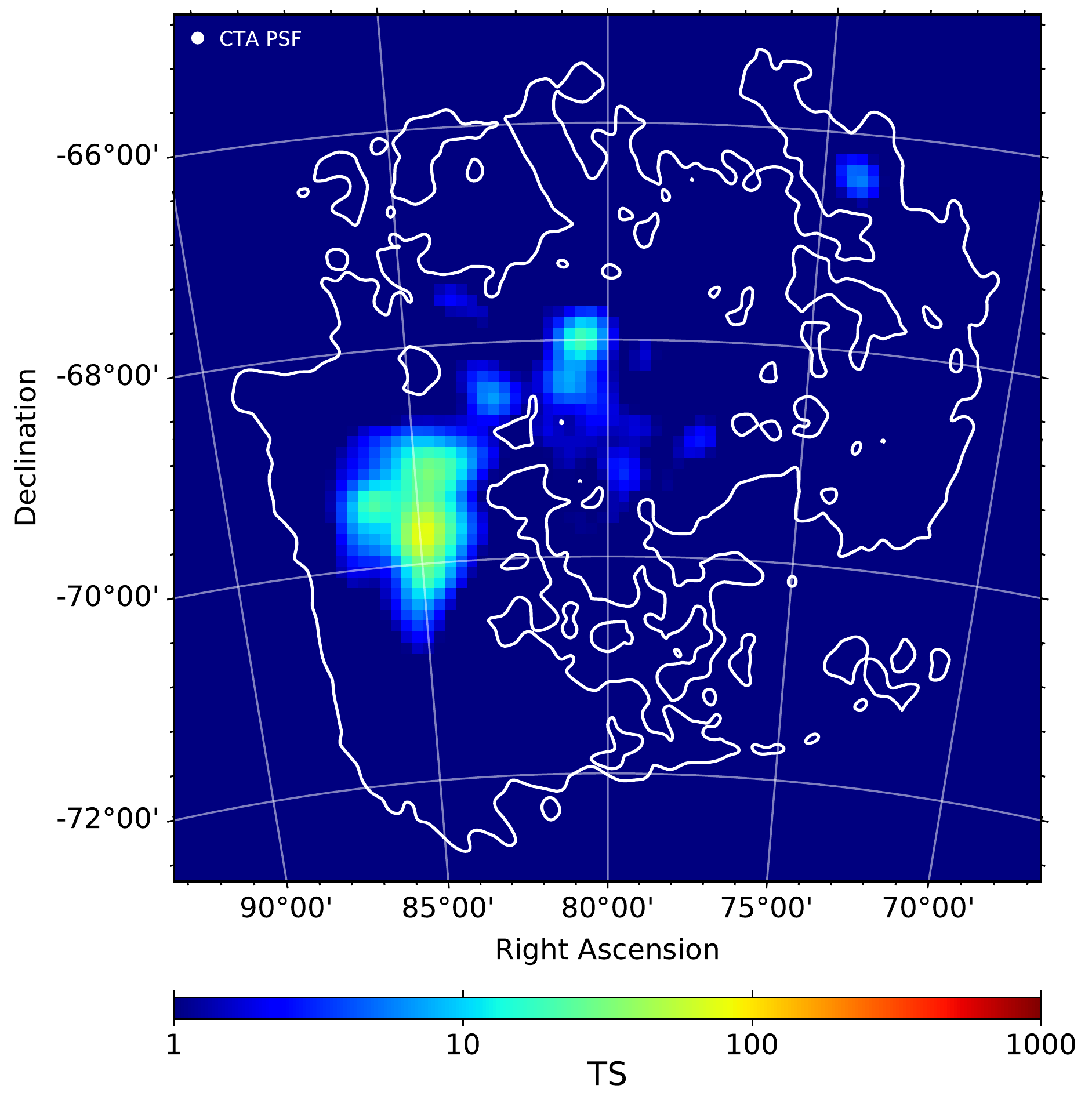}
\includegraphics[width=0.8\columnwidth]{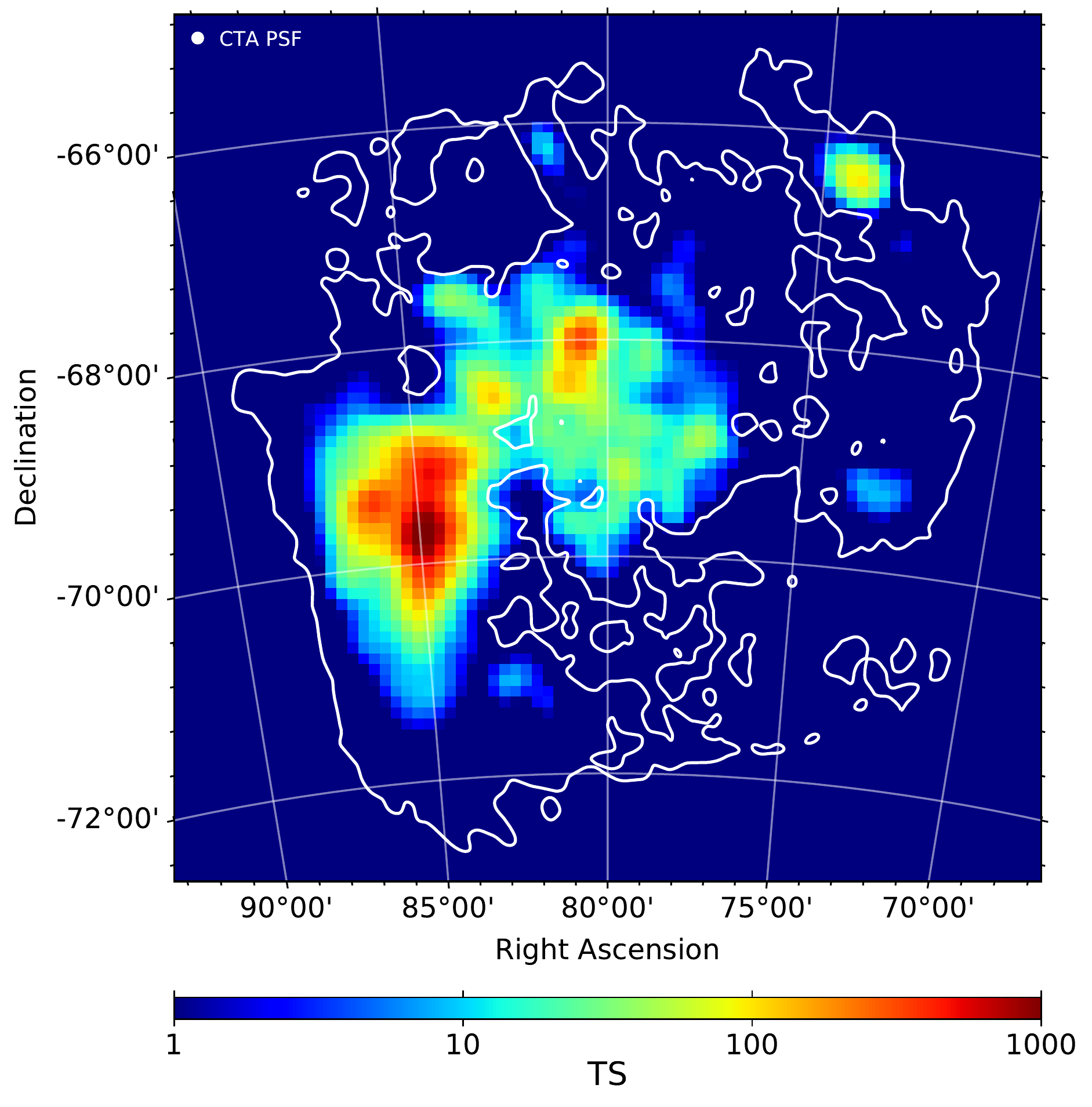}
\caption{Average TS maps for three galactic pion-decay emission models: the baseline setup (top panel) and hardened variants obtained by scaling the energy dependence with power laws of index +0.2 and +0.5 (middle and bottom panel, respectively) These are referred to as BASE, HARD020, and HARD050 in the text. The TS maps were obtained from Asimov data sets using as test source a 2D Gaussian model with fixed $\sigma=0.1\deg$ associated to a power-law spectral model with fixed photon index 2.5.}
\label{fig:res:ism:piontsmaps}
\end{center}
\end{figure}

Figure \ref{fig:res:sens:pionic} illustrates that model HARD020, with properties similar to those inferred for central regions of the Milky Way, could be detectable in individual energy bins over most of the 0.1-10\tev range, while model HARD050, representative of the few starburst galaxies studied over the GeV-TeV range, could be spectrally resolved over the entire energy band, up to almost 100\tev. In terms of broadband detection, in the ideal case of a full spatial-spectral maximum-likelihood approach with the true source model, models BASE, HARD020, and HARD050 would be detected with average TS values of about 60, 350, and 6650 respectively. Despite such significance levels, however, recovering the emission on large, galactic scales may not be trivial and will most likely be restricted to just a few regions of the galaxy. Presenting a complete and realistic data analysis aimed at extracting large-scale emission of unknown true distribution is beyond the scope of this paper. Instead, we illustrate in Fig. \ref{fig:res:ism:piontsmaps} the layout of significant diffuse emission by showing TS maps obtained from Asimov data sets by fitting a test source consisting of a 2D Gaussian with fixed $\sigma=0.1\deg$ for the spatial part, and a power-law with fixed photon index 2.5 for the spectral part, on a regular grid of $80 \times 80$ positions spanning the whole field. For model BASE, extended emission over several tenths of a degree is recovered at only one position in the galaxy, in the molecular ridge south of 30 Doradus, and with limited significance peaking at $TS \sim 20$. Model HARD020 offers the perspective of more accessible and widespread emission, with significant degree-scale emission in two regions, the molecular ridge and the molecular cloud complexes towards star-forming region N44, with TS values peaking at $\sim100$. Last, model HARD050 makes it possible to detect interstellar emission over a region spanning about one-third of the galaxy and covering many gas-rich and active star-forming locations. Should this model be a good description of the \gls{lmc}, the separation of diffuse emission from populations of sources in these sites will become a major issue. 

The most prominent feature in the TS maps is the so-called molecular ridge. It is a cloud complex stretching $\sim$2\,kpc south of 30 Doradus, at right ascension 84\deg\ and from declination -69.5 \deg\ to -71\deg. It consists in dense clouds bathed in lower-density molecular gas \citep{Fukui:2008}. Together with an arc-like structure delineating the southeastern edge of the galaxy (less visible in our model map), the region contains an estimated 35\% of the total CO-traced molecular gas mass of the galaxy \citep{Mizuno:2001}, and about 20\% of the total amount of atomic gas (which would then be the dominant gas phase by mass in this area, since there is about eight times more atomic gas than molecular gas in the galaxy; see Sect. \ref{sec:model:ism}) \citep{Luks:1992}. The existence of so much interstellar gas packed in this region offers a useful way to probe the background level of \glspl{cr} in the galaxy, ensuring an efficient production of gamma-rays over a relatively small patch of the sky that fits into one \gls{cta} field of view, thus restricting the data analysis challenge to the search for a moderately extended source. On the other hand, such a region may be particularly prone to ionization-dependent transport effects, like those investigated in \citet{Bustard:2021} for GeV \glspl{cr}. Fast transport in cold dense gas, if also applicable to TeV \glspl{cr}, would tend to lower the pion-decay emission of these gas phases, although \citet{Bustard:2021} seem to suggest that the impact is mostly on emission layout and not so much on total luminosity, which would mitigate the problem for a distant source like the \gls{lmc}.

\subsection{The 30 Doradus star-forming region} \label{sec:res:30dor}

As mentioned in Sect. \ref{sec:model:ism}, 30 Doradus was not included as a \gls{cr} injection site in our model for large-scale interstellar emission, mainly because the region as such was not detected as a peculiar or outstanding source in Fermi-LAT observations and not detected at all in the HE.S.S. survey, probably due to its relatively young age (see discussion and arguments in appendix \ref{app:crsrc}). There is, however, a huge potential for particle acceleration and gamma-ray production in 30 Doradus, so emission should be present at some level and we here investigate under which conditions it could be accessible to the \gls{cta} survey.

We focus on the region as a whole, in the sense of a site delivering a large amount of mechanical power, a fraction of which can be tapped for particle acceleration, irrespective of how exactly the latter is achieved. In interpreting Suzaku X-ray observations of the 30 Doradus nebula, the authors of \citet{Cheng:2021} came up with an estimated $8.3 \times 10^{52}$\,erg of total mechanical energy injected by the central OB association NGC~2070, more precisely by a stellar population made up of two sub-groups with estimated ages 2 and 4\,Myr \citep{Sabbi:2012}, although more extended star formation up to 6-8\,Myr is also proposed \citep{Schneider:2018}. In this framework, most of the energy was released from stellar winds and only $1.3 \times 10^{52}$\,erg is expected to result from \gls{sne}. Bearing in mind the uncertainty on the actual star formation history, this corresponds to a mechanical luminosity of $\sim 10^{39}$\punit during a few Myr, similar to the value inferred for the most massive star-forming regions in the Galaxy \citep{Aharonian:2019}.

We implemented the \gls{sfr} emission model described in Sect. \ref{sec:model:sfr} in the specific case of 30 Doradus. We considered six possibilities for diffusion suppression within 100\,pc of the \gls{sfr} center -- 1, 3, 10, 30, 100, 300 times smaller than the average interstellar value over large galactic scales -- and for each value we computed the corresponding diffusion kernel and the associated pion decay emission for the actual gas distribution in the region. At this point, only the morphology of each model matters, and the normalization is arbitrary. In a first study, we performed an iterative series of simulations and analyses of survey observations of the \gls{sfr}, with instrumental background as the only other source component, renormalizing the \gls{sfr} source model between iterations until the process converges towards a global detection with $TS=25$ over the full energy range. The eventual renormalization factor of the \gls{sfr} model sets the minimum \gls{cr} injection luminosity, hence gamma-ray emission level, that would allow detecting the \gls{sfr} for a given diffusion suppression factor. In a second study, we perform the same kind of iterative search, aiming this time at the minimum renormalization of the model that would allow its significant differentiation from the reference case of diffusion without suppression (i.e., \glspl{cr} released by the \gls{sfr} diffuse out with the average galactic coefficient). We adopted $TS=16$ as the minimum significance level in that case. 

\begin{figure}
\begin{center}
\includegraphics[width=\columnwidth]{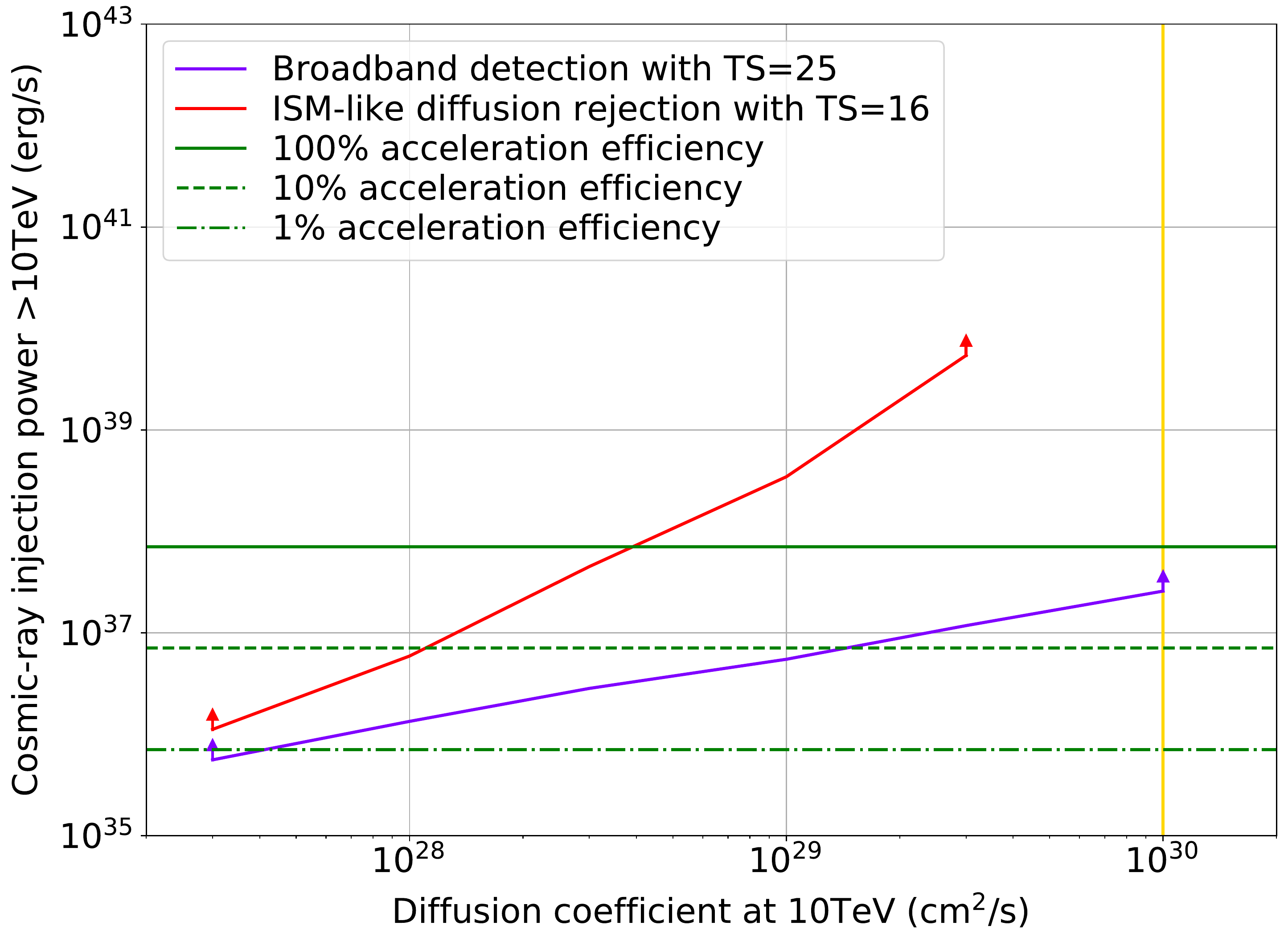}
\caption{Integrated sensitivity to pion-decay emission from \glspl{cr} injected at the center of \gls{sfr} 30 Doradus and diffusing away, expressed as the minimum \gls{cr} injection power above 10\tev, as a function of the 10\tev diffusion coefficient in the $<100$\,pc vicinity. The purple curve shows the requirement to achieve a global detection with TS=25 over the full energy range, while the red curve shows the requirement to achieve both detection and significant rejection of fast diffusion typical of the average galactic disk at a level of TS=16. The green curves display the level of $>10$\tev \gls{cr} injection in 30 Doradus, for different efficiencies in converting the $\sim 10^{39}$\punit mechanical luminosity into $>1$\gev \glspl{cr}, and the yellow vertical line marks the large-scale average value of the diffusion coefficient, for reference.}
\label{fig:res:30dor:const}
\end{center}
\end{figure}

The results are shown in Fig. \ref{fig:res:30dor:const}. If diffusion around 30 Doradus is not suppressed, the required \gls{cr} injection power for global $TS=25$ detection corresponds to about 40\% of the assumed $\sim 10^{39}$\punit\ mechanical power of the cluster (under the hypothesis of a power-law injection spectrum with index 2.25 extending from 1\gev to a cutoff energy of 1\pev). This rather high value is not unmotivated theoretically for \glspl{sfr} \citep{Bykov:2001} and it remains allowed in terms of energetics: the energy census in 30 Doradus and other \gls{sb}s (30 Doradus C or DEM L192 for instance) reveals that half or more of the injected energy is not found in the form of kinetic or thermal energy in the \gls{sb}s, hence lost to some channels not accounted for in classical \gls{sb} models \citep{Weaver:1977,MacLow:1988}; \gls{cr} acceleration with high efficiencies is one possibility \citep{Butt:2008}, although there are several other competing and well-motivated mechanisms \citep{Cheng:2021}. On the other hand, constraints on acceleration from stellar winds in several Galactic clusters can be as low as 1\% \citep{Maurin:2016}, although most objects in this study are much smaller and younger than 30 Doradus and thus not directly comparable. As diffusion suppression increases, \glspl{cr} are more confined in the vicinity of the source, which raises gamma-ray emission from the region and diminishes the energetics required for detection. For diffusion suppression factors of a few hundreds, 30 Doradus would be detectable for acceleration efficiencies below 1\%. Such a level of diffusion suppression is not unreasonable theoretically and seems indicated in similar Galactic \glspl{sfr} \citep{Aharonian:2019}.

\begin{figure}
\begin{center}
\includegraphics[width=\columnwidth]{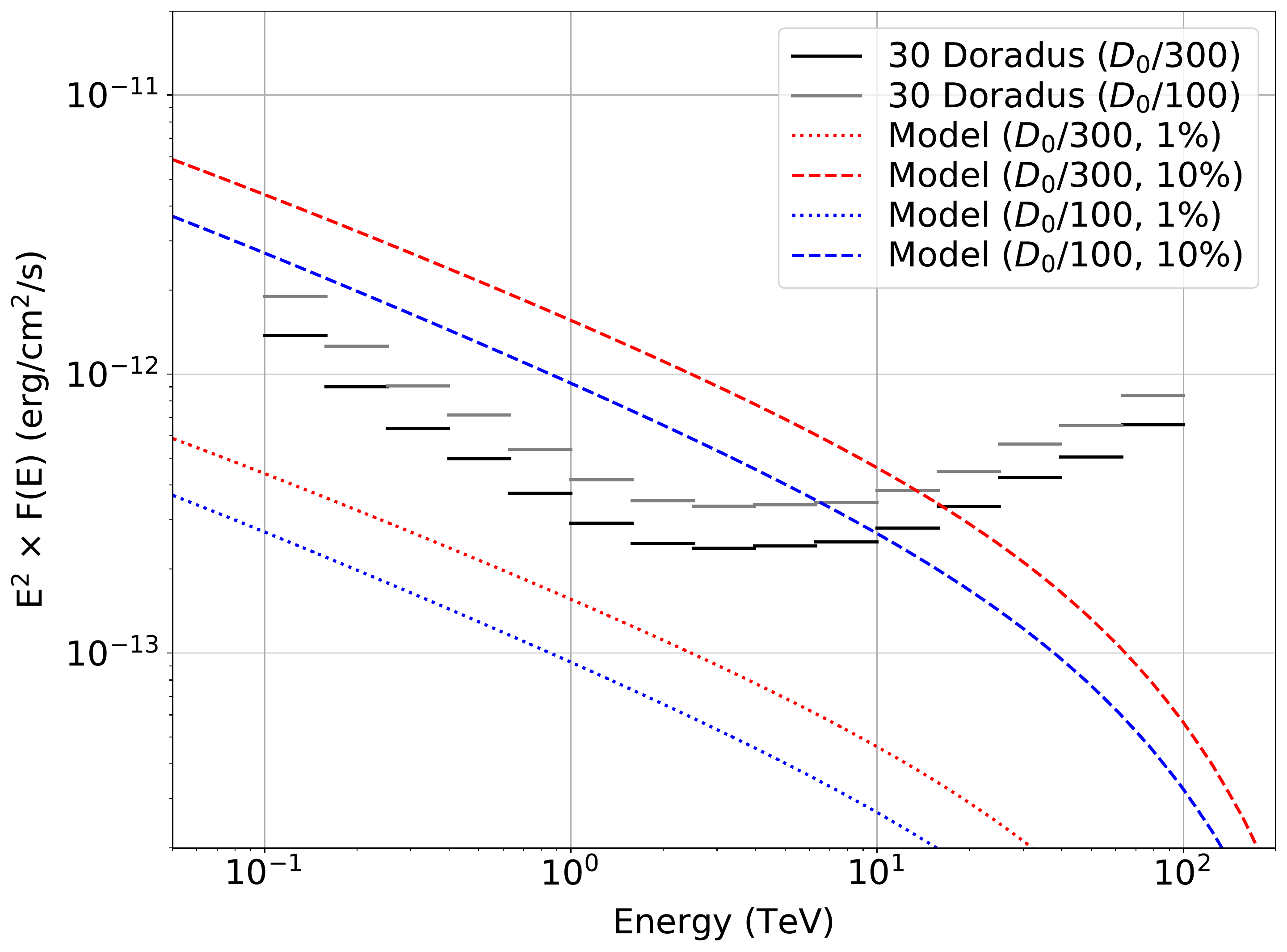}
\caption{Spectral sensitivity to pion-decay emission from \glspl{cr} injected at the center of \gls{sfr} 30 Doradus and diffusing away in a $<100$\,pc region where diffusivity is suppressed by a factor 100 or 300 with respect to the average galactic value. Overlaid are model spectra for 1\% and 10\% efficiencies in converting the assumed $\sim 10^{39}$\punit mechanical luminosity of 30 Doradus into $>1$\gev \glspl{cr}.}
\label{fig:res:30dor:spec}
\end{center}
\end{figure}

Figure \ref{fig:res:30dor:spec} shows the spectral sensitivities to the specific emission morphologies resulting from the highest levels of diffusion suppression considered here, 100 and 300. They are compared to model spectra for 1\% and 10\% efficiencies in converting the assumed $\sim 10^{39}$\punit mechanical luminosity of 30 Doradus into $>1$\gev \glspl{cr}. It appears that even with such efficient confinement in the vicinity of the \gls{sfr}, acceleration efficiencies of 10\% at least would be required to detect the source up to 10\tev, where the turnover due to the 1\pev cutoff in the proton spectrum barely starts to be discernible. It therefore seems that the survey sensitivity may not be enough to investigate the possible role of \glspl{sfr} in the production of the highest-energy galactic cosmic rays, unless the injection is harder than assumed here (power-law index 2.25).

While prospects for simple detection of 30 Doradus as an \gls{sfr} are encouraging, its identification as such may be more challenging. Even more challenging would be the inference of the \gls{cr} injection power, which depends on the actual diffusion coefficient \citep{Aharonian:2019}. Solid evidence serving both goals would be the significant detection of a spatial intensity distribution characteristic of central injection and suppressed diffusion (see Fig. \ref{fig:model:sfr:emiss}). A complete and realistic assessment of such a prospect, including decomposing the emission into a radial profile while simultaneously determining foreground and background interstellar emission from the galaxy, is beyond the scope of the present work. Instead, we quantified the requirement for being at least able to differentiate the emission profile characteristic of diffusion suppression at a given level from no diffusion suppression at all. Figure \ref{fig:res:30dor:const} shows that the requirements on the level of injection then become more demanding. In the case of diffusion suppression by a factor of a hundred or more, acceleration efficiencies in the range 1-10\% are required for the \gls{sfr} to be detected and identified as having a specific emission profile not compatible with fast diffusion typical of the average galactic disk. This does not necessarily ensure a recovery of the diffusion coefficient true value with high precision, though, but it could be used at the very least to set an upper limit on the diffusion coefficient in the vicinity of the source. For lower diffusion suppression factors, below a few tens, the requirements for morphological separation ramp up to prohibitive values of the injection power, in excess of 100\%. In this range, source detection remains possible but it would be impossible to constrain the intensity distribution to a specific diffusion profile.

Last, we emphasise that the above prospects should be considered as optimistic because they were derived from simulations including 30 Doradus as only astrophysical source. In the process of detecting and studying such an extended target, the unknown interstellar emission from the galactic disk would come as a nuisance parameter. While our baseline model for it suggests it could hardly be detectable, and thus not affect too much the above estimates, alternative models could bring the level of large-scale interstellar emission to much higher values that could jeopardize a proper identification of 30 Doradus as a gamma-ray emitting \gls{sfr} (see Sect. \ref{sec:res:ism}).

\subsection{Source population} \label{sec:res:pop}

\begin{figure}
\begin{center}
\includegraphics[width=\columnwidth]{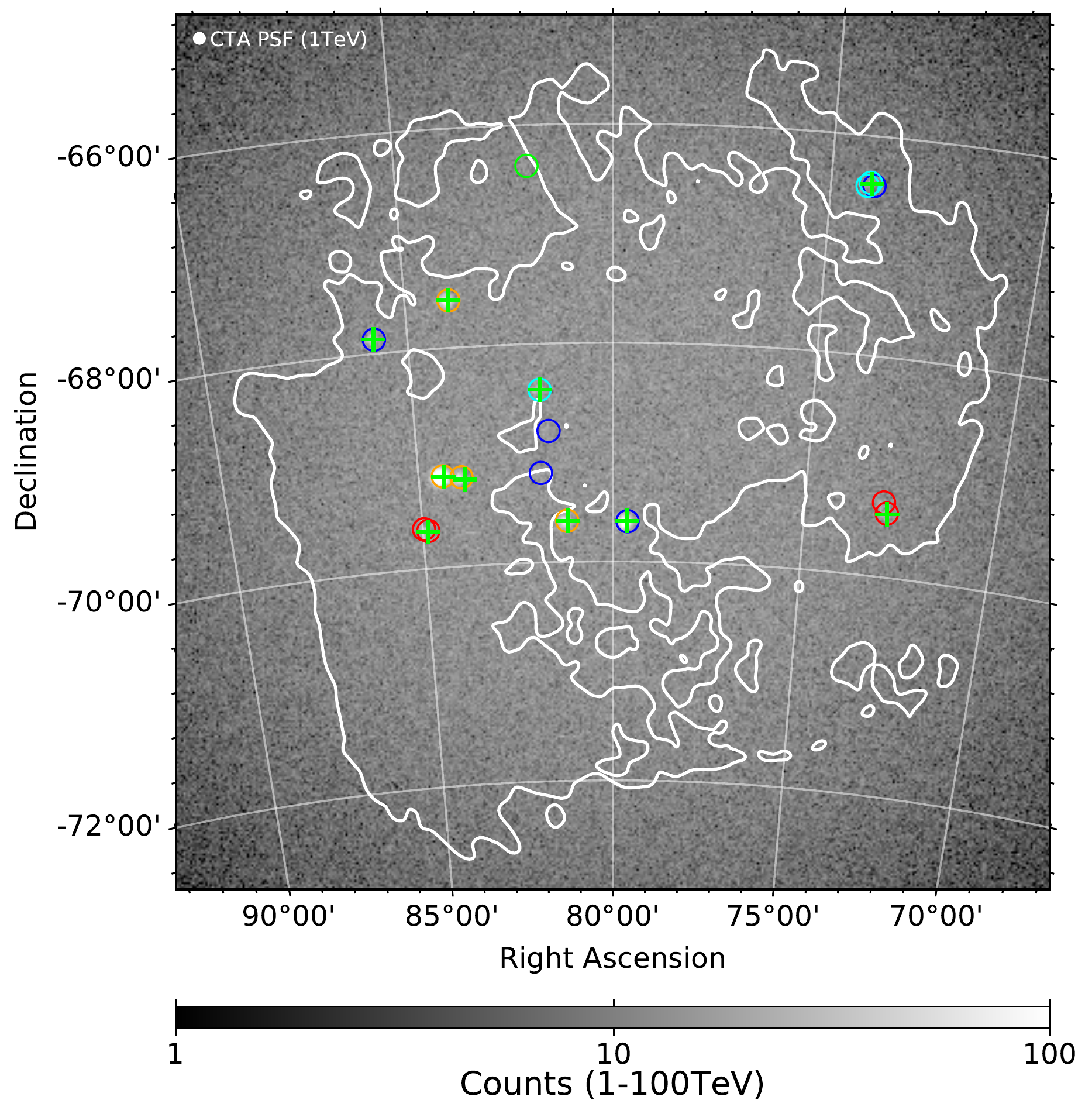}
\caption{Distribution of point-like sources found in a blind search for a significance threshold of $5\sigma$ (green crosses). The colored circles are the positions of the most luminous true sources in the model, with 1-10\tev fluxes $>2.0 \times 10^{-14}$\feunit: \glspl{snr} in green, \glspl{isnr} in cyan, \glspl{pwn} in blue, pulsar halos in red, and known sources in orange (same marker size for all such objects, irrespective of their true or mock physical extent). The background image in gray scale is a 1-100\tev counts map.}
\label{fig:res:pop:chart}
\end{center}
\end{figure}

The realization of the source population model introduced in Sect. \ref{sec:model:pop} contains 71 \glspl{snr}, 10 \glspl{isnr}, 91 \glspl{pwn}, and 167 halos around mature pulsars, so a total of 339 objects. In Sect. \ref{sec:res:sens}, we determined that the typical point source sensitivity reached in the \gls{cta} survey is $1.6-2.9 \times 10^{-14}$\feunit\ in the 1-10\tev range, with variations due to position in the field and spectral shape. As illustrated in Fig. \ref{fig:model:pop:lum}, this would allow the detection of at most a dozen objects, and only half of that if we are conservative and assume the higher end of the sensitivity range. We show below that the latter conservative estimate seems more appropriate to what can actually be achieved.

For \glspl{snr} and \glspl{isnr}, our model suggests that the survey would give access to a handful of sources making up the very high end of the luminosity distribution only, those objects resulting from high explosion energies and/or high acceleration efficiencies and/or very dense target fields (gas or photons). Less than 10\% of the \glspl{pwn} population could be probed, which could however double the number of known such objects in the \gls{lmc}. Currently only three have been identified: N~157B, already detected as TeV source; B0540-69, which holds promise for detection with \gls{cta} owing to its highly powerful Crab-twin pulsar \citep{MartinTorres:2014}; and B0453-685 \citep{Gaensler:2003,McEntaffer:2012}. In the framework of our model, a comparable number of pulsar halos would be detected in the survey, despite the optimistic assumption that all middle-aged pulsars past the \gls{pwn} stage do experience such an evolutionary phase \citep[see][for arguments on the possible rarity of pulsar halos]{Martin:2022}.

We implemented a blind search for point-like sources in one simulation of the survey based on the full model, i.e., including the four known sources, pion-decay and inverse-Compton large-scale emission templates, plus all \glspl{snr}, \glspl{isnr}, \glspl{pwn}, and pulsar halos from the realization of our population model. This was done using the \textit{cssrcdetect} tool from the ctools package, which implements a peak detection algorithm in a significance map. By trial and error, comparing the output to the true population of sources, we found that the optimum parameters are: (i) a counts map in the 1-100\tev range, for the calculation of the significance; (ii) an averaging radius of 0.5\deg, to compute the mean value and standard deviation in each pixel; (iii) an exclusion radius of 0.2\deg, to exclude counts around a previously detected source; (iv) a correlation kernel consisting in a disk with radius of 0.05\deg, to smooth the input counts map.

Using a significance threshold of $5\sigma$ leads to the detection of ten sources, the four known ones and six mock objects. The layout of these sources is illustrated in Fig. \ref{fig:res:pop:chart}, which displays also the distribution of the 13 most luminous objects in our mock population model, with 1-10\tev fluxes $>2.0 \times 10^{-14}$\feunit. Among the 13 brightest mock sources, three are not recovered when using a $5\sigma$ detection threshold because they are the faintest of the subset and have unfavorable spectra yielding the lowest photon fluxes. They are detected when lowering the threshold to $3\sigma$, but at this point a non-negligible number of spurious detections appear, at positions devoid of true sources in our model. Leaving known sources aside, Fig. \ref{fig:res:pop:chart} reveals that half of the detected sources encompasses isolated mock objects, while the other half is positionally coincident with clusters of bright sources. In these locations, one source is detected where there are in reality two or three overlapping true sources strong enough to be detected individually each. This gives a first taste of the issue of source confusion in the \gls{lmc}. Whether these could be separated by more dedicated studies, exploiting spectral differences and/or the improved angular resolution at higher energies and/or events subclasses with better incident direction reconstruction, is beyond the scope of this work.

We assessed whether the recovered sources have a significant extension and found only marginal evidence for extension in two sources, with TS values of $12-14$. We also checked that adopting a different hypothesis for source distribution, relaxing the scaling of the membership probability with $\rm H\RomNumCap{2}$ region luminosity so as to yield sources more uniformly spread over all $\rm H\RomNumCap{2}$ regions instead of clustered inside the most prominent ones, does not profoundly alter the picture. The limited effect is due to the fact that the 138 $\rm H\RomNumCap{2}$ regions used in this work tend to be themselves clustered over a rather small fraction of the galaxy area.

\begin{figure}
\begin{center}
\includegraphics[width=\columnwidth]{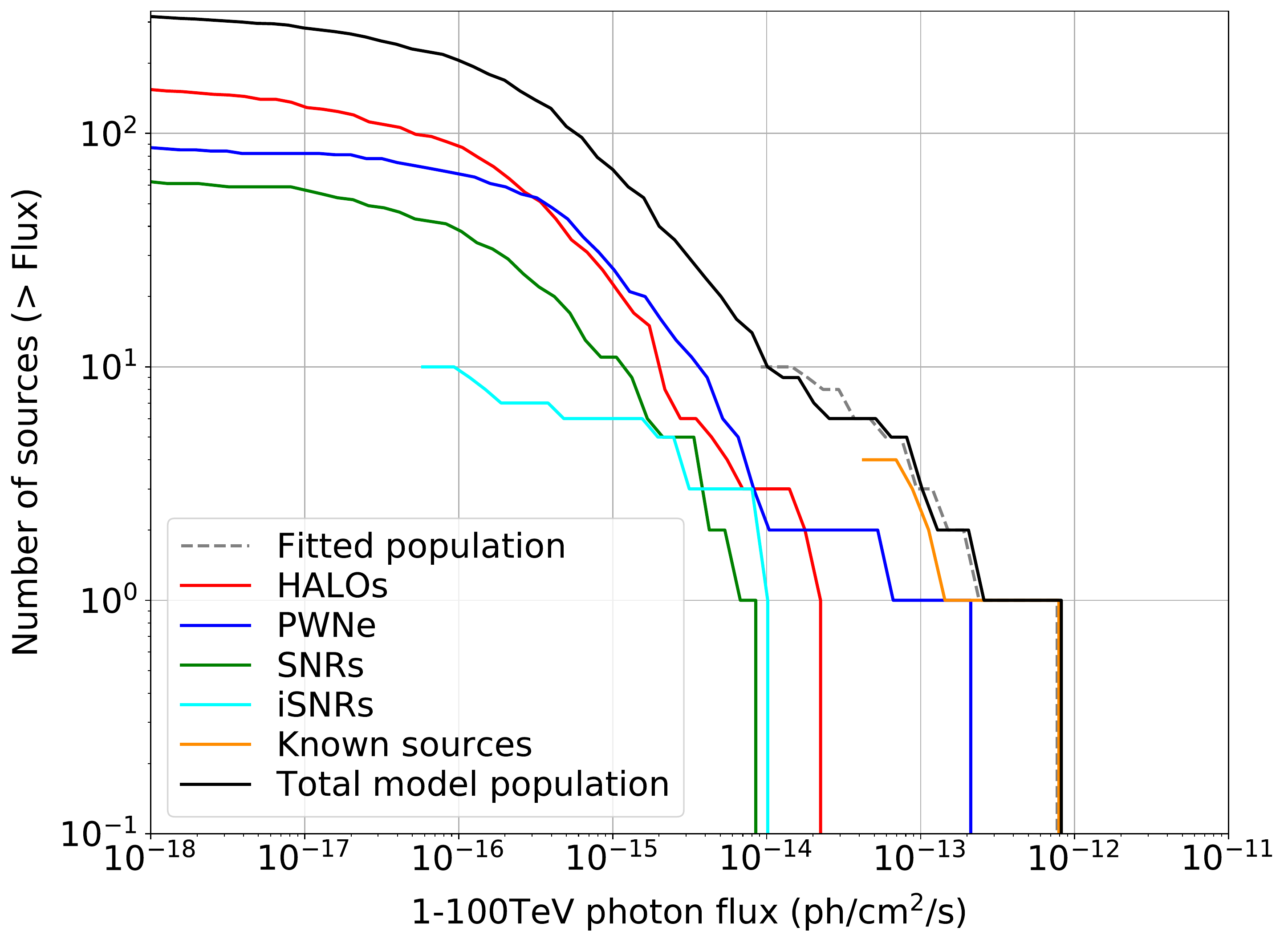}
\caption{Flux distribution of the fitted $>5\sigma$ sources detected in the blind search, compared to the true flux distributions of the various source populations in the model.}
\label{fig:res:pop:lognlogs}
\end{center}
\end{figure}

We completed the blind search process by fitting the detected $>5\sigma$ sources over the full 0.1-100\tev range, assuming for each source a simple power-law model. The positions of all sources were optimized in the process. Apart from the sources found in the blind search, only the true model for the instrumental background was fitted to the data (i.e., no models for large-scale interstellar emission). The ten sources are detected with TS values ranging from a few tens to a few thousands. The flux distribution of the fitted sources is shown in Fig. \ref{fig:res:pop:lognlogs} and demonstrates an overall pretty satisfactory recovery of the true source fluxes, despite using a very simple procedure. The effect of source confusion in a limited number of cases is visible as a small deviation at the lowest detected fluxes.

\subsection{Dark Matter sensitivity curves} \label{sec:res:dm}

In this section, we calculate forecasts for the detection of an additional emission component due to the annihilation of \gls{dm} particles. The goal is to explore whether \gls{cta} will be able to observe the annihilation of \gls{wimp} in the \gls{lmc}, and if detection (or failure thereof) can be useful in the identification of \gls{dm} candidates. In doing so, we incorporated the fact that the prospects for such detection or constraint depend on the degree of contamination of the observations by classical sources.

In line with the common use in the literature, we do not aim here at forecasts for a specific \gls{dm} candidate but rather at sampling the range of signals that should be expected within a ``standard'' \gls{wimp} scenario. While treating $\sigv$ and $\mdm$ as independent variables, we sample the ``single particle'' spectrum of the annihilation, namely the gamma--ray emission produced in an annihilation process in which the two \gls{dm} particles annihilate in different types of standard model primaries and generate a subsequent cascade of high-energy photons. 

Using the \gls{dm} profiles and spectra described in previous sections, we performed a likelihood analysis to calculate upper limits, as described in Sect. \ref{sec:simana:fit}, using the function $\textit{ctulimit}$ from the ctools software package. The Asimov dataset for the full baseline emission model, used in the analyses presented in previous sections, is here used as a background from which we try to disentangle the \gls{dm} emission. Each \gls{dm} model is included in the fitted model separately and we calculate the differential flux upper limit, the integrated flux upper limit over the energy range 0.1 GeV-\gls{dm} particle mass, and the integrated energy flux limit.

\begin{figure}
\includegraphics[width=\columnwidth]{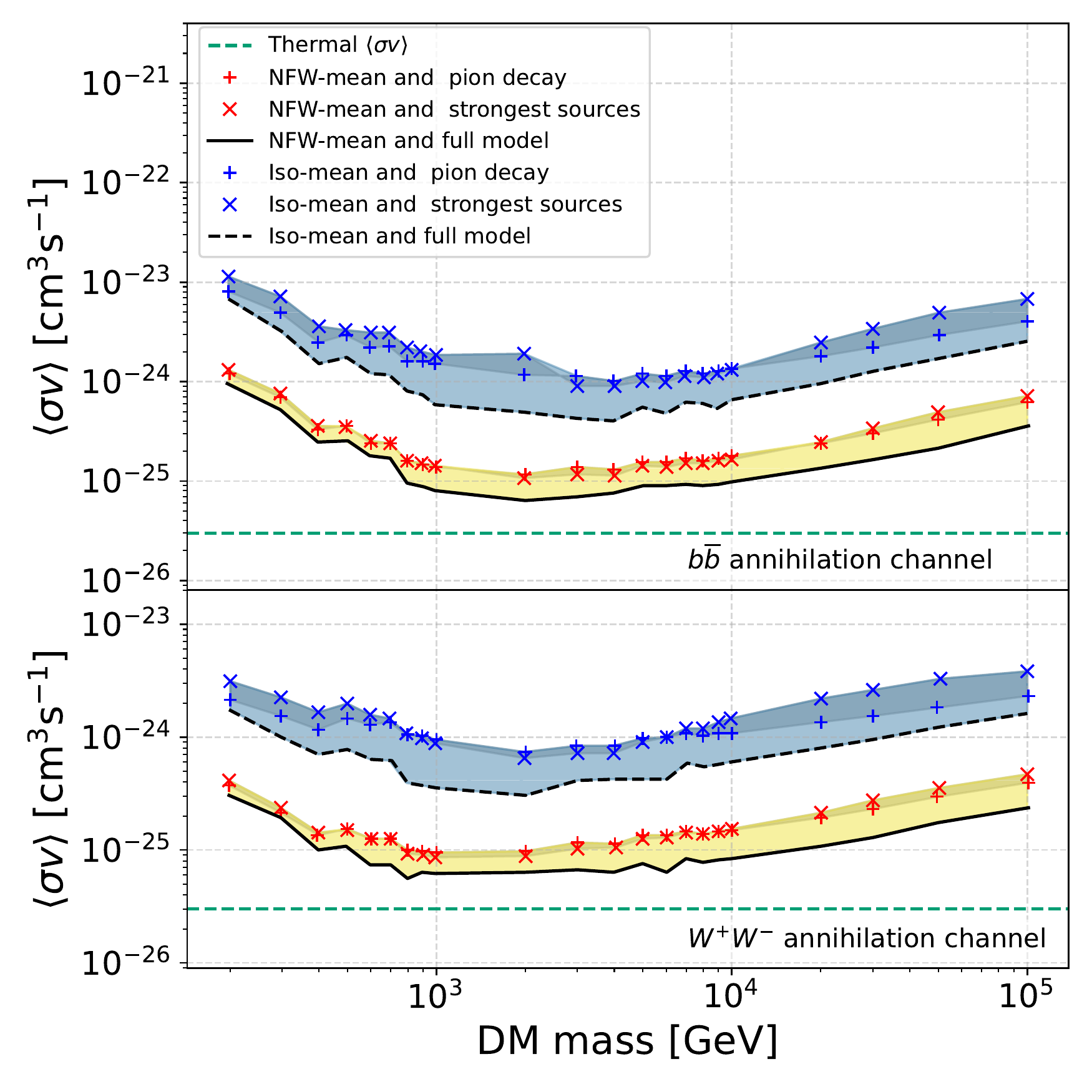}
\caption{Sensitivity bands in terms of velocity-averaged annihilation cross section as a function of \gls{dm} particle mass, resulting from the mean density profiles listed in Table \ref{tab:dmprofiles} and the $b\overline b$ and $W^+W^-$ annihilation channels, for different \gls{lmc} emission models.}
\label{fig:dmsensicurves1}
\end{figure}

\begin{figure}
\includegraphics[width=\columnwidth]{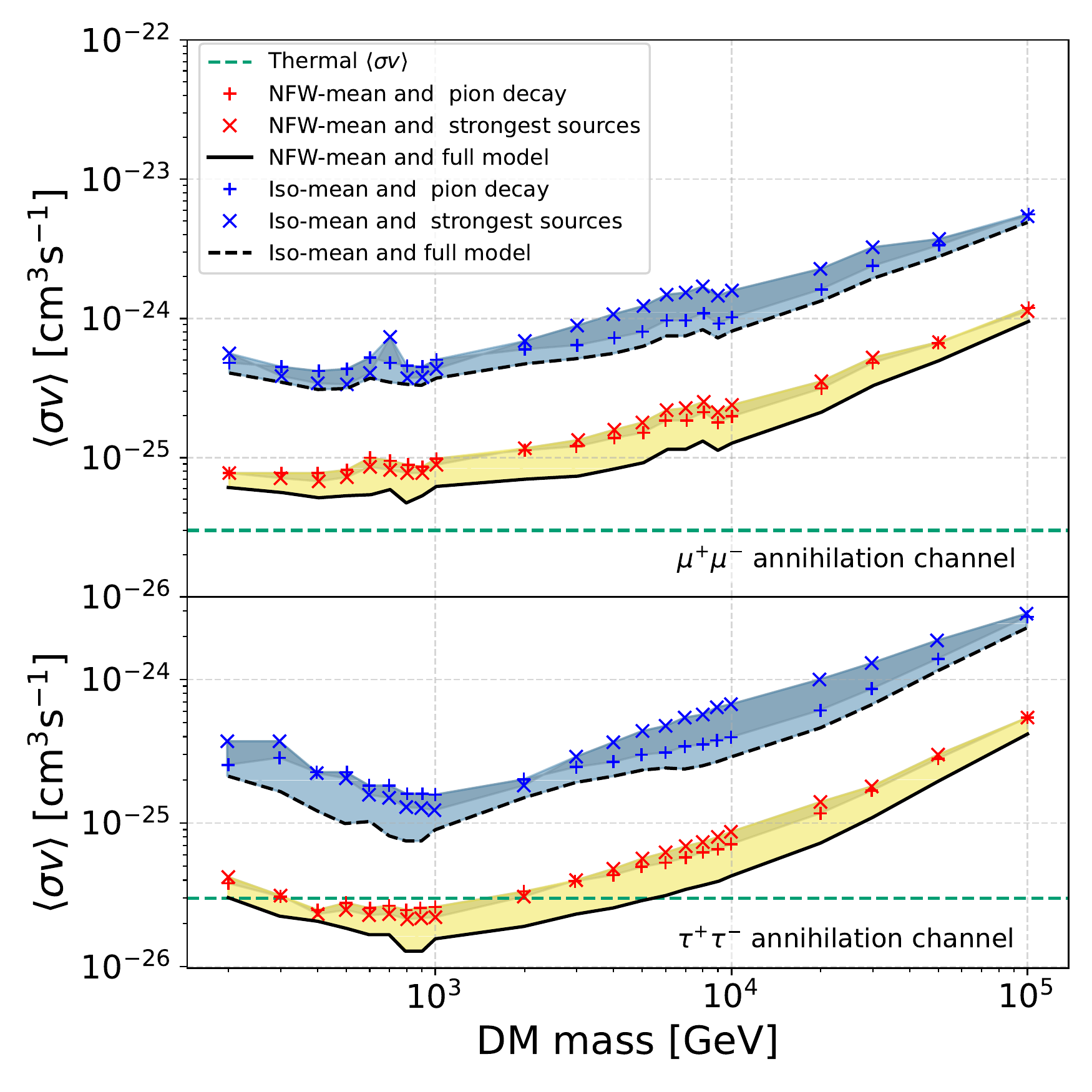}
\caption{Same as Fig. \ref{fig:dmsensicurves1} for the $\mu^+ \mu^-$ and $\tau^+ \tau^-$ annihilation channels.}
\label{fig:dmsensicurves2}
\end{figure}

\begin{figure}
\includegraphics[width=\columnwidth]{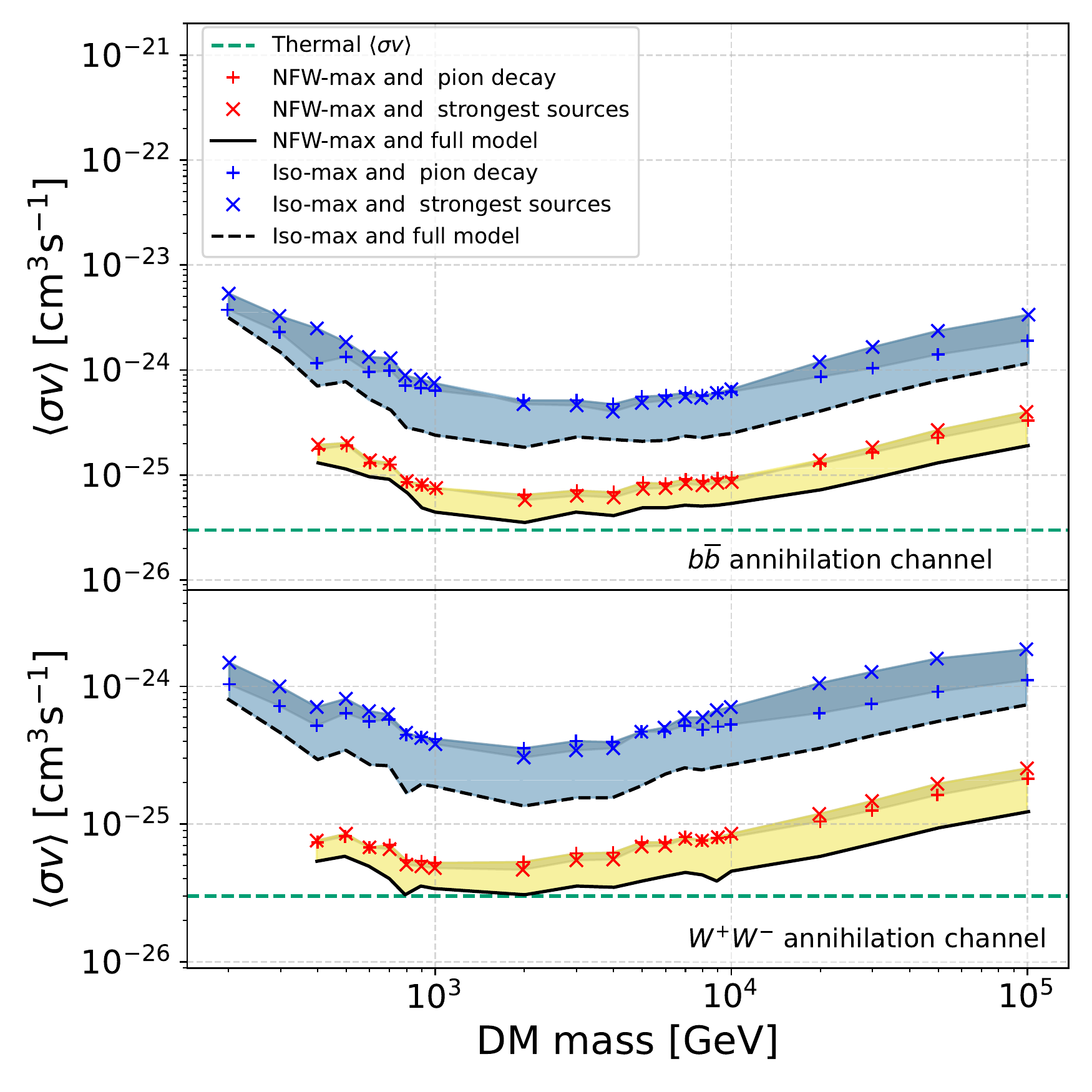}
\caption{Sensitivity bands in terms of velocity-averaged annihilation cross section as a function of \gls{dm} particle mass, resulting from the max density profiles listed in Table \ref{tab:dmprofiles} and the annihilation channels: $b\overline b$, $W^+W^-$, for different \gls{lmc} emission models.}
\label{fig:dmsensicurves5}
\end{figure}

\begin{figure}
\includegraphics[width=\columnwidth]{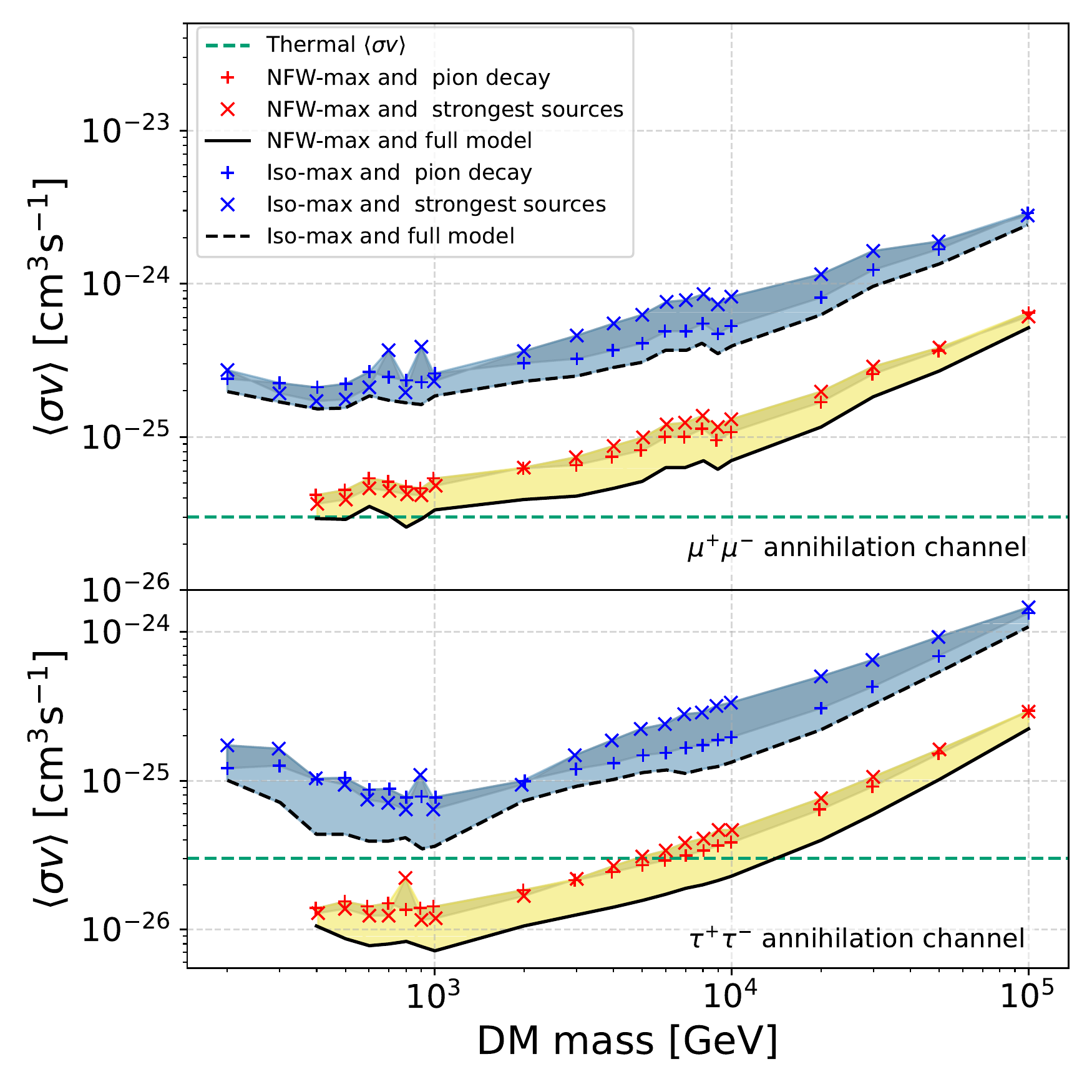}
\caption{Same as Fig. \ref{fig:dmsensicurves5} for the $\mu^+ \mu^-$ and $\tau^+ \tau^-$ annihilation channels.}
\label{fig:dmsensicurves6}
\end{figure}

The limit in terms of $\sigv$ can be extracted from the resulting integrated flux limit, using Eq. \ref{eq:dmflux}, obtaining the corresponding value of $\sigv$, for the specific J-factor and particle mass. 
The sensitivity curves of $\sigv$ versus \gls{dm} particle masses are shown in Figs. \ref{fig:dmsensicurves1} to \ref{fig:dmsensicurves6}. Three curves are shown in the plots for each annihilation channel and profile, corresponding to different scenarios regarding the gamma-ray background. The black lines represent the sensitivity curves obtained when including all the components of the \gls{lmc} emission model in the fit. The crosses and blades represent the results reducing the fitted model to just the pion-decay template or the strongest point sources, respectively. In these cases, the full emission model is used in the observations simulation, but just a subset of the components are included in the fitted emission model on top of which we search for a \gls{dm} signal.

The sensitivity curve is lower when fitting the full emission model, pointing to a better sensitivity to a \gls{dm} signal when all true components of the emission are used in the analysis. This illustrates the advantage of having an extended object that can be spatially resolved, such as the \gls{lmc}, the emission of which can be carefully decomposed. When using the full model, the baryonic emission components are more tightly constrained and better separated from the \gls{dm} template. Alternatively, when using only a subset of the components (pion decay template or strongest sources), their correlation with other true components of the baryonic emission leads to a higher \gls{dm} signal being required for significant detection. We therefore present the sensitivity curves in the shape of shaded bands, to illustrate the uncertainty on our knowledge of the actual composition of the \gls{lmc} emission.

For the majority of models, the sensitivity bands lie above the canonical thermal cross section, which defines the range of \gls{wimp} models which allow to recover the current \gls{dm} density in the universe, and are considered canonical \gls{wimp}. However, for channels with a higher integrated flux, like $\tau^+\tau^-$, and for the maximal version of the \gls{nfw} profile, the sensitivity bands show that \gls{cta} could be able to reach below the canonical cross-section parameter space in a range of masses around 1 TeV. A recent study on \gls{dm} detection in the \gls{lmc} using radio data \citep{Regis:2021} already excluded masses below 480\,GeV for the $b \overline b$, $W^+ W^-$, $\tau^+\tau^-$, $\mu^+ \mu^-$ channels, but according to our results, \gls{cta} could extend those limits up to a few TeV, at least for some of the studied \gls{dm} candidate models. These results are ideal because, among other things, we are assuming a perfect knowledge of the gamma-ray background emission, but still they point out that \gls{cta} observations of the \gls{lmc} could be very useful to constrain the \gls{dm} annihilation emission models. 

Our forecast sensitivity is compatible with those inferred with a similar technique for other targets, for instance the \gls{gc} as studied in \citet{Acharyya:2021}.
Indeed, the possibility to detect \gls{dm} in the \gls{lmc} is weaker than in the \gls{gc}, owing to the combined effect of the larger distance and shallower distribution (smaller J-factor) of the first with respect to the latter.

\section{Summary and conclusions} \label{sec:conclu}

We simulated the observation and analysis of a 340h survey of the \gls{lmc} with \gls{cta}, based on an emission model comprising components for the four already known TeV point sources, the galaxy-scale interstellar emission, and mock populations of \glspl{snr}, \glspl{pwn}, and pulsar halos. We also assessed prospects for the detectability of young remnant SN~1987A and star-forming region 30 Doradus, and we derived the constraints that can be obtained on the nature of dark matter from observing the massive halo of the \gls{lmc}.

Known point sources PWN N~157B, SNR N~132D, HMXB LMC P3, and SB 30 Doradus C will be detected with very high significances over most of the 0.1-100\tev range. The sensitivity of the \gls{cta} survey will allow for fine spectral studies and provide a meaningful extension of the spectra down to $\sim$100\gev, thus allowing a reliable connection to the range probed with Fermi-LAT, and up to or beyond 10\tev. Such a broad spectral coverage will be instrumental in characterizing the global particle acceleration efficiency, the respective contribution of different particle populations to the emission, and the maximum energies reached in these accelerators.

The young remnant from SN~1987A has not yet been detected in \gls{he}/\gls{vhe} gamma-rays, so quantitative prospects are very much dependent on the assumed model for particle acceleration and its recent and future evolution. A simple model, informed by current constraints on the shock dynamics and the sub-GeV electron population, suggests that \gls{cta} could detect hadronic emission from a population of emitting nuclei whose spectrum flattens towards high energies up to power-law indices $\sim2.0$. A softer emission, as observed in many Galactic \glspl{snr}, could be within reach of the survey pending a flux increase by at least a factor $3-4$ over $\sim2015-2035$, which would trigger additional observations to get a meaningful spectral characterization.

Beyond known point sources or promising candidates, the survey should allow the detection of an additional half a dozen sources, typically with 1-10\tev fluxes $>2-3 \times 10^{-14}$\feunit. Our source population model suggests that this sample would be dominated by pulsar-powered objects, and a couple of interacting \glspl{snr}. Source confusion may be an issue even for the small fraction of the source population accessible to the survey, with half of the simulated detections being found in locations where $2-3$ bright sources overlap. Most sources will be detected as point-like objects, which will likely complicate their classification based on \gls{vhe} observations alone.

Interstellar emission on the galaxy scale in the LMC is constrained by Fermi-LAT observations at 10-100\gev, but its properties in the \gls{vhe} remain essentially unexplored. Our baseline model for interstellar emission predicts that pion-decay dominates inverse-Compton radiation by at least an order of magnitude over the entire \gls{cta} range. If its spectrum extends from 10\gev to higher energies with a rather soft photon index $\sim2.7$, it will remain mostly out of reach of the survey, except for the low-significance detection of emission over less than half a degree from the molecular ridge south of 30 Doradus. If the \gls{cr} population in the \gls{lmc} exhibits a spectral hardening above a few 100\gev, as inferred from diffuse emission of the (innermost regions of the) \gls{mw} and observed in the local flux of primary \gls{cr} nuclei, and if this trend extends up to 10-100\tev, then degree-scale interstellar pion decay emission could be detected in the direction of major molecular cloud complexes, with high significance over the 0.1-10\tev range. 

Star-forming region 30 Doradus was not detected as a prominent and specific source in current GeV and TeV observations, but it has a clear potential for efficient production of \glspl{cr} and \gls{he}/\gls{vhe} radiation. In the framework of continuous injection of accelerated particles by 30 Doradus, followed by diffusion away from it, the region could be detected for acceleration efficiency of at least 40\% of the assumed $\sim 10^{39}$\punit\ mechanical power of the central stellar clusters. If diffusion in the $<100$\,pc vicinity of the source is suppressed, the requirement on acceleration efficiency is relaxed down to below 1\% for diffusion suppression factors of a few hundreds. Detailed studies of such a target, including in particular a characterization of its intensity distribution, will however be challenging given the distance to the source and the need to separate it from foreground and background interstellar emission from the galaxy. Moreover, the survey sensitivity may not be sufficient to investigate the possible role of \glspl{sfr} in producing PeV-scale galactic \gls{crs}.

Finally, prospects for the detection of a possible $\gamma$-ray signal from \gls{dm} annihilation were computed, harnessing the extended nature of the \gls{lmc}, crucial to disentangle \gls{dm} emission from a baryonic background, and its relatively high J-factor, comparable to other popular \gls{dm} candidates such as dwarf spheroidals. 
Several \gls{dm} density profiles, annihilation channels and particle masses have been tested on top of the baseline baryonic emission model for the \gls{lmc} to compute the velocity-averaged annihilation cross section required for detection. The majority of models lie about five to ten times above the canonical thermal cross section, benchmark for self-annihilation in the case of \gls{wimp} \gls{dm} being a thermal relic. 
However, by adopting NFW density profiles, maximizing the \gls{dm} density and still complying with the LMC rotation curves constraints, and annihilation channels yielding large integrated flux, like the $\tau^+\tau^-$ channel, the computed sensitivity reaches below the thermal cross section at $\sim$TeV energies, meaning these \gls{dm} models could be detected or excluded by \gls{cta}, making the \gls{lmc} a worthy candidate for \gls{dm} searches. 

The work introduced in this article provides a first quantitative assessment of the prospects opened by a deep survey of the \gls{lmc} with \gls{cta}. It is rather optimistic in many assumptions, for instance data analysis methods relying exclusively on full spatial and spectral maximum likelihood approaches with perfect knowledge of the instrumental background distribution in the field of view. The material developed during this work should serve as a starting point for more detailed studies, for instance assessing the impact of alternative data analysis methods or refining the prospects for specific sources such as SN~1987A by comparing more advanced and dedicated models to the sensitivity curves provided here.

\section*{Acknowledgements}

We gratefully acknowledge financial support from the following agencies and organizations:
State Committee of Science of Armenia, Armenia;
The Australian Research Council, Astronomy Australia Ltd, The University of Adelaide, Australian National University, Monash University, The University of New South Wales, The University of Sydney, Western Sydney University, Australia;
Federal Ministry of Education, Science and Research, and Innsbruck University, Austria;
Conselho Nacional de Desenvolvimento Cient\'{\i}fico e Tecnol\'{o}gico (CNPq), Funda\c{c}\~{a}o de Amparo \`{a} Pesquisa do Estado do Rio de Janeiro (FAPERJ), Funda\c{c}\~{a}o de Amparo \`{a} Pesquisa do Estado de S\~{a}o Paulo (FAPESP), Funda\c{c}\~{a}o de Apoio \`{a} Ci\^encia, Tecnologia e Inova\c{c}\~{a}o do Paran\'a - Funda\c{c}\~{a}o Arauc\'aria, Ministry of Science, Technology, Innovations and Communications (MCTIC), Brasil;
Ministry of Education and Science, National RI Roadmap Project DO1-153/28.08.2018, Bulgaria;
The Natural Sciences and Engineering Research Council of Canada and the Canadian Space Agency, Canada;
CONICYT-Chile grants CATA AFB 170002, ANID PIA/APOYO AFB 180002, ACT 1406, FONDECYT-Chile grants, 1161463, 1170171, 1190886, 1171421, 1170345, 1201582, Gemini-ANID 32180007, Chile;
Croatian Science Foundation, Rudjer Boskovic Institute, University of Osijek, University of Rijeka, University of Split, Faculty of Electrical Engineering, Mechanical Engineering and Naval Architecture, University of Zagreb, Faculty of Electrical Engineering and Computing, Croatia;
Ministry of Education, Youth and Sports, MEYS  LM2015046, LM2018105, LTT17006, EU/MEYS CZ.02.1.01/0.0/0.0/16\_013/0001403, CZ.02.1.01/0.0/0.0/18\_046/0016007 and CZ.02.1.01/0.0/0.0/16\_019/0000754, Czech Republic; 
Academy of Finland (grant nr.317636 and 320045), Finland;
Ministry of Higher Education and Research, CNRS-INSU and CNRS-IN2P3, CEA-Irfu, ANR, Regional Council Ile de France, Labex ENIGMASS, OCEVU, OSUG2020 and P2IO, France;
Max Planck Society, BMBF, DESY, Helmholtz Association, Germany;
Department of Atomic Energy, Department of Science and Technology, India;
Istituto Nazionale di Astrofisica (INAF), Istituto Nazionale di Fisica Nucleare (INFN), MIUR, Istituto Nazionale di Astrofisica (INAF-OABRERA) Grant Fondazione Cariplo/Regione Lombardia ID 2014-1980/RST\_ERC, Italy;
ICRR, University of Tokyo, JSPS, MEXT, Japan;
Netherlands Research School for Astronomy (NOVA), Netherlands Organization for Scientific Research (NWO), Netherlands;
University of Oslo, Norway;
Ministry of Science and Higher Education, DIR/WK/2017/12, the National Centre for Research and Development and the National Science Centre, UMO-2016/22/M/ST9/00583, Poland;
Slovenian Research Agency, grants P1-0031, P1-0385, I0-0033, J1-9146, J1-1700, N1-0111, and the Young Researcher program, Slovenia; 
South African Department of Science and Technology and National Research Foundation through the South African Gamma-Ray Astronomy Programme, South Africa;
The Spanish groups acknowledge the Spanish Ministry of Science and Innovation and the Spanish Research State Agency (AEI) through grants AYA2016-79724-C4-1-P, AYA2016-80889-P, AYA2016-76012-C3-1-P, BES-2016-076342, FPA2017-82729-C6-1-R, FPA2017-82729-C6-2-R, FPA2017-82729-C6-3-R, FPA2017-82729-C6-4-R, FPA2017-82729-C6-5-R, FPA2017-82729-C6-6-R, PGC2018-095161-B-I00, PGC2018-095512-B-I00, PID2019-107988GB-C22; the “Centro de Excelencia Severo Ochoa” program through grants no. SEV-2016-0597, SEV-2016-0588, SEV-2017-0709, CEX2019-000920-S; the “Unidad de Excelencia Mar\'ia de Maeztu” program through grant no. MDM-2015-0509; the “Ram\'on y Cajal” programme through grants RYC-2013-14511, RYC-2017-22665; and the MultiDark Consolider Network FPA2017-90566-REDC. They also acknowledge the Atracci\'on de Talento contract no. 2016-T1/TIC-1542 granted by the Comunidad de Madrid; the “Postdoctoral Junior Leader Fellowship” programme from La Caixa Banking Foundation, grants no.~LCF/BQ/LI18/11630014 and LCF/BQ/PI18/11630012; the “Programa Operativo” FEDER 2014-2020, Consejer\'ia de Econom\'ia y Conocimiento de la Junta de Andaluc\'ia (Ref. 1257737), PAIDI 2020 (Ref. P18-FR-1580) and Universidad de Ja\'en; “Programa Operativo de Crecimiento Inteligente” FEDER 2014-2020 (Ref.~ESFRI-2017-IAC-12), Ministerio de Ciencia e Innovaci\'on, 15\% co-financed by Consejer\'ia de Econom\'ia, Industria, Comercio y Conocimiento del Gobierno de Canarias; the Spanish AEI EQC2018-005094-P FEDER 2014-2020; the European Union’s “Horizon 2020” research and innovation programme under Marie Skłodowska-Curie grant agreement no. 665919; and the ESCAPE project with grant no. GA:824064;
Swedish Research Council, Royal Physiographic Society of Lund, Royal Swedish Academy of Sciences, The Swedish National Infrastructure for Computing (SNIC) at Lunarc (Lund), Sweden;
State Secretariat for Education, Research and Innovation (SERI) and Swiss National Science Foundation (SNSF), Switzerland;
Durham University, Leverhulme Trust, Liverpool University, University of Leicester, University of Oxford, Royal Society, Science and Technology Facilities Council, UK;
U.S. National Science Foundation, U.S. Department of Energy, Argonne National Laboratory, Barnard College, University of California, University of Chicago, Columbia University, Georgia Institute of Technology, Institute for Nuclear and Particle Astrophysics (INPAC-MRPI program), Iowa State University, the Smithsonian Institution, Washington University McDonnell Center for the Space Sciences, The University of Wisconsin and the Wisconsin Alumni Research Foundation, USA. This research has made use of the CTA instrument response functions provided by the CTA Consortium and Observatory, see \url{http://www.cta-observatory.org/science/ctao-performance/} for more details.

The research leading to these results has received funding from the European Union's Seventh Framework Programme (FP7/2007-2013) under grant agreements No~262053 and No~317446.
This project is receiving funding from the European Union's Horizon 2020 research and innovation programs under agreement No~676134. 
We acknowledge financial support from the French Agence Nationale de la Recherche under reference ANR-19-CE31-0014 (GAMALO project) and the Italian grant 2017W4HA7S “NAT-NET: Neutrino and Astroparticle Theory Network” (PRIN 2017) funded by the Italian Ministero dell’Istruzione, dell’Università e della Ricerca (MIUR), and Iniziativa Specifica TAsP of INFN.

This work made use of the SIMBAD database, operated at CDS, Strasbourg, France, of NASA's Astrophysics Data System Bibliographic Services, and of the Southern H-Alpha Sky Survey Atlas (SHASSA), which is supported by the National Science Foundation. The results presented were produced with the aid of the APLpy, Astropy, Matplotlib, NumPy, SciPy, and Naima open-access software tools.

\section*{Data Availability}

The full galaxy emission model developed in this work can be provided on request by the corresponding authors of the article. The tools for simulating the survey observations and analyzing the resulting data are publicly available at \url{http://cta.irap.omp.eu/ctools/index.html}, and the instrument response functions can be downloaded from \url{https://zenodo.org/record/5499840\#.Y9D4nvGZMbY}.



\bibliographystyle{mnras}
\bibliography{Biblio/DM.bib,Biblio/LMC.bib,Biblio/Halo.bib,Biblio/ISM.bib,Biblio/ISRF.bib,Biblio/Pulsars.bib,Biblio/DataAnalysis.bib,Biblio/Physics.bib,Biblio/Starbursts.bib,Biblio/Fermi.bib,Biblio/CosmicRayEscape.bib,Biblio/CosmicRayAcceleration.bib,Biblio/CosmicRayTransport.bib,Biblio/CosmicRayMeasurements.bib,Biblio/GalacticDiffuseEmission.bib,Biblio/Superbubbles.bib,Biblio/StarFormingRegions.bib,Biblio/CTA.bib}



\appendix

\section{Cosmic-ray source distribution}
\label{app:crsrc}

Models for interstellar emission on a galactic scale were computed under the assumption of diffusion-loss transport in a uniform model of the \gls{ism} of \gls{crs} steadily injected from an ensemble of point sources. As a tracer for sites of \gls{cr} injection related to the massive star population, we used a selection of $\rm H\RomNumCap{2}$ regions from the catalog of \citet{Pellegrini:2012}, which covers the entire galaxy and provides an indirect mean to determine the membership distribution of young star clusters. 

We restricted our selection to regions with H$\alpha$ luminosities above $10^{37}$\punit, a limit below which the H$\alpha$ luminosity function flattens as a result of stochastic ionizing populations, hence a regime where our assumption of steady \gls{cr} injection would be less and less valid. For reference, \citet{Pellegrini:2012} indicate that an H$\alpha$ luminosity of $10^{37}$\punit\ corresponds to the Orion Nebula, which harbours a single O6.5 V star. Above this value, the luminosity function is close to a power law with a slope of -1.8 and extends up to the tremendous $10^{39.66}$\punit\ luminosity of the 30 Doradus region that is ionised by hundreds of O stars in cluster R136a. This object is literally extraordinary, even beyond \gls{lmc} and within the Local Group, and actually dominates the output of the galaxy. In our model, we handled it separately from the rest of star clusters for three reasons: (i) including it as any other star cluster in a distribution of \gls{cr} sources would result in \gls{cr} injection and the related gamma-ray emission to be strongly concentrated in 30 Doradus, which may too favourably bias detection prospects; (ii) although there is some age spread over the region, most OB stars are young with ages $<$5Myr \citep{Schneider:2018}, so it may well be that only very few \gls{sne} exploded in the recent past, hence a limited \gls{cr} injection \citep[see also][about the very recent increase in star formation of 30 Doradus]{Harris:2009}; (iii) supporting the previous point, 30 Doradus is not conspicuous at GeV energies, at least not in proportion to its H$\alpha$ emission \citep{Ackermann:2016}. 

For all 138 remaining regions in our sample, we converted H$\alpha$ luminosity into ionizing luminosities, applying a correction for LyC radiation escape using the morphological classification and escape fraction determined by \citet{Pellegrini:2012}. We took ionizing luminosity as a measure of the richness of each star cluster, to which we assumed \gls{cr} injection power is proportional. There are several caveats to this approach: while some proportionality between ionizing and \gls{cr} injection powers can be expected in the limit of continuous and high star formation rate and \gls{cr} acceleration by \gls{sne} only, the relationship is most likely subject to variations when one includes effects such as finite sampling of the initial mass function, stellar age spread, collective acceleration processes given the actual stellar cluster substructure, etc. Another caveat is that such a relation does not hold for the contribution from thermonuclear \gls{sne}, whose rate is relatively high in the \gls{lmc} compared to that integrated over larger volumes and durations in the local Universe \citep[by about a factor of two; see][]{Maggi:2016}. This is most likely an effect of the actual star formation history of the \gls{lmc} over the past 2\,Gyr, which leads to a relatively high rate of thermonuclear \gls{sne} now. The spatial distribution of thermonuclear \gls{sne} will be less appropriately traced by $\rm H\RomNumCap{2}$ regions, but we checked that adopting a less concentrated and more uniform distribution of \gls{cr} sources does not affect our conclusions on the detectability of large-scale interstellar emission.

\section{Cosmic-ray propagation}
\label{app:crprop}

For large-scale interstellar emission, we consider that \gls{cr} transport away from injection sites proceeds by spatial diffusion limited by energy losses, in a medium with homogeneous and isotropic properties. We solved the diffusion-loss equation for a point-like and stationary source following \citet{Atoyan:1995}. The spectral density at radius $r$ from the source and time $t$ since injection start is obtained from:
\begin{align}
\label{eq:srckern}
\frac{\dop{N}}{\dop{E} \dop{V}}(r,E,t) = \int_{t_{\rm inj}}^{t}{\frac{\dot{E}(E_0)}{\dot{E}(E)} \frac{Q(E_0)}{\pi^{3/2}r_{\rm diff}^3} e^{-r^2/r^2_{\rm diff}} \dop{t_0}}
\end{align}
where $E_0$ is the initial particle energy at injection time $t_0$ and integration runs over injection history. The earliest possible injection time $t_{\rm inj}$ is computed from the maximum cooling time from cutoff energy $E_{\rm cut}$ down to current energy $E$:
\begin{align}
&t_{\rm inj}(E,t) = {\rm max}([0,t-t_{\rm cool}(E_{\rm cut},E)]) \\
&t_{\rm cool}(E_{\rm cut},E) = \int_{E}^{E_{cut}}{\frac{\dop{e}}{\dot{E}(e)}}
\end{align}
while the diffusion radius $r_{\rm diff}$ is computed from diffusion coefficient $D$ and energy loss rate $\dot{E}$:
\begin{align}
r_{\rm diff}(E_0,E) = 2 \left[ \int_{E}^{E_0}{\frac{D(e)}{\dot{E}(e)}\dop{e}} \right]^{1/2}
\end{align}
Proton and electron spatial distributions around the stationary source are computed by integrating over an injection duration of $t=100$\,Myr, instead of computing an exact steady-state solution. This accounts for the fact that the star formation history of the \gls{lmc} was not steady over recent times, and in particular exhibits a drop in star formation at 100\,Myr in most regions (but this has limited impact at the very high energies probed with \gls{cta}, at which particles diffuse on time scales smaller than this assumed injection duration). The above solution implicitly assumes a zero density at infinity. If the extent of the diffusion region across the galaxy has a finite value, it should in principle be possible to retrieve a similar predicted emission by assuming a smaller diffusion coefficient (for diffusion-dominated transport).

\section{Interstellar radiation fields}
\label{app:isrf}

The model for the \gls{isrf} was developed from the work of \citet{Paradis:2011}, in which the broadband infrared dust emission of the \gls{lmc} was linearly decomposed into gas phases, eventually yielding dust emissivity spectra $Q_Y(\nu)$ per unit column density for each phase $Y$. The level of stellar radiation heating the dust was obtained from fits of these emissivities with predictions from the \textit{DustEM} dust emission model under two different assumptions for the stellar field: dust in the molecular and atomic phases is exposed to a radiation field characteristic of the solar neighbourhood, $R^{\rm Mathis}(\nu)$, while dust in the ionized phase is heated by a radiation field more appropriate to the vicinity of massive star clusters, $R^{\rm GALEV}(\nu)$. For our purposes, we retained for our baseline model the results for the regime defined as ``typical $\rm H\RomNumCap{2}$ regions" in the article (case 2), which covers most of the \gls{lmc} disk and is thus appropriate for an average \gls{isrf} model on large scales. 

In practice, infrared emissivities were scaled by gas column densities defined below, while stellar radiation fields were renormalized by the fitting factors $r$ given in Table 2 of \citet{Paradis:2011} and further scaled by filling factors $f_X$ for their respective gas phase X, ionised $i$ or neutral $n$ (atomic and molecular are grouped because the filling factor of the molecular phase is small compared to the atomic and ionised phases). The interstellar radiation spectral energy density $U(\nu)$ is the sum of components arising from stars and dust, plus the cosmic microwave background. It reads:
\begin{align}
&U(\nu) = U_{{\rm dust}}(\nu) + U_{{\rm stars}}(\nu) + U_{{\rm CMB}}(\nu) \\
&U_{{\rm dust}}(\nu) = \frac{4\pi}{c} \left[ N_{\rm H\RomNumCap{1}} Q_{\rm H\RomNumCap{1}}(\nu) + N_{\rm H_2} Q_{\rm H_2}(\nu) + N_{\rm H\RomNumCap{2}} Q_{\rm H\RomNumCap{2}}(\nu) \right] \\
&U_{{\rm stars}}(\nu) = \frac{4\pi}{c} \left[ r_{\rm i} f_{\rm i} R^{\rm GALEV}(\nu) + r_{\rm n} f_{\rm n} R^{\rm Mathis}(\nu) \right]
\end{align}
with $f_{\rm i} =  f_{\rm n} = 0.5$ \citep{DeAvillez:2004}. $U_{{\rm CMB}}$ is the radiation spectral energy density of the cosmic microwave background. The average hydrogen column densities for the atomic, molecular, and ionised phases are obtained from the gas masses and gas disk geometry assumed in Sect. \ref{sec:model:ism}:
\begin{align}
&N_{\rm H\RomNumCap{1}} = 1.85 \times 10^{21}\,{\rm H}\,{\rm cm}^{-2} \\
&N_{\rm H_2} = 2.43 \times 10^{20}\,{\rm H}\,{\rm cm}^{-2} \\
&N_{\rm H\RomNumCap{2}} = 5.48 \times 10^{19}\,{\rm H}\,{\rm cm}^{-2}
\end{align}

An alternative \gls{isrf} model, as part of the so-called ``gas-rich \gls{ism} model", was also developed assuming an average neutral gas column density ten times the average given above, while the ionized gas column density becomes $6.18 \times 10^{20}$\cunit, as computed following \citet{Paradis:2011}, using electron density $n_e=3.98$\vunit, and an H$\alpha$ intensity of 113.3 Rayleigh corresponding to the limit between ``typical $\rm H\RomNumCap{2}$ regions" and ``very bright $\rm H\RomNumCap{2}$ regions'' in the article. 

\section{Harder interstellar emission}
\label{app:hard}

Analysis of the Galactic diffuse emission observed with Fermi-LAT in the 0.1-100\gev range suggests a progressive increase in density and hardening of the spectrum of $\sim10-100$\gev \glspl{cr} as we move from outer to inner regions of the Milky Way, with a peak at galactocentric radii of a few kpc, the molecular ring position, which is also where the largest density of \gls{cr} sources are expected to be found \citep{Ackermann:2012,Acero:2016,Pothast:2018}. The typical spectral hardening of $\sim$0.2-0.4 in power-law index can be explained by position-dependent transport properties \citep{Evoli:2012,Gaggero:2015,Recchia:2016,Cerri:2017}. In several of these works, \gls{cr} source density is an important if not the main driver of \gls{cr} transport properties, either directly, in producing efficient self-confinement, or indirectly, in generating perpendicular outflows and/or magnetic field topologies, with the result that regions harbouring a larger number of \gls{cr} sources are associated with harder gamma-ray emission. With a supernova activity $\sim$10 times lower than the Milky Way in a volume $\sim30-40$ times smaller, the above effects could to some extent be at play in the \gls{lmc} and produce an average pion-decay emission harder than assumed in our baseline model. 

It is not clear, however, how this extrapolates to higher energies, especially those probed with \gls{cta}. Theoretically, different scenarios for the origin of the inferred hard spectrum of $\sim10-100$\gev \glspl{cr} in the inner Galaxy predict different behaviours at higher energies. Self-confinement, strongly suppressing diffusion and allowing advection to become a more dominant transport process \citep{Recchia:2016}, is expected to cease above $\sim100$\gev, where the \gls{cr} flux is too weak to excite significant turbulence; higher-energy particles would thus enter a regime of diffusive transport in externally driven turbulence \citep{Blasi:2012}, which would soften their spectrum compared to a more advection-dominated regime at lower energies. Alternatively, anisotropic diffusion, with predominant parallel diffusion off the plane in the inner regions, would preserve a hard spectrum in the inner Galactic regions at higher energies. 

Observationally, evidence for a hardening seems to extend to at least sub-TeV gamma-rays \citep{Pothast:2018,Neronov:2020}. At higher energies, the body of available information grows rapidly \citep{Abdo:2008,Abramowski:2014,Bartoli:2015,Amenomori:2021}, but instrumental and data analysis challenges make it difficult to firmly establish the spectrum of diffuse emission of purely interstellar origin (the main difficulties being background rejection, proper determination of individual source extension, and uncertain contribution from unresolved sources). The authors of \citet{Neronov:2020} inferred a hard gamma-ray spectrum at sub-TeV energies even in outer regions of the Galaxy, and \citet{Pothast:2018} reports a systematically harder emission at all Galactocentric radii in the $\ge 30$\gev range. This suggests the possibility that hard spectra of \glspl{cr} may be rather universal throughout most of the Galactic disk, which seems consistent with the hardening above $200-300$\,GV observed in the spectra of the local flux of primary \gls{cr} nuclei \citep{Aguilar:2015a,Aguilar:2015b,Aguilar:2017,Aguilar:2020}, at least up to about 10\tev \citep{An:2019}.

Another piece of evidence for hard gamma-ray emission on galactic scales, hence possibly a hard interstellar population of \glspl{cr}, comes from starburst galaxies. Hard spectra in the GeV range were measured \citep{Ajello:2020} and extend all the way up to the TeV range for a few starburst galaxies \citep{Acciari:2009,Abdalla:2018e}. These are admittedly extreme in their interstellar conditions and not representative of the \gls{lmc}; yet the very origin of the hard spectra remains unclear: they could be due to advection being the dominant transport mechanism up to very high energies \citep{Peretti:2019}, or to a diffusion scheme specific to the actual interstellar conditions \citep{Krumholz:2020}. To what extent one or the other scenario applies to the \gls{lmc} is unknown. About possible advection in a galactic wind, a large-scale multiphase outflow was detected in the \gls{lmc} on both near and far sides of the galaxy \citep{Barger:2016}, with a velocity $\sim 100$\kms\ that is on the low end of characteristic velocities for starburst-driven wind originating in star formation \citep{Veilleux:2005,Sturm:2011}. Such an outflow can actually be driven by the \gls{cr} population of the galaxy, \citep[see][for one among many recent developments on the topic]{Bustard:2020}, but whatever its origin, it has the potential to shape the spatial and spectral distribution of \glspl{cr} in the galaxy depending on the actual values of other parameters governing \gls{cr} transport in the \gls{ism}.

\section*{Affiliations}

\begin{enumerate}[label=$^{\arabic*}$,ref=\arabic*]
\item University of Alabama, Tuscaloosa, Department of Physics and Astronomy, Gallalee Hall, Box 870324 Tuscaloosa, AL 35487-0324, USA\label{AFFIL::UAlabamaTuscaloosa}
\item Laboratoire Lagrange, Universit\'e C\^ote d{\textquoteright}Azur, Observatoire de la C\^ote d{\textquoteright}Azur, CNRS, Blvd de l'Observatoire, CS 34229, 06304 Nice Cedex 4, France\label{AFFIL::OCotedAzur}
\item Laboratoire Leprince-Ringuet, CNRS/IN2P3, \'Ecole polytechnique, Institut Polytechnique de Paris, 91120 Palaiseau, France\label{AFFIL::LLREcolePolytechnique}
\item Departament de F{\'\i}sica Qu\`antica i Astrof{\'\i}sica, Institut de Ci\`encies del Cosmos, Universitat de Barcelona, IEEC-UB, Mart{\'\i} i Franqu\`es, 1, 08028, Barcelona, Spain\label{AFFIL::ICCUB}
\item Instituto de Astrof{\'\i}sica de Andaluc{\'\i}a-CSIC, Glorieta de la Astronom{\'\i}a s/n, 18008, Granada, Spain\label{AFFIL::IAACSIC}
\item Instituto de F{\'\i}sica Te\'orica UAM/CSIC and Departamento de F{\'\i}sica Te\'orica, Universidad Aut\'onoma de Madrid, c/ Nicol\'as Cabrera 13-15, Campus de Cantoblanco UAM, 28049 Madrid, Spain\label{AFFIL::IFTUAMCSIC}
\item Pontificia Universidad Cat\'olica de Chile, Av. Libertador Bernardo O'Higgins 340, Santiago, Chile\label{AFFIL::UPontificiaCatolicadeChile}
\item Gran Sasso Science Institute (GSSI), Viale Francesco Crispi 7, 67100 L{\textquoteright}Aquila, Italy and INFN-Laboratori Nazionali del Gran Sasso (LNGS), via G. Acitelli 22, 67100 Assergi (AQ), Italy\label{AFFIL::GSSIandINFNAquila}
\item INAF - Osservatorio Astrofisico di Arcetri, Largo E. Fermi, 5 - 50125 Firenze, Italy\label{AFFIL::OArcetri}
\item T\"UB\.ITAK Research Institute for Fundamental Sciences, 41470 Gebze, Kocaeli, Turkey\label{AFFIL::Tubitak}
\item INFN Sezione di Napoli, Via Cintia, ed. G, 80126 Napoli, Italy\label{AFFIL::INFNNapoli}
\item INFN Sezione di Padova and Universit\`a degli Studi di Padova, Via Marzolo 8, 35131 Padova, Italy\label{AFFIL::UPadovaandINFN}
\item Institute for Cosmic Ray Research, University of Tokyo, 5-1-5, Kashiwa-no-ha, Kashiwa, Chiba 277-8582, Japan\label{AFFIL::UTokyoICRR}
\item Kapteyn Astronomical Institute, University of Groningen, Landleven 12, 9747 AD, Groningen, The Netherlands\label{AFFIL::UGroningen}
\item Department of Physics, Chemistry \& Material Science, University of Namibia, Private Bag 13301, Windhoek, Namibia\label{AFFIL::UNamibia}
\item Centre for Space Research, North-West University, Potchefstroom, 2520, South Africa\label{AFFIL::NWU}
\item Universit\"at Hamburg, Institut f\"ur Experimentalphysik, Luruper Chaussee 149, 22761 Hamburg, Germany\label{AFFIL::UHamburg}
\item School of Physics and Astronomy, Monash University, Melbourne, Victoria 3800, Australia\label{AFFIL::UMonash}
\item Department of Astronomy, University of Geneva, Chemin d'Ecogia 16, CH-1290 Versoix, Switzerland\label{AFFIL::UGenevaISDC}
\item Universit\'e Paris-Saclay, Universit\'e Paris Cit\'e, CEA, CNRS, AIM, F-91191 Gif-sur-Yvette Cedex, France\label{AFFIL::CEAIRFUDAp}
\item Department of Physics, Graduate School of Science, The University of Tokyo, 7-3-1 Hongo, Bunkyo-ku, Tokyo 113-0033, Japan\label{AFFIL::UTokyoGSS}
\item Research Center for the Early Universe, School of Science, The University of Tokyo, 7-3-1 Hongo, Bunkyo-ku, Tokyo 113-0033, Japan\label{AFFIL::UTokyoRCEUSS}
\item IPARCOS-UCM, Instituto de F{\'\i}sica de Part{\'\i}culas y del Cosmos, and EMFTEL Department, Universidad Complutense de Madrid, E-28040 Madrid, Spain\label{AFFIL::UCMAltasEnergias}
\item Faculty of Science and Technology, Universidad del Azuay, Cuenca, Ecuador.\label{AFFIL::UAzuay}
\item Deutsches Elektronen-Synchrotron, Platanenallee 6, 15738 Zeuthen, Germany\label{AFFIL::DESY}
\item Centro Brasileiro de Pesquisas F{\'\i}sicas, Rua Xavier Sigaud 150, RJ 22290-180, Rio de Janeiro, Brazil\label{AFFIL::CBPF}
\item Instituto de Astrof{\'\i}sica de Canarias and Departamento de Astrof{\'\i}sica, Universidad de La Laguna, La Laguna, Tenerife, Spain\label{AFFIL::IAC}
\item Institut f\"ur Theoretische Physik, Lehrstuhl IV: Plasma-Astroteilchenphysik, Ruhr-Universit\"at Bochum, Universit\"atsstra{\ss}e 150, 44801 Bochum, Germany\label{AFFIL::UBochum}
\item Center for Astrophysics | Harvard \& Smithsonian, 60 Garden St, Cambridge, MA 02138, USA\label{AFFIL::CfAHarvardSmithsonian}
\item CIEMAT, Avda. Complutense 40, 28040 Madrid, Spain\label{AFFIL::CIEMAT}
\item Max-Planck-Institut f\"ur Physik, F\"ohringer Ring 6, 80805 M\"unchen, Germany\label{AFFIL::MPP}
\item INFN Sezione di Perugia and Universit\`a degli Studi di Perugia, Via A. Pascoli, 06123 Perugia, Italy\label{AFFIL::UPerugiaandINFN}
\item Pidstryhach Institute for Applied Problems in Mechanics and Mathematics NASU, 3B Naukova Street, Lviv, 79060, Ukraine\label{AFFIL::IAPMMLviv}
\item Univ. Savoie Mont Blanc, CNRS, Laboratoire d'Annecy de Physique des Particules - IN2P3, 74000 Annecy, France\label{AFFIL::LAPPUSavoieMontBlanc}
\item Center for Astrophysics and Cosmology (CAC), University of Nova Gorica, Nova Gorica, Slovenia\label{AFFIL::UNovaGoricaCAC}
\item INAF - Osservatorio Astronomico di Roma, Via di Frascati 33, 00040, Monteporzio Catone, Italy\label{AFFIL::ORoma}
\item ETH Z\"urich, Institute for Particle Physics and Astrophysics, Otto-Stern-Weg 5, 8093 Z\"urich, Switzerland\label{AFFIL::ETHZurich}
\item INFN Sezione di Bari, via Orabona 4, 70126 Bari, Italy\label{AFFIL::INFNBari}
\item Politecnico di Bari, via Orabona 4, 70124 Bari, Italy\label{AFFIL::PolitecnicoBari}
\item INAF - Osservatorio Astronomico di Palermo {\textquotedblleft}G.S. Vaiana{\textquotedblright}, Piazza del Parlamento 1, 90134 Palermo, Italy\label{AFFIL::OPalermo}
\item Nicolaus Copernicus Astronomical Center, Polish Academy of Sciences, ul. Bartycka 18, 00-716 Warsaw, Poland\label{AFFIL::NicolausCopernicusAstronomicalCenter}
\item IRFU, CEA, Universit\'e Paris-Saclay, B\^at 141, 91191 Gif-sur-Yvette, France\label{AFFIL::CEAIRFUDPhP}
\item Centre for Advanced Instrumentation, Department of Physics, Durham University, South Road, Durham, DH1 3LE, United Kingdom\label{AFFIL::UDurham}
\item INAF - Osservatorio di Astrofisica e Scienza dello spazio di Bologna, Via Piero Gobetti 93/3, 40129  Bologna, Italy\label{AFFIL::OASBologna}
\item University of Geneva - D\'epartement de physique nucl\'eaire et corpusculaire, 24 rue du G\'en\'eral-Dufour, 1211 Gen\`eve 4, Switzerland\label{AFFIL::UGenevaDPNC}
\item CCTVal, Universidad T\'ecnica Federico Santa Mar{\'\i}a, Avenida Espa\~na 1680, Valpara{\'\i}so, Chile\label{AFFIL::UTecnicaFedericoSantaMaria}
\item The Henryk Niewodnicza\'nski Institute of Nuclear Physics, Polish Academy of Sciences, ul. Radzikowskiego 152, 31-342 Cracow, Poland\label{AFFIL::IFJ}
\item INAF - Osservatorio Astronomico di Capodimonte, Via Salita Moiariello 16, 80131 Napoli, Italy\label{AFFIL::OCapodimonte}
\item Aix Marseille Univ, CNRS/IN2P3, CPPM, Marseille, France\label{AFFIL::CPPMUAixMarseille}
\item Universit\'e Paris Cit\'e, CNRS, CEA, Astroparticule et Cosmologie, F-75013 Paris, France\label{AFFIL::APCUParisCite}
\item University of the Witwatersrand, 1 Jan Smuts Avenue, Braamfontein, 2000 Johannesburg, South Africa\label{AFFIL::UWitwatersrand}
\item INFN Sezione di Torino, Via P. Giuria 1, 10125 Torino, Italy\label{AFFIL::INFNTorino}
\item Dipartimento di Fisica - Universit\'a degli Studi di Torino, Via Pietro Giuria 1 - 10125 Torino, Italy\label{AFFIL::UTorino}
\item Palack\'y University Olomouc, Faculty of Science, Joint Laboratory of Optics of Palack\'y University and Institute of Physics of the Czech Academy of Sciences, 17. listopadu 1192/12, 779 00 Olomouc, Czech Republic\label{AFFIL::UOlomouc}
\item INAF - Osservatorio Astrofisico di Catania, Via S. Sofia, 78, 95123 Catania, Italy\label{AFFIL::OCatania}
\item University of Oxford, Department of Physics, Clarendon Laboratory, Parks Road, Oxford, OX1 3PU, United Kingdom\label{AFFIL::UOxford}
\item INAF - Istituto di Astrofisica Spaziale e Fisica Cosmica di Milano, Via A. Corti 12, 20133 Milano, Italy\label{AFFIL::IASFMilano}
\item LUTH, GEPI and LERMA, Observatoire de Paris, Universit\'e PSL, Universit\'e Paris Cit\'e, CNRS, 5 place Jules Janssen, 92190, Meudon, France\label{AFFIL::ObservatoiredeParis}
\item INAF - Istituto di Radioastronomia, Via Gobetti 101, 40129 Bologna, Italy\label{AFFIL::RadioastronomiaINAF}
\item INAF - Istituto Nazionale di Astrofisica, Viale del Parco Mellini 84, 00136 Rome, Italy\label{AFFIL::INAF}
\item Instituto de Astronomia, Geof{\'\i}sico, e Ci\^encias Atmosf\'ericas - Universidade de S\~ao Paulo, Cidade Universit\'aria, R. do Mat\~ao, 1226, CEP 05508-090, S\~ao Paulo, SP, Brazil\label{AFFIL::IAGUSaoPaulo}
\item Instituto de F{\'\i}sica de S\~ao Carlos, Universidade de S\~ao Paulo, Av. Trabalhador S\~ao-carlense, 400 - CEP 13566-590, S\~ao Carlos, SP, Brazil\label{AFFIL::IFSCUSaoPaulo}
\item Max-Planck-Institut f\"ur Kernphysik, Saupfercheckweg 1, 69117 Heidelberg, Germany\label{AFFIL::MPIK}
\item Universit\'a degli Studi di Napoli {\textquotedblleft}Federico II{\textquotedblright} - Dipartimento di Fisica {\textquotedblleft}E. Pancini{\textquotedblright}, Complesso universitario di Monte Sant'Angelo, Via Cintia - 80126 Napoli, Italy\label{AFFIL::UNapoli}
\item INFN Sezione di Bari and Universit\`a degli Studi di Bari, via Orabona 4, 70124 Bari, Italy\label{AFFIL::UandINFNBari}
\item Institut f\"ur Astronomie und Astrophysik, Universit\"at T\"ubingen, Sand 1, 72076 T\"ubingen, Germany\label{AFFIL::IAAT}
\item Universit\'e Bordeaux, CNRS, LP2I Bordeaux, UMR 5797, 19 Chemin du Solarium, F-33170 Gradignan, France\label{AFFIL::LP2I}
\item Department of Astronomy and Astrophysics, University of Chicago, 5640 S Ellis Ave, Chicago, Illinois, 60637, USA\label{AFFIL::UChicagoDAA}
\item LAPTh, CNRS, USMB, F-74940 Annecy, France\label{AFFIL::LAPTh}
\item Institut f\"ur Physik \& Astronomie, Universit\"at Potsdam, Karl-Liebknecht-Strasse 24/25, 14476 Potsdam, Germany\label{AFFIL::UPotsdam}
\item Escola de Artes, Ci\^encias e Humanidades, Universidade de S\~ao Paulo, Rua Arlindo Bettio, CEP 03828-000, 1000 S\~ao Paulo, Brazil\label{AFFIL::EACHUSaoPaulo}
\item Astronomical Observatory of Taras Shevchenko National University of Kyiv, 3 Observatorna Street, Kyiv, 04053, Ukraine\label{AFFIL::AstObsofUKyiv}
\item University of California, Davis, One Shields Ave., Davis, CA 95616, USA\label{AFFIL::UCaliforniaDavis}
\item RIKEN, Institute of Physical and Chemical Research, 2-1 Hirosawa, Wako, Saitama, 351-0198, Japan\label{AFFIL::RIKEN}
\item Western Sydney University, Locked Bag 1797, Penrith, NSW 2751, Australia\label{AFFIL::UWesternSydney}
\item INAF - Istituto di Astrofisica e Planetologia Spaziali (IAPS), Via del Fosso del Cavaliere 100, 00133 Roma, Italy\label{AFFIL::IAPS}
\item Department of Physics, Nagoya University, Chikusa-ku, Nagoya, 464-8602, Japan\label{AFFIL::UNagoya}
\item INFN Sezione di Pisa, Edificio C {\textendash} Polo Fibonacci, Largo Bruno Pontecorvo 3, 56127 Pisa\label{AFFIL::INFNPisa}
\item INFN Sezione di Roma Tor Vergata, Via della Ricerca Scientifica 1, 00133 Rome, Italy\label{AFFIL::INFNRomaTorVergata}
\item Alikhanyan National Science Laboratory, Yerevan Physics Institute, 2 Alikhanyan Brothers St., 0036, Yerevan, Armenia\label{AFFIL::NSLAlikhanyan}
\item INFN Sezione di Catania, Via S. Sofia 64, 95123 Catania, Italy\label{AFFIL::INFNCatania}
\item Universidade Federal Do Paran\'a - Setor Palotina, Departamento de Engenharias e Exatas, Rua Pioneiro, 2153, Jardim Dallas, CEP: 85950-000 Palotina, Paran\'a, Brazil\label{AFFIL::UFPR}
\item N\'ucleo de Astrof{\'\i}sica e Cosmologia (Cosmo-ufes) \& Departamento de F{\'\i}sica, Universidade Federal do Esp{\'\i}rito Santo (UFES), Av. Fernando Ferrari, 514. 29065-910. Vit\'oria-ES, Brazil\label{AFFIL::UFES}
\item Astrophysics Research Center of the Open University (ARCO), The Open University of Israel, P.O. Box 808, Ra{\textquoteright}anana 4353701, Israel\label{AFFIL::OpenUniversityofIsrael}
\item Department of Physics, The George Washington University, Washington, DC 20052, USA\label{AFFIL::GWUWashingtonDC}
\item FZU - Institute of Physics of the Czech Academy of Sciences, Na Slovance 1999/2, 182 21 Praha 8, Czech Republic\label{AFFIL::FZU}
\item National Institute of Technology, Ichinoseki College, Hagisho, Ichinoseki, Iwate 021-8511, Japan\label{AFFIL::NITIchinoseki}
\item Universidad Nacional Aut\'onoma de M\'exico, Delegaci\'on Coyoac\'an, 04510 Ciudad de M\'exico, Mexico\label{AFFIL::UNAMMexico}
\item Department of Physics and Astronomy and the Bartol Research Institute, University of Delaware, Newark, DE 19716, USA\label{AFFIL::UDelaware}
\item Universit\"at Innsbruck, Institut f\"ur Astro- und Teilchenphysik, Technikerstr. 25/8, 6020 Innsbruck, Austria\label{AFFIL::UInnsbruck}
\item Dipartimento di Scienze Fisiche e Chimiche, Universit\`a degli Studi dell'Aquila and GSGC-LNGS-INFN, Via Vetoio 1, L'Aquila, 67100, Italy\label{AFFIL::UandINFNAquila}
\item Astronomical Observatory, Jagiellonian University, ul. Orla 171, 30-244 Cracow, Poland\label{AFFIL::UJagiellonian}
\item Friedrich-Alexander-Universit\"at Erlangen-N\"urnberg, Erlangen Centre for Astroparticle Physics, Nikolaus-Fiebiger-Str. 2, 91058 Erlangen, Germany\label{AFFIL::UErlangenECAP}
\item Astronomical Institute of the Czech Academy of Sciences, Bocni II 1401 - 14100 Prague, Czech Republic\label{AFFIL::ASU}
\item Faculty of Science, Ibaraki University, Mito, Ibaraki, 310-8512, Japan\label{AFFIL::UIbaraki}
\item Institut de Fisica d'Altes Energies (IFAE), The Barcelona Institute of Science and Technology, Campus UAB, 08193 Bellaterra (Barcelona), Spain\label{AFFIL::IFAEBIST}
\item Institut de Recherche en Astrophysique et Plan\'etologie, CNRS-INSU, Universit\'e Paul Sabatier, 9 avenue Colonel Roche, BP 44346, 31028 Toulouse Cedex 4, France\label{AFFIL::IRAPUToulouse}
\item Dept. of Physics and Astronomy, University of Leicester, Leicester, LE1 7RH, United Kingdom\label{AFFIL::ULeicester}
\item Sorbonne Universit\'e, CNRS/IN2P3, Laboratoire de Physique Nucl\'eaire et de Hautes Energies, LPNHE, 4 place Jussieu, 75005 Paris, France\label{AFFIL::LPNHEUSorbonne}
\item Universit\`a degli studi di Catania, Dipartimento di Fisica e Astronomia {\textquotedblleft}Ettore Majorana{\textquotedblright}, Via S. Sofia 64, 95123 Catania, Italy\label{AFFIL::UCatania}
\item Finnish Centre for Astronomy with ESO, University of Turku, Finland, FI-20014 University of Turku, Finland\label{AFFIL::UTurku}
\item Department of Physics, Humboldt University Berlin, Newtonstr. 15, 12489 Berlin, Germany\label{AFFIL::UBerlin}
\item INFN Sezione di Trieste and Universit\`a degli Studi di Trieste, Via Valerio 2 I, 34127 Trieste, Italy\label{AFFIL::UandINFNTrieste}
\item Escuela Polit\'ecnica Superior de Ja\'en, Universidad de Ja\'en, Campus Las Lagunillas s/n, Edif. A3, 23071 Ja\'en, Spain\label{AFFIL::UJaen}
\item Anton Pannekoek Institute/GRAPPA, University of Amsterdam, Science Park 904 1098 XH Amsterdam, The Netherlands\label{AFFIL::UAmsterdam}
\item Saha Institute of Nuclear Physics, Bidhannagar, Kolkata-700 064, India\label{AFFIL::SahaInstitute}
\item Dipartimento di Fisica e Chimica {\textquotedblleft}E. Segr\`e{\textquotedblright}, Universit\`a degli Studi di Palermo, Via Archirafi 36, 90123, Palermo, Italy\label{AFFIL::UPalermo}
\item INFN and Universit\`a degli Studi di Siena, Dipartimento di Scienze Fisiche, della Terra e dell'Ambiente (DSFTA), Sezione di Fisica, Via Roma 56, 53100 Siena, Italy\label{AFFIL::USienaandINFN}
\item Department of Physics, Columbia University, 538 West 120th Street, New York, NY 10027, USA\label{AFFIL::BarnardCollegeColumbiaUniversity}
\item Department of Physics, Yamagata University, Yamagata, Yamagata 990-8560, Japan\label{AFFIL::UYamagata}
\item University of Bia{\l}ystok, Faculty of Physics, ul. K. Cio{\l}kowskiego 1L, 15-245 Bia{\l}ystok, Poland\label{AFFIL::UBiaystok}
\item Department of Physics, Tokai University, 4-1-1, Kita-Kaname, Hiratsuka, Kanagawa 259-1292, Japan\label{AFFIL::UTokai}
\item Charles University, Institute of Particle \& Nuclear Physics, V Hole\v{s}ovi\v{c}k\'ach 2, 180 00 Prague 8, Czech Republic\label{AFFIL::UPrague}
\item Astronomical Observatory of Ivan Franko National University of Lviv, 8 Kyryla i Mephodia Street, Lviv, 79005, Ukraine\label{AFFIL::AstObsofULviv}
\item Institute for Space{\textemdash}Earth Environmental Research, Nagoya University, Furo-cho, Chikusa-ku, Nagoya 464-8601, Japan\label{AFFIL::UNagoyaISEE}
\item Kobayashi{\textemdash}Maskawa Institute for the Origin of Particles and the Universe, Nagoya University, Furo-cho, Chikusa-ku, Nagoya 464-8602, Japan\label{AFFIL::UNagoyaKMI}
\item Department of Physics and Astronomy, University of California, Los Angeles, CA 90095, USA\label{AFFIL::UCLA}
\item Graduate School of Technology, Industrial and Social Sciences, Tokushima University, Tokushima 770-8506, Japan\label{AFFIL::UTokushima}
\item Cherenkov Telescope Array Observatory, Saupfercheckweg 1, 69117 Heidelberg, Germany\label{AFFIL::CTAOHeidelberg}
\item INAF - Istituto di Astrofisica Spaziale e Fisica Cosmica di Palermo, Via U. La Malfa 153, 90146 Palermo, Italy\label{AFFIL::IASFPalermo}
\item University of Pisa, Largo B. Pontecorvo 3, 56127 Pisa, Italy \label{AFFIL::UPisa}
\item University of Rijeka, Faculty of Physics, Radmile Matejcic 2, 51000 Rijeka, Croatia\label{AFFIL::URijeka}
\item INAF - Osservatorio Astronomico di Padova, Vicolo dell'Osservatorio 5, 35122 Padova, Italy\label{AFFIL::OPadova}
\item International Institute of Physics, Universidade Federal do Rio Grande do Norte, 59078-970, Natal, RN, Brasil\label{AFFIL::URioGrandedoNorteIIP}
\item Departamento de F{\'\i}sica, Universidade Federal do Rio Grande do Norte, 59078-970, Natal, RN, Brasil\label{AFFIL::URioGrandedoNortePhys}
\item Landessternwarte, Zentrum f\"ur Astronomie  der Universit\"at Heidelberg, K\"onigstuhl 12, 69117 Heidelberg, Germany\label{AFFIL::LSW}
\item Centre for Astro-Particle Physics (CAPP) and Department of Physics, University of Johannesburg, PO Box 524, Auckland Park 2006, South Africa\label{AFFIL::UJohannesburg}
\item Departamento de Astronom{\'\i}a, Universidad de Concepci\'on, Barrio Universitario S/N, Concepci\'on, Chile\label{AFFIL::UdeConcepcion}
\item INAF - Osservatorio Astronomico di Brera, Via Brera 28, 20121 Milano, Italy\label{AFFIL::OBrera}
\item Main Astronomical Observatory of the National Academy of Sciences of Ukraine, Zabolotnoho str., 27, 03143, Kyiv, Ukraine\label{AFFIL::ObsNASUkraine}
\item Space Technology Centre, AGH University of Science and Technology, Aleja Mickiewicza, 30, 30-059, Krak\'ow, Poland\label{AFFIL::AGHCracowSTC}
\item Academic Computer Centre CYFRONET AGH, ul. Nawojki 11, 30-950, Krak\'ow, Poland\label{AFFIL::CYFRONETAGH}
\item Cherenkov Telescope Array Observatory gGmbH, Via Gobetti, Bologna, Italy\label{AFFIL::CTAOBologna}
\item Department of Physical Science, Hiroshima University, Higashi-Hiroshima, Hiroshima 739-8526, Japan\label{AFFIL::UHiroshima}
\item Institute of Space Sciences (ICE, CSIC), and Institut d'Estudis Espacials de Catalunya (IEEC), and Instituci\'o Catalana de Recerca I Estudis Avan\c{c}ats (ICREA), Campus UAB, Carrer de Can Magrans, s/n 08193 Cerdanyola del Vall\'es, Spain\label{AFFIL::ICECSIC}
\item INAF - Osservatorio Astrofisico di Torino, Strada Osservatorio 20, 10025  Pino Torinese (TO), Italy\label{AFFIL::OTorino}
\item School of Physical Sciences, University of Adelaide, Adelaide SA 5005, Australia\label{AFFIL::UAdelaide}
\item Department of Physical Sciences, Aoyama Gakuin University, Fuchinobe, Sagamihara, Kanagawa, 252-5258, Japan\label{AFFIL::UAoyamaGakuin}
\item School of Physics and Astronomy, Sun Yat-sen University, Zhuhai, China\label{AFFIL::USunYatsen}
\end{enumerate}


\bsp	
\label{lastpage}
\end{document}